\documentclass[letterpaper,10pt,twocolumn,twoside,journal]{IEEEtran}
\usepackage{times}
\usepackage{setspace}
\usepackage{url}
\spacing{1}
\usepackage[utf8]{inputenc}
\usepackage[T1]{fontenc}
\usepackage{graphicx}		% For \includegraphics
\usepackage{wrapfig}
\usepackage[format=plain,font=footnotesize,labelfont=bf,labelsep=period]{caption}
\usepackage{sidecap} % For sidecaptions for figures
\usepackage{subfig}
\usepackage[export]{adjustbox}
\usepackage[font=small]{caption}
\usepackage{float}

\usepackage{amsmath} % assumes amsmath package installed
\usepackage{amssymb}  % assumes amsmath package installed
\usepackage{amsthm}
\usepackage{mathtools}
\usepackage[normalem]{ulem}
\usepackage{paralist}	% For enumerations with better titled numbers
\usepackage[space]{grffile} % Space in filenames
\usepackage{color}
\let\labelindent\relax
\usepackage{enumitem}
\usepackage{bm}
\usepackage{cancel}
\usepackage{hhline}
\usepackage{cite}
\usepackage{multirow}

\newtheorem{theorem}{Theorem}

\theoremstyle{definition}
\newtheorem{definition}{Definition}
\theoremstyle{remark}
\newtheorem{remark}{Remark}
\theoremstyle{definition}

\theoremstyle{definition}

\newtheorem{example}{Example}

\newcommand{\R}{\mathbb{R}}
\newcommand{\C}{\mathcal{C}}

\newcommand{\K}{\mathcal{K}}

\definecolor{blue}{RGB}{38,38,134}
\definecolor{darkblue}{RGB}{0,0,102}
\definecolor{lightblue}{RGB}{77,77,148}

\definecolor{gold}{RGB}{234, 170, 0}
\definecolor{metallic_gold}{RGB}{139, 111, 78}

\renewcommand{\cal}[1]{\mathcal{ #1 }}
\newcommand{\mb}[1]{\mathbf{ #1 }}

\newcommand{\der}[2]{\frac{\mathrm{d} #1 }{\mathrm{d} #2 }}
\newcommand{\derp}[2]{\frac{\partial #1 }{\partial #2 }}

\DeclareMathOperator*{\argmin}{argmin}

\usepackage{xfrac}

\usepackage{dblfloatfix}

\title{Control Barrier Functions and Input-to-State Safety with Application to Automated Vehicles}

\author{Anil Alan$^{1}$, Andrew J. Taylor$^{2}$, Chaozhe R. He$^{1,3}$, Aaron D. Ames$^{2}$, and G\'abor Orosz$^{1,4}$ 
\thanks{This research is supported in part by the National Science Foundation, 
CPS Award \#1932091.}
\thanks{$^{1}$A. Alan, C. R. He, and G. Orosz are with the Department of Mechanical Engineering, University of Michigan, Ann Arbor, MI 48109, USA ${\tt\small \{anilalan,hchaozhe, orosz\}@umich.edu}$}%
\thanks{$^{2}$A. J. Taylor and A. D. Ames are with the Department of Computing \& Mathematical Sciences, California Institute of Technology, Pasadena, CA 91125, USA ${\tt\small \{ajtaylor, ames\}@caltech.edu}$}%
\thanks{$^{3}$C. R. He is also with {Plus.ai Inc.}, Cupertino, CA 95014, USA 
${\tt\small chaozhe.he@plus.ai}$  }
\thanks{$^{4}$G. Orosz is also with the Department of Civil and Environmental Engineering, University of Michigan, Ann Arbor, MI 48109, USA}%
}

\begin{document}

\maketitle
\thispagestyle{empty}
% \pagestyle{plain}

%%%%%%%%%%%%%%%%%%%%%%%%%%%%%%%%%%%%%%%%%%%%%%%%%%%%%%%%%%%%%%%%%%%%%%%%%%%%%%%%%
\begin{abstract}
Balancing safety and performance is one of the predominant challenges in modern control system design. Moreover, it is crucial to robustly ensure safety without inducing unnecessary conservativeness that degrades performance. In this work we present a constructive approach for safety-critical control synthesis via \textit{Control Barrier Functions} (CBF). By filtering a hand-designed controller via a CBF, we are able to attain performant behavior while providing rigorous guarantees of safety. In the face of disturbances, robust safety and performance are simultaneously achieved through the notion of \textit{Input-to-State Safety} (ISSf). We take a tutorial approach by developing the CBF-design methodology in parallel with an inverted pendulum example, making the challenges and sensitivities in the design process concrete. To establish the capability of the proposed approach, we consider the practical setting of safety-critical design via CBFs for a \textit{connected automated vehicle} (CAV) in the form of a class-8 truck without a trailer. Through experimentation we see the impact of unmodeled disturbances in the truck's actuation system on the safety guarantees provided by CBFs. We characterize these disturbances and using ISSf, produce a robust controller that achieves safety without conceding performance. We evaluate our design both in simulation, and for the first time on an automotive system, experimentally.

\end{abstract}
%%%%%%%%%%%%%%%%%%%%%%%%%%%%%%%%%%%%%%%%%%%%%%%%%%%%%%%%%%%%%%%%%%%%%%%%%%%%%%%%%

%%%%%%%%%%%%%%%%%%%%%%%%%%%%%%%%%%%%%%%%%%%%%%%%%%%%%%%%%%%%%%%%%%%%%%%%%%%%%%%%%
%%%%%%%%%%%%%%%%%%%%%%%%%%%%%%%%%%%%%%%%%%%%%%%%%%%%%%%%%%%%%%%%%%%%%%%%%%%%%%%%%

\begin{figure*}[t]
	\centering
	\includegraphics[width=\linewidth]{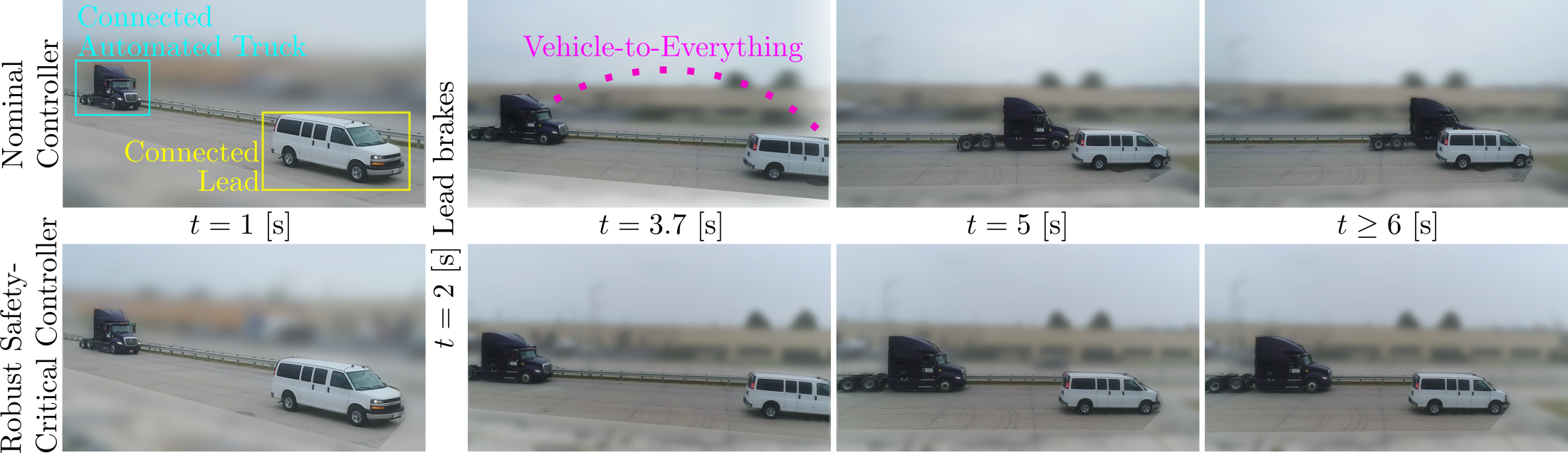}
	\caption{Experimental configuration for heavy-duty CAV problem. (Top) Controller design without robustifying element yields safety violation and collision. (Bottom) Robust safety-critical controller ensures CAV brakes early and aggressively enough to maintain safe distance.}
	\label{fig:intro_truck}
\end{figure*}

%%%%%%%%%%%%%%%%%%%%%%%%%%%%%%%%%%%%%%%%%%%%%%%%%%%%%%%%%%%%%%%%%%%%%%%%%%%%%%%%%
\section{Introduction} \label{sec:intro}
%%%%%%%%%%%%%%%%%%%%%%%%%%%%%%%%%%%%%%%%%%%%%%%%%%%%%%%%%%%%%%%%%%%%%%%%%%%%%%%%%

% Safety is of critical importance, but need to balance with design for performance. Control Barrier Functions practical applications 

\par Safety is an ever more pressing requirement for modern control systems as they are deployed into increasingly complex real-world environments. Simultaneously, meeting performance requirements is a major driving factor in control system design. As these two objectives may naturally oppose each other, it is necessary to consider an \textit{active} approach for enforcing safety that impacts performance only when it is critical for the safety of the system \cite{gurriet2018online, hobbs2021run}. \textit{Control Barrier Functions} (CBFs) have been demonstrated to be a powerful tool for constructively synthesizing controllers that yield strong performance and intervene only when safety is at risk of being compromised \cite{ames2014control, ames2017control, ames2019control}. The utility of CBFs has been confirmed by their experimental application on real-world control systems, including mobile robots \cite{xu2017realizing, gurriet2018online}, robotic swarms \cite{wang2017safety}, autonomous aerial vehicles \cite{wang2018safe}, robotic arms \cite{singletary2022safety}, robotic manipulators \cite{cortez2019control}, quadrupedal robots \cite{grandia2021multi}, and bipedal robots \cite{csomay2021episodic}, as well as simulation results on automotive systems \cite{ames2014control}, autonomous naval vehicles \cite{thyri2020reactive}, and spacecraft \cite{mote2021optimization}. The variety in this collection of results indicates that CBFs capture fundamental concepts underlying the notion of safety, irrespective of a specific domain, and suggests that CBFs are a valuable tool to consider in the process of modern control system design.

% CBFs overview
% \cite{ames2014control, ames2017control, ames2019control}

% Mobile robots
% \cite{xu2017realizing, gurriet2018online}

% Swarms
% \cite{wang2017safety}
% \cite{borrmann2015control}

% Quadrotors
% \cite{wang2018safe}

% Grasping 
% \cite{cortez2019control}

% Manipulators
% \cite{singletary2019online}

% Quadrupeds
% \cite{grandia2021multi}

% Bipeds
% \cite{csomay2021episodic}

% Naval vehicles
% \cite{thyri2020reactive}

%% Just simulation
% Bipeds - \cite{hsu2015control, nguyen2020dynamics}

% Brief theory discussion

\par One of the appealing features of the CBF-based methodology for safety-critical control synthesis is the relatively intuitive nature of the theoretical safety guarantees they endow a system with. The study of \textit{set invariance}, or the state of a system remaining within a prescribed set, has long been of interest in the study of dynamic systems \cite{nagumo1942lage} and control \cite{blanchini2008set}. The foundational work in \cite{prajna2004safety} proposed the notion of a \textit{barrier function} as a tool for checking the invariance of a set given a model of the system dynamics. In simple terms\footnote{We recommend the reader to \cite{konda2020characterizing} for a comprehensive mathematical study of the connections between barrier functions and set invariance.}, a barrier function takes positive values for states inside a set, and is zero on the boundary of the set. If the time derivative of the barrier function is positive on the boundary of the set, the value of barrier must grow and the system thus must remain in the set. This idea was quickly adapted to the context of control synthesis, yielding CBFs and a means to constructively achieve set invariance. Synthesis was first proposed through structured controllers \cite{wieland2007constructive}, but was later expanded using convex optimization to produce \textit{safety-filters} that minimally modify a hand-designed controller to ensure safety \cite{ames2014control, ames2017control, gurriet2018online}. The combination of intuitive theoretical concepts with relatively simple control synthesis techniques promoted rapid development of CBFs, including formulations for higher-order systems \cite{nguyen2016exponential, xiao2019control, tan2021high} and discrete-time systems \cite{agrawal2017discrete}, as well as constructive tools for synthesizing CBFs \cite{wang2018permissive,robey2020learning,clark2021verification} and methods for sets with complex geometries \cite{glotfelter2018boolean, notomista2021safety}. 

% Robustness

\par Inherent in the theoretical safety guarantees provided by CBFs is a dependence on the model of the system dynamics, thus raising subsequent questions of robustness. Resulting works have explored robustness to disturbances \cite{xu2015robustness, jankovic2018robust, takano2018robust, kolathaya2018input, garg2021robust, alan2021safe, choi2021robust}, measurement errors \cite{dean2020guaranteeing}, unmodeled dynamics \cite{seiler2021control}, and sector-bounded uncertainties \cite{buch2021robust}. The early work in \cite{xu2015robustness} noticed a robustness to disturbances inherent in CBFs, which drawing inspiration from the notion of Input-to-State Stability frequently seen when considering robust stabilization of nonlinear systems \cite{sontag2008input}, was formalized into the idea of \textit{Input-to-State Safety} (ISSf) in \cite{kolathaya2018input}. Instead of trying to keep a specific set invariant in the presence of disturbances as in \cite{jankovic2018robust, takano2018robust}, which may induce conservativeness and degrade the performance of a controller, ISSf quantifies how the set kept invariant grows in the presence of disturbances. Moreover, it provides a simple modification for CBF-based controllers to control this growth, which was extended in \cite{alan2021safe} to permit greater performance while maintaining meaningful safety guarantees. As we demonstrate in this work, this paradigm for robust safety naturally lends itself to the design-test-redesign process, as the growth of the invariant set can be tuned to satisfy safety requirements while meeting performance metrics.  

% Connected Automated Vehicles

\par Despite the fact that CBFs were initially presented as a tool for safety-critical control synthesis for automotive systems \cite{ames2014control, xu2015robustness}, they have yet to be experimentally realized on them. A primary challenge in using CBFs to ensure safety for a complex system such as a full-scale \textit{connected automated vehicle} (CAV) lies in addressing discrepancies between the system model and the real-world system. In the context of a heavy-duty CAV, a significant portion of these discrepancies arise due to simplified models of the complex interactions within the CAV braking elements \cite{sridhar2021antilock}, and manifest as disturbances in the input applied to the system. Accounting for these complicated interactions in the controller design may greatly increase the intricacy of the resulting controller, but completely ignoring them may yield safety violations under critical conditions such as a harsh brake from a preceding vehicle as seen in Fig.~\ref{fig:intro_truck} (top). Thus, balancing the complexity of the model used in design with the need to satisfy safety requirements is a challenging yet appropriate setting to deploy robust CBF-based control design.

% Contributions
 
\par There are two main contributions in this paper. The first is a tutorial presentation of a robust safety-critical design methodology using CBFs and ISSf. Concepts are introduced in parallel with an inverted pendulum example, thus providing a concrete context for readers to quickly establish an understanding of the relevant details in safety-critical control synthesis. We provide an appropriate level of theoretical discussion to clearly state the theoretical safety guarantees achieved with this control paradigm, but focus predominantly on the practical challenges and trade-offs encountered in safety-critical control design. Compared to the original works \cite{ames2014control, ames2017control} and overview work \cite{ames2019control} on CBFs, we believe that this presentation provides a more approachable introduction to the topic of safety-critical control synthesis for practitioners. Moreover, all details necessary to exactly recreate the simulation results in the inverted pendulum example are provided. 

The second contribution of this work is a more practical application of the presented safety-critical control design methodology that considers a heavy-duty CAV, seen in Figure \ref{fig:intro_truck}. We highlight the entire process of safety-critical control design including system modeling, specification of safety requirements via a CBF, nominal performance-based controller design, simulation, and experimental testing on a full-scale automated class-8 tractor. The impacts of unmodeled disturbances seen in experimental results are quantified and used to robustify the safety-critical controller, which is subsequently implemented in simulation and experimentally. We believe that combined tutorial presentation and the proposed design-test-redesign process on a challenging real-world system is precisely the approach necessary to advance CBF-based control design from the academic setting to a tool useful for the practicing control engineer.

% Organization

\par The organization of this paper is as follows. In Section \ref{sec:background} we present the safety-critical control problem, review CBFs, and explore how a nominal controller may be modified via CBFs to endow a system with theoretical safety guarantees. In Section \ref{sec:issf} we introduce disturbances into the input to the system, and explore how these impact theoretical safety guarantees through the lens of ISSf. Moreover, we present a simple framework for robustly modifying a CBF and the resulting controller design to provide a measure of control over how these safety guarantees degrade. In Section \ref{sec:truck} the connected automated vehicle problem is presented considering an automated heavy-duty vehicle. A CBF specified to encode safety for the CAV and a hand-designed nominal controller are incorporated into a safety-critical controller that is evaluated in simulation and verified to ensure safety. In Section \ref{sec:exp} we deploy the controller experimentally, and see how unmodeled disturbances lead to degradation of safety guarantees. We characterize these disturbances and robustify the controller design, and lastly verify the ability of this controller to meet safety requirements both in simulation and experiments.

% Control Barrier Functions theory overview, extensions to higher-order systems, discrete systems, synthesis, non-smooth systems,

% \cite{konda2020characterizing}
% \cite{wang2018permissive,robey2020learning,clark2021verification}
% \cite{glotfelter2017nonsmooth, glotfelter2018boolean, notomista2021safety}

%  MPC+CBFs, STL+CBFs, stochastic systems
% \cite{rosolia2021multirate, zeng2020safety}
% \cite{lindemann2019control,charitidou2021barrier}
% \cite{clark2019control, yaghoubi2020risk, wang2021chance}

% adaptive+learning

% \cite{taylor2019adaptive, lopez2020robust, xiao2021adaptive, cheng2019end, taylor2020learning, choi2020reinforcement, cohen2020approximate}

%%%%%%%%%%%%%%%%%%%%%%%%%%%%%%%%%%%%%%%%%%%%%%%%%%%%%%%%%%%%%%%%%%%%%%%%%%%%%%%%%
\section{Safety-Critical Control}     \label{sec:background}
%%%%%%%%%%%%%%%%%%%%%%%%%%%%%%%%%%%%%%%%%%%%%%%%%%%%%%%%%%%%%%%%%%%%%%%%%%%%%%%%%
In this section we provide a review of safety and Control Barrier Functions (CBFs). These definitions will be used in the formulation of the safety-critical control problem. To make these concepts more concrete, we apply them to an inverted pendulum system.

\subsection{Control Barrier Functions}
Consider the nonlinear control affine system:
\begin{equation}
    \label{eq:eom}
\dot{\mb{x}} = \mb{f}(\mb{x})+\mb{g}(\mb{x})\mb{u},
\end{equation}
with state ${\mb{x}\in\R^n}$, input ${\mb{u}\in\R^m}$, and continuous functions ${\mb{f}:\R^n\to\R^n}$ and ${\mb{g}:\R^n\to\R^{n\times m}}$. Systems described by such equations often appear in robotics, aerospace, power electronics, and automotive systems.

\begin{example}
Consider a control system for an inverted pendulum as depicted in Fig.~\ref{fig:flow_diag}, and described by the model:
\begin{equation}
\label{eq:ivp_model}
\frac{\mathrm{d}}{\mathrm{d}t}\begin{bmatrix} \theta \\ \dot{\theta} \end{bmatrix} = \underbrace{\begin{bmatrix} \dot{\theta} \\ \frac{g}{l}\sin\theta \end{bmatrix}}_{\mb{f}(\mb{x})} + \underbrace{\begin{bmatrix} 0 \\ \frac{1}{ml^2} \end{bmatrix}}_{\mb{g}(\mb{x})} u, 
\end{equation}
with pendulum angle ${\theta\in\R}$ and angular velocity ${\dot{\theta}\in\R}$ defining the state $\mb{x} = [\theta,\dot{\theta}]^\top$, and parameters given by the mass ${m}$, length ${l}$, and gravitational acceleration constant ${g}$. In this example we will use the parameter values ${m=2}$~[kg], ${l=1}$~[m] and ${g=10}$~[m/s$^2$]. The single input ${u\in\R}$ is a torque applied at the pendulum base.
\end{example}

The input $\mb{u}$ is often specified via a state-feedback controller ${\mb{k}:\R^n\to\R^m}$, 
yielding the closed-loop system dynamics:
\begin{equation}
    \label{eq:cloop}
    \dot{\mb{x}} =   \mb{f}(\mb{x})+\mb{g}(\mb{x})\mb{k}(\mb{x}).
\end{equation}
We assume that for any initial condition ${\mb{x}_0 \triangleq \mb{x}(0) \in \R^n}$, there exists a unique solution $\mb{x}(t)$ to \eqref{eq:cloop} for ${t\geq 0}$, such that the system is forward complete \cite{perko2013differential}. 
The notion of safety is formalized by specifying a \textit{safe set} in the state space which the state of the system must remain in to be considered safe. This can be utilized in many practical applications such as distance-keeping \cite{ames2014control}, lane-keeping \cite{xu2017realizing}, and collision avoidance \cite{wang2017safety} of automated vehicles. In particular, consider a set ${\C\subset \R^n}$ defined as the 0-superlevel set of a continuously differentiable function ${h:\R^n \to \R}$, yielding:
\begin{align}
    \C &= \left\{\mb{x} \in \R^n : h(\mb{x}) \geq 0\right\}, \label{eq:safeset}\\
    \partial\C &= \{\mb{x} \in \R^n : h(\mb{x}) = 0\},\label{eq:safesetboundary}\\
    \textrm{Int}(\C) &= \{\mb{x} \in \R^n : h(\mb{x}) > 0\},\label{eq:safesetinterior}
\end{align}
where $\partial\C$ and $\textrm{Int}(\C)$ are the \textit{boundary} and \textit{interior}, respectively, of the set $\C$. We refer to $\C$ as the \textit{safe set}. 
This construction motivates the following definitions of forward invariance and safety:

\begin{figure}[t]
    \centering
    \includegraphics[trim=0 0 0 0,clip, width=0.25\linewidth]{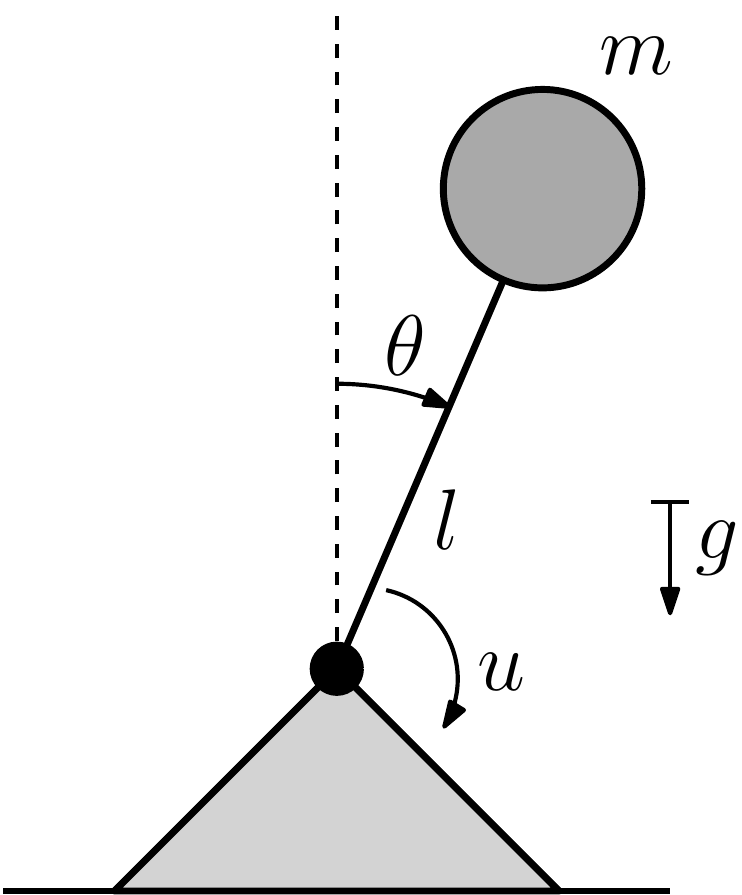}
    \caption{Schematic of an inverted pendulum control system.}
    \label{fig:flow_diag}
    \vspace{-4.5 mm}
\end{figure}

\begin{figure}[b]
\vspace{-4.5 mm}
    \centering
       {\includegraphics[width =0.4\textwidth, valign =t]{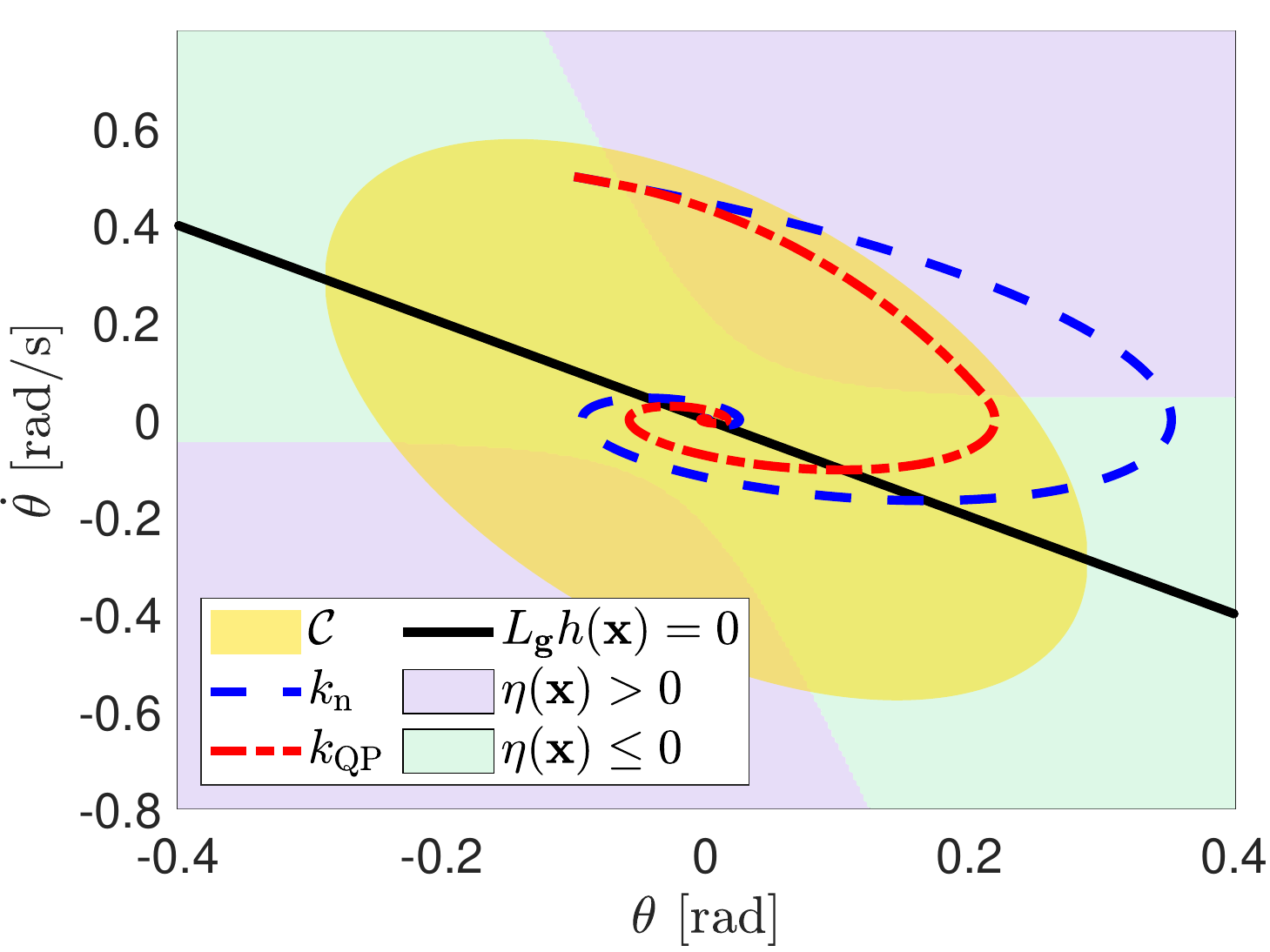}}
       \caption{Simulation results for the inverted pendulum system. The gold ellipse is the safe set $\C$ as defined in \eqref{eq:ivp_C}. The black line is the set where $L_\mb{g}h(\mb{x})=0$ as defined in \eqref{eq:ivp_Lgh0}. The dashed blue line is the trajectory of the system evolving under $k_{\rm n}$ as defined in \eqref{eq:ivp_kn}, which leaves the safe set $\C$. The green and purple regions indicate where the controller $k_{\rm n}$ meets and fails to meet the CBF condition, respectively. The dashed red line is the trajectory of the system evolving under $k_{\rm QP}$ as defined in \eqref{eq:switchstrucmax}-\eqref{eq:switchstrucmin}, which remains inside the safe set $\C$.}
      \label{fig:ivtpndsim}
\end{figure}

\begin{definition}[\textit{Forward Invariance \& Safety}]
A set ${\C\subset\R^n}$ is \textit{forward invariant} if for every ${\mb{x}_0\in\C}$, the solution ${\mb{x}(t)}$ to \eqref{eq:cloop} satisfies ${\mb{x}(t) \in \C}$ for all ${t\geq0}$. The system \eqref{eq:cloop} is \textit{safe} with respect to the set $\C$ if the set $\C$ is forward invariant.
\end{definition}

\begin{example}
A set $\C$ that we wish to keep safe for the inverted pendulum that restricts the angular position and velocity is given by the 0-superlevel set of the function:
\begin{equation}
\label{eq:ivp_h}
    h(\theta,\dot{\theta}) = 1 - \frac{\theta^2}{a^2} - \frac{\dot{\theta}^2}{b^2}-\frac{\theta\dot{\theta}}{ab},
\end{equation}
with parameters ${a,b>0}$. In this example we will use the parameter values ${a = 0.25}$~[rad] and ${b=0.5}$~[rad/s]. The resulting set:
\begin{equation}
\label{eq:ivp_C}
    \C=\left\{ \begin{bmatrix} \theta \\ \dot{\theta} \end{bmatrix} \in \R^2  \left|  1 - \frac{\theta^2}{a^2} - \frac{\dot{\theta}^2}{b^2} - \frac{\theta\dot{\theta}}{ab} \geq 0 \right. \right\},
\end{equation}
is an ellipse as depicted in Fig.~\ref{fig:ivtpndsim} by a gold region.
\end{example}

\begin{figure}[t]
    \centering
    \includegraphics[ width=0.32\textwidth]{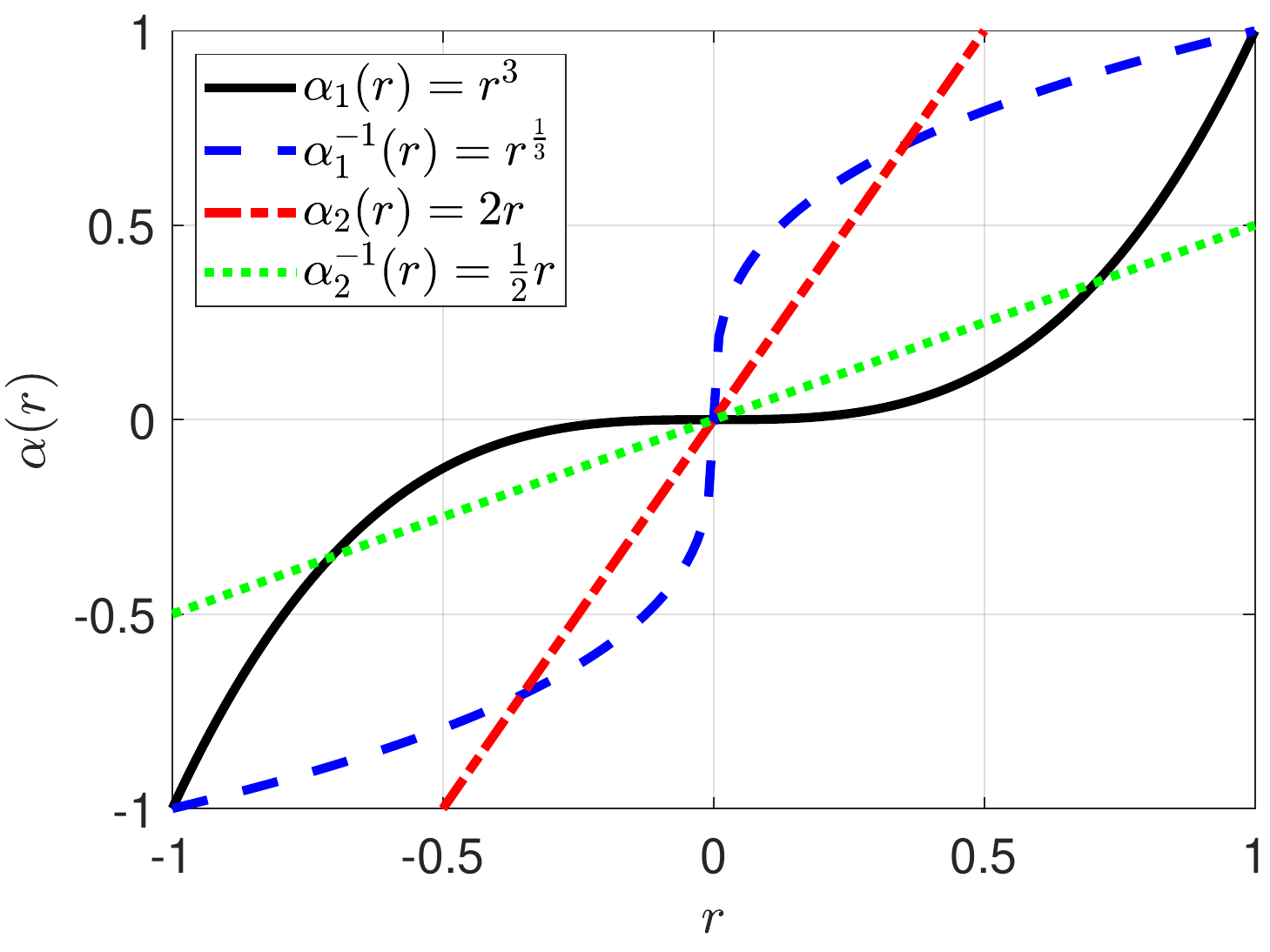}
    \caption{Visualization of class $\cal{K}_\infty^{\rm e}$ functions and their inverses.}
    \label{fig:classK_alt}
    \vspace{-3.5 mm}
\end{figure}

Before defining Control Barrier Functions, we review the following definitions \cite{kellett2014compendium, ames2019control}. 
We denote a continuous function ${\alpha:\R_{\geq 0}\to\R_{\geq 0}}$ as \textit{class $\cal{K}_\infty$} (${\alpha\in\cal{K}_\infty}$) if ${\alpha(0)=0}$, $\alpha$ is strictly increasing and ${\lim_{r\to\infty}\alpha(r)=\infty}$. As an example, any function in the form $\alpha(r)=r^c$ where $c>0$ is class $\cal{K}_\infty$. Note that differentiability is not required for a class $\cal{K}_\infty$ function. Similarly, a continuous function ${\alpha:\R\to\R}$ is said to belong to \textit{extended class $\cal{K}_\infty$} (${\alpha\in\cal{K}_{\infty}^{\rm e}}$) if ${\alpha(0)=0}$, $\alpha$ is strictly increasing, and ${\lim_{r\to\infty}\alpha(r)=\infty}$ and ${\lim_{r\to-\infty}\alpha(r)=-\infty}$. The previous example of ${\alpha(r)=r^c}$ is class $\cal{K}_{\infty}^{\rm e}$ for ${c=1,3,5,\hdots}$ The inverses of class $\cal{K}_\infty$ and class $\cal{K}_{\infty}^{\rm e}$ functions belong to class $\cal{K}_\infty$ and class $\cal{K}_{\infty}^{\rm e}$, respectively. Examples of these functions and their inverses are depicted in Fig.~\ref{fig:classK_alt}. We may use these functions to define Control Barrier Functions.

\begin{definition}[\textit{Control Barrier Function}, \cite{ames2017control}]
Let ${\C\subset\R^n}$ be the 0-superlevel set of a continuously differentiable function ${h:\R^n\to\R}$.
The function $h$ is a \textit{Control Barrier Function} (CBF) for the system \eqref{eq:eom} on $\C$ if there exists ${\alpha\in\K_{\infty}^{\rm e}}$ such that for all ${\mb{x}\in\R^n}$:
%\cut{$\mb{x}\in\C$}:
\begin{equation}
\label{eq:cbf}
    \sup_{\mb{u}\in\R^m}   \Bigg[  \overbrace{ \underbrace{ \derp{h}{\mb{x}}(\mb{x})\mb{f}(\mb{x})}_{L_\mb{f}h(\mb{x})}+\underbrace{\derp{h}{\mb{x}}(\mb{x})\mb{g}(\mb{x})}_{L_\mb{g}h(\mb{x})}\mb{u} 
    }^{\dot{h}(\mb{x},\mb{u})} \Bigg]  > -\alpha(h(\mb{x})).
\end{equation}
\end{definition}

%\begin{figure*}[t]
%\centering
%    \begin{subfloat}
%        {\includegraphics[scale =0.45, valign =t ]{Figures/IP_diagram.pdf}}
%    \end{subfloat}
%    \hfill
%    \hspace*{0.1 cm}
%    \begin{subfloat}
%        {\includegraphics[scale = 0.49, angle = 270]{Figures/flow_diagram.pdf}}
%    \end{subfloat}
%\caption{Schematic of an inverted pendulum (left) and the safety-critical control structure in block diagram (right).}
%\label{fig:flow_diag}
%\end{figure*}

\noindent An equivalent way to express \eqref{eq:cbf} is given in \cite{jankovic2018robust} as:
\begin{equation}
\label{eq:cbf_alt}
    L_\mb{g}h(\mb{x}) = \mb{0} \implies  L_\mb{f}h(\mb{x}) +\alpha(h(\mb{x})) > 0.
\end{equation}
This expression is often an easier condition to evaluate in certifying that a given function is a CBF.

\begin{example}
The function $h$ given as in \eqref{eq:ivp_h} is a CBF for the inverted pendulum system \eqref{eq:ivp_model} on $\C$. 
To see this, consider a function ${\alpha\in\K_{\infty}^{\rm e}}$ defined as ${\alpha(r)=\alpha_{\rm c} r}$ with ${\alpha_{\rm c}>0}$ satisfying ${\alpha_{\rm c} \leq b/a}$. In this example we will take the parameter value ${\alpha_{\rm c} = 0.2}$~[1/s]. Checking the CBF condition defined in \eqref{eq:cbf_alt}, we see that:
\begin{equation}
\label{eq:ivp_Lgh0}
   L_\mb{g}h(\theta_0,\dot{\theta}_0) = 0 \implies \dot{\theta}_0 = -\frac{b}{2a}\theta_0.
\end{equation}
This equation defines a line as depicted in Fig.~\ref{fig:ivtpndsim} by a solid black line. We have that on this line:
\begin{equation*}
  L_\mb{f}h(\theta_0,\dot{\theta}_0) + \alpha(h(\theta_0,\dot{\theta}_0)) = \alpha_{\rm c} + \frac{3}{4a^2}\left(\frac{b}{a}- \alpha_{\rm c} \right) \theta_0^2 > 0,
\end{equation*}
such that the condition \eqref{eq:cbf_alt} is met for our choice of $\alpha_{\rm c}$.

We note that if we consider a set, denoted by $\tilde{\C}$ and defined as the 0-superlevel set of a function ${\tilde{h}:\R^2\to\R}$ given by:
\begin{equation}
    \tilde{h}(\theta,\dot{\theta}) = 1- \frac{\theta^2}{a^2}-\frac{\dot{\theta}^2}{b^2},
\end{equation}
which does not include the term ${\theta\dot{\theta}/ab}$, then the function $\tilde{h}$  is not a CBF for the system \eqref{eq:ivp_model} on $\tilde{\C}$. To see this, note that:
\begin{equation}
{L_\mb{g}\tilde{h}(\theta_0,\dot{\theta}_0) = 0} \implies {\dot{\theta}_0 = 0}.
\end{equation}
In turn, we have that for any ${\alpha\in\K_\infty^{\rm e}}$:
\begin{equation}
L_\mb{f}\tilde{h}(\theta_0,\dot{\theta}_0) + \alpha(\tilde{h}(\theta_0,\dot{\theta}_0)) = \alpha\left(1-\theta_0^2/a^2\right).
\end{equation}
The condition \eqref{eq:cbf_alt} is not satisfied for ${\vert\theta_0\vert\geq a}$ (including ${\vert\theta_0\vert=a}$, which would be in $\partial\tilde{\C}$ and thus in the safe set). Thus it is important to choose the safe set and design the CBF to be compatible with the system dynamics, eliminating points where the CBF condition is not met.
\end{example}

Given a CBF $h$ for \eqref{eq:eom} on $\C$ and a corresponding function ${\alpha\in\cal{K}_{\infty}^{\rm e}}$, we can consider the point-wise set of all control values that satisfy \eqref{eq:cbf}:
\begin{equation}
    K_{\rm CBF}(\mb{x}) = \left\{\mb{u}\in\R^m ~\left|~ \dot{h}(\mb{x},\mb{u})\geq-\alpha(h(\mb{x})) \right. \right\}.
\label{eq:Kcbf}
\end{equation}
One of the main theoretical result for CBFs relates controllers taking values in the set $K_{\rm CBF}$ to the safety of \eqref{eq:cloop} on $\C$:
\begin{theorem}[\cite{ames2017control, konda2020characterizing}]
\label{thm:cbf}
Let $\C\subset\R^n$ be the 0-superlevel set of a continuously differentiable function ${h:\R^n\to\R}$. If $h$ is a CBF for \eqref{eq:eom} on $\C$, then the set ${K_{\rm CBF}(\mb{x})}$ is non-empty for each ${\mb{x}\in\R^n}$, and for any continuous controller ${\mb{k}:\R^n\to\R^m}$ such that ${\mb{k}(\mb{x})\in K_{\rm CBF}(\mb{x})}$ for all ${\mb{x}\in\R^n}$, the system \eqref{eq:cloop} is safe with respect to the set $\C$.
\end{theorem}
Proofs of Theorem~\ref{thm:cbf} may be found in \cite{ames2017control, konda2020characterizing}. We note the distinction of a strict inequality in the CBF condition in \eqref{eq:cbf} and a non-strict inequality in \eqref{eq:Kcbf}. As studied in \cite{jankovic2018robust}, satisfaction of the strict inequality in \eqref{eq:cbf} and \eqref{eq:cbf_alt} is a property of the function $h$ and the dynamics $\mb{f}$ and $\mb{g}$, but not does depend on a specific controller. This property is useful in establishing regularity properties of controllers synthesized with the CBF $h$. In particular, it imposes requirements on the function $h$ when ${L_\mb{g}h(\mb{x})=\mb{0}}$, as seen in the preceding example. In contrast, enforcing safety via Theorem \ref{thm:cbf} only require that the inputs specified by a given controller meet the non-strict inequality in \eqref{eq:Kcbf} (such an input's existence is implied by the CBF condition in \eqref{eq:cbf}).

\subsection{Safety-Critical Controller}
It is often possible to design a controller that achieves a desired degree of performance, but for which it is difficult to verify necessary safety requirements are met.

\begin{example}
Consider a continuous controller ${k_{\rm n}:\R^2\to\R}$ for the inverted pendulum model \eqref{eq:ivp_model} that stabilizes the pendulum to an upright position, given by the feedback linearization \cite{sastry1999nonlinear} or computed torque controller \cite{murray1994mathematical} of the form:
\begin{equation}
\label{eq:ivp_kn}
    k_{\rm n}(\theta,\dot{\theta}) = ml^2\left(-\frac{g}{l}\sin\theta-K_{\rm p}\theta-K_{\rm d}\dot{\theta}\right),
\end{equation}
with controller gains ${K_{\rm p}, K_{\rm d}>0}$. This controller yields the closed-loop system:
\begin{equation}
    \label{eq:ivp_cloop}
    \frac{\mathrm{d}}{\mathrm{d}t}\begin{bmatrix} \theta \\ \dot{\theta} \end{bmatrix} = \begin{bmatrix} 0 & 1 \\ -K_{\rm p} & -K_{\rm d}
    \end{bmatrix}\begin{bmatrix}\theta \\ \dot{\theta} \end{bmatrix},
\end{equation}
such that the upright equilibrium ${\mb{x}^*=\begin{bmatrix}  0,0 \end{bmatrix}^\top}$ is exponentially stable. In this example we will use the parameter values ${K_{\rm p}=0.6}$~[1/s$^2$] and ${K_{\rm d}=0.6}$~[1/s]. We use numerical integration to determine a solution trajectory from the initial condition ${\mb{x}(0)=[-0.1, ~ 0.5]^\top\in\C}$. This trajectory is depicted in Fig.~\ref{fig:ivtpndsim} by a dashed blue curve. Although the controller $k_{\rm n}$ stabilizes the system to the upright position, in doing so it causes the state of the system to leave the safe set $\C$. 
\end{example}

CBFs provide a means for modifying a controller to ensure it explicitly enforces the safety of the system. Suppose that we have a continuous controller ${\mb{k}_{\rm n}:\R^n\to\R^m}$, referred to as the \textit{nominal controller}, that does not necessarily ensure the closed-loop system \eqref{eq:cloop} is safe with respect to the set $\C$, but achieves a desired degree of performance. Furthermore, suppose that we have a CBF $h$ for \eqref{eq:eom} on $\C$ with corresponding function ${\alpha\in\cal{K}_{\infty}^{\rm e}}$. The goal of maintaining the performance of the nominal controller $\mb{k}_{\rm{n}}$ while ensuring the safety of the system \eqref{eq:cloop} with respect to the set $\C$ motivates an optimization-based safety-critical controller ${\mb{k}_{\rm QP}:\R^n\to\R^m}$ defined as:
\begin{align}
\label{eq:SC-QP}
%\tag{SC-QP}
\mb{k}_{\rm QP}(\mb{x}) =  \,\,\underset{\mb{u}\in\R^m}{\argmin}  &  \quad \frac{1}{2} \| \mb{u}-\mb{k}_{\rm n}(\mb{x}) \|_2^2  \\
\mathrm{s.t.} \quad & \quad L_\mb{f}h(\mb{x}) +L_\mb{g}h(\mb{x})\mb{u}
\geq -\alpha(h(\mb{x})). \nonumber
\end{align}
This controller takes the same value as the nominal controller if the nominal controller meets the requirements for safety specified by the CBF $h$, i.e., ${\mb{k}_{\rm QP}(\mb{x})=\mb{k}_{\rm n}(\mb{x})}$ if ${\mb{k}_{\rm n}(\mb{x})\in K_{\rm CBF}(\mb{x})}$. If the nominal controller does not meet the safety requirements, i.e., $\mb{k}_{\rm n}(\mb{x})\notin K_{\rm CBF}(\mb{x})$, the input is chosen to meet the safety requirement with the smallest deviation from the value of $\mb{k}_{\rm n}$. The following theorem describes the feasibility 
of the optimization problem defining this controller, and provides a closed-form solution for the optimization problem:

\begin{theorem}
\label{thm:equiv}
Let $\C$ be the 0-superlevel set of a continuously differentiable function ${h:\R^n\to\R}$, and let ${\mb{k}_{\rm n}:\R^n\to\R^m}$ be a continuous controller. If $h$ is a CBF for \eqref{eq:eom} on the set $\C$ with corresponding function ${\alpha\in\K_{\infty}^{\rm e}}$, then the optimization problem in \eqref{eq:SC-QP} is feasible for any ${\mb{x}\in\R^n}$ and has a closed-form solution given by:
\begin{equation}
\label{eq:SC-CF}
%\tag{SC-CF}
\mb{k}_{\rm QP}(\mb{x}) = \mb{k}_{\rm n}(\mb{x}) + \max\left\{0, \eta(\mb{x}) \right\}L_\mb{g}h(\mb{x})^\top,    
\end{equation}
where the function ${\eta:\R^n\to\R}$ is defined as:
\begin{equation}
\label{eq:lambdastar}
    \eta(\mb{x}) = \begin{cases}  -\frac{L_\mb{f}h(\mb{x})+L_\mb{g}h(\mb{x})\mb{k}_{\rm n}(\mb{x})+\alpha(h(\mb{x}))}{\Vert L_{\mb{g}}h(\mb{x}) \Vert_2^2} \quad & \textrm{if}~ L_\mb{g}h(\mb{x}) \neq \mb{0}, \\  0 & \textrm{if}~ L_\mb{g}h(\mb{x}) = \mb{0}.
    \end{cases}
\end{equation}
Furthermore, $\mb{k}_{\rm QP}$ is continuous and ${\mb{k}_{\rm QP}(\mb{x})\in K_{\textrm{\rm CBF}}(\mb{x})}$ for all ${\mb{x}\in\R^n}$.
\end{theorem}

A proof of this theorem is provided in the appendix. The function $\eta$ only takes positive values ($\eta(\mb{x})>0$) when the nominal controller does not meet safety requirements:
\begin{equation*}
    L_\mb{f}h(\mb{x}) + L_\mb{g}h(\mb{x})\mb{k}_{\rm n}(\mb{x})+\alpha(h(\mb{x})) < 0,
\end{equation*}
and thus the nominal controller $\mb{k}_{\rm n}$ is only modified when it does not satisfy safety requirements. The second case in the definition of the function $\eta$ is presented to resolve the singularity that occurs at ${L_\mb{g}h(\mb{x})=\mb{0}}$ when the closed-form solution \eqref{eq:SC-CF} is implemented. We note that, as stated in Theorem \ref{thm:equiv}, the controller $\mb{k}_{\rm QP}$ is continuous, and thus, this singularity does not produce a large jump in the input. It may even be ignored if the controller is implemented as the optimization problem in \eqref{eq:SC-QP} and numerically solved.

\begin{remark}
For a single input ${(m=1)}$, if ${L_\mb{g}h(\mb{x}) > 0}$ for a particular $\mb{x}\in\R^n$, the controller \eqref{eq:SC-CF} can be expressed as:
\begin{equation}
\label{eq:switchstrucmax}
k_{\rm QP}(\mb{x}) = \max\left\{k_{\rm n}(\mb{x}), -\frac{L_\mb{f}h(\mb{x})+\alpha(h(\mb{x}))}{L_\mb{g}h(\mb{x})}\right\}.
\end{equation}
Similarly, if ${L_\mb{g}h(\mb{x}) < 0}$ for a particular $\mb{x}\in\R^n$, the controller \eqref{eq:SC-CF} reduces to:
\begin{equation}
\label{eq:switchstrucmin}
k_{\rm QP}(\mb{x}) = \min\left\{k_{\rm n}(\mb{x}), -\frac{L_\mb{f}h(\mb{x})+\alpha(h(\mb{x}))}{L_\mb{g}h(\mb{x})}\right\}.
\end{equation}
These controllers can be switched between based on the sign of $L_\mb{g}h(\mb{x})$, with ${k_{\rm QP}(\mb{x})=k_{\rm n}(\mb{x})}$ when ${L_\mb{g}h(\mb{x})=0}$.
\end{remark}

\begin{example}
We deploy the switching controller ${k_{\rm QP}:\R^2\to\R}$ defined in \eqref{eq:switchstrucmax}-\eqref{eq:switchstrucmin} for the inverted pendulum system using the nominal controller ${k_{\rm n}:\R^2\to\R}$ defined in \eqref{eq:ivp_kn}. We use numerical integration to determine a solution trajectory from the initial condition ${\mb{x}(0)=[-0.1, ~ 0.5]^\top\in\C}$. This trajectory is depicted in Fig.~\ref{fig:ivtpndsim} by a dashed red curve.
We see that the controller $k_{\rm QP}$ ensures that the solution trajectory remains within the safe set $\C$ by deviating from the nominal controller in the purple region as specified by \eqref{eq:switchstrucmax}-\eqref{eq:switchstrucmin}.
\end{example}

%%%%%%%%%%%%%%%%%%%%%%%%%%%%%%%%%%%%%%%%%%%%%%%%%%%%%%%%%%%%%%%%%%%%%%%%%%%%%%%%%
\vspace{-0 mm}
\section{Robustness to Disturbance}  \label{sec:issf}
\vspace{-0 mm}
%%%%%%%%%%%%%%%%%%%%%%%%%%%%%%%%%%%%%%%%%%%%%%%%%%%%%%%%%%%%%%%%%%%%%%%%%%%%%%%%%
A challenge frequently encountered when deploying model-based controllers onto real-world systems is a mismatch between the commanded input and the input actually received by the system. This mismatch can arise due to actuator dynamics, actuator delays, input quantization, input saturation, or noise. In the case when a state feedback controller $\mb{k}$ is utilized, any error in state measurements can cause further variation from the ideal control effort.
These imperfections in how control inputs affect the system can lead to degradation in the safety guarantees attained by the safety-critical controller \eqref{eq:SC-CF}. 

In this part we consider a system with an input disturbance:
\begin{equation}
\label{eq:eom_dist}
    \dot{\mb{x}} = \mb{f}(\mb{x}) + \mb{g}(\mb{x})(\mb{u}+\mb{d}(t)),
\end{equation}
where ${\mb{d} : \R_{\geq 0} \to \R^m}$ reflects a time varying disturbance modifying the input $\mb{u}$ (such that the input the system actually receives is ${\mb{u}+\mb{d}(t)}$). We assume that the disturbance is bounded and piecewise continuous\footnote{We take this definition as in \cite{khalil2002nonlinear}, with the existence of one-sided limits.} in time. This is a practical assumption, and determining such bounds on the disturbance is an important step of the control design. This assumption also allows us to define:
\begin{equation}
   {\Vert\mb{d}\Vert_\infty = \sup_{t\geq 0}\Vert\mb{d}(t)\Vert_2 < \infty}.
\end{equation}
Given a continuous controller ${\mb{k}:\R^n\to\R^m}$, we may also introduce the notion of a disturbed closed-loop system:
\begin{equation}
\label{eq:cloop_dist}
    \dot{\mb{x}} = \mb{f}(\mb{x})+\mb{g}(\mb{x})(\mb{k}(\mb{x})+\mb{d}(t)).
\end{equation}
As before, we assume that for any initial condition ${\mb{x}_0\triangleq\mb{x}(0)\in\R^n}$ and any bounded and piecewise continuous disturbance signal ${\mb{d}:\R_{\geq0}\to\R^m}$, there exists a unique solution $\mb{x}(t)$ to \eqref{eq:cloop_dist} for ${t\geq 0}$.

\begin{figure}[t]
    \centering
      {\includegraphics[trim=0 260 0 260,clip,width=.35\textwidth, valign =t]{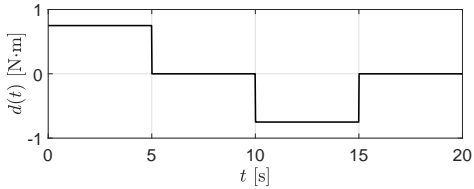}}
      \caption{Disturbance signal for the inverted pendulum system example as defined in \eqref{eq:ivp_dist}.}
      \label{fig:ivtpnddistsig}
      \vspace{-4 mm}
\end{figure}

\begin{example}
We will consider an example disturbance signal for the inverted pendulum specified as:
\begin{equation}
\label{eq:ivp_dist}
    d(t) = M(1-s(t-5)-s(t-10)+s(t-15))
\end{equation}
where ${M\geq 0}$ and ${s:\R\to\R}$ is the \textit{heaviside function}:
\begin{equation}
\label{eq:heaviside}
    s(\tau) = \begin{cases} 0 \qquad &\textrm{if}~\tau<0, \\
              1 \qquad &\textrm{if}~\tau\geq 0. \end{cases}
\end{equation}
With this disturbance we have ${\Vert d \Vert_\infty = M}$. In this example we use the parameter value ${M = 0.75}$~[N$\cdot$m] and the corresponding disturbance signal is depicted in Fig.~\ref{fig:ivtpnddistsig}.
\end{example}

\vspace{-5 mm}
\subsection{Input-to-State Safety}
In the presence of an input disturbance, \textit{Input-to-State Safe Control Barrier Functions (ISSf-CBFs)} provide a tool for designing controllers with a formal safety guarantee \cite{kolathaya2018input, alan2021safe}.
First, we present the notion of input-to-state safety (ISSf) which captures the intuition that it may no longer be possible to render the set $\C$ forward invariant (and thus safe) in the presence of disturbances. Instead, a larger set that scales proportionally with the disturbance may instead be rendered forward invariant. Specifically, consider the set ${\C_\delta\subset\R^n}$ defined as:
\begin{align}
    \C_{\delta} &= \left\{\mb{x} \in \R^n : h(\mb{x})+\gamma(h(\mb{x}),\delta) \geq 0\right\}, \label{eq:Cdelta}\\
    \partial\C_{\delta} &= \{\mb{x} \in \R^n : h(\mb{x})+\gamma(h(\mb{x}),\delta) = 0\},
    \label{eq:BoundCdelta}\\
    \textrm{Int}(\C_{\delta}) &= \{\mb{x} \in \R^n : h(\mb{x})+\gamma(h(\mb{x}),\delta) > 0\},\label{eq:IntCdelta}
\end{align}
with ${\gamma:\R \times \R_{\geq 0} \to \R_{\geq0}}$ satisfying
${\gamma(a,\cdot) \in \K_\infty}$ for all ${a\in\R}$. This implies ${\C_{\delta} = \C}$ when ${\delta=0}$. We also require ${\gamma(\cdot,b)}$ to be continuously differentiable for all ${b\in\R_{\geq 0}}$. We have that $\partial\C_{\delta}$ and $\textrm{Int}(\C_{\delta})$ are the \textit{boundary} and \textit{interior}, respectively, of the set $\C_{\delta}$. With this construction in mind, we have the following definition:

\begin{definition}[\textit{Input-to-State Safety (ISSf)}]
Let $\C\subset\R^n$ be the 0-superlevel set of a continuously differentiable function ${h:\R^n\to\R}$. The system \eqref{eq:cloop_dist} is \textit{input-to-state safe} (ISSf) with respect to the set $\C$ if there exists ${\gamma:\R \times \R_{\geq 0} \to \R_{\geq0}}$ satisfying ${\gamma(a,\cdot) \in \K_\infty}$ for all ${a\in\R}$ and ${\gamma(\cdot,b)}$ continuously differentiable for all ${b\in\R_{\geq 0}}$ such that for all ${\delta\geq 0}$ and ${\mb{d}:\R_{\geq 0}\to\R^m}$ satisfying ${\Vert\mb{d}\Vert_\infty \leq \delta}$, the set $\C_{\delta}$ defined by \eqref{eq:Cdelta}-\eqref{eq:IntCdelta} is forward invariant. If the system \eqref{eq:cloop_dist} is input-to-state safe with respect to the set $\C$, the set $\C$ is referred to as an \textit{input-to-state safe set (ISSf set)}.
\end{definition}

Similar to how Control Barrier Functions were defined in Sec.~\ref{sec:background}, we now define \textit{Input-to-State Safe Control Barrier Functions} as a tool for robust safety-critical control synthesis:

\begin{definition}[\textit{Input-to-State Safe Control Barrier Function (ISSf-CBF)}]
Let ${\C\subset\R^n}$ be the 0-superlevel set of a continuously differentiable function ${h:\R^n\to\R}$. The function $h$ is an \textit{Input-to-State Safe Control Barrier Function (ISSf-CBF)}  for \eqref{eq:eom_dist} on $\C$ if there exists an ${\alpha\in\K_{\infty}^{\rm e}}$ and a continuously differentiable function ${\epsilon:\R\to\R_{>0}}$ such that for all ${\mb{x}\in\R^n}$:
\begin{equation}
\label{eq:issf-cbf}
   \sup_{\mb{u}\in\R^m}  \left[ L_\mb{f}h(\mb{x}) + L_\mb{g}h(\mb{x})\mb{u} \right] >
-\alpha(h(\mb{x})) + \frac{\Vert L_{\mb{g}}h(\mb{x}) \Vert_2^2}{\epsilon(h(\mb{x}))}.%\nonumber
\end{equation}
\end{definition}

Given an ISSf-CBF $h$ for \eqref{eq:eom_dist} and corresponding functions ${\alpha\in\K_{\infty}^{\rm e}}$ and ${\epsilon:\R\to\R_{>0}}$, we can consider the point-wise set of all control values that satisfy \eqref{eq:issf-cbf}:
\begin{align}
\label{eqn:KISSf}
\resizebox{1\hsize}{!}{
    $K_{\rm ISSf}(\mb{x}) = \left\{\mb{u}\in\R^m ~\left|~ \dot{h}(\mb{x},\mb{u})\geq-\alpha(h(\mb{x})) + \frac{\Vert L_\mb{g}h(\mb{x}) \Vert_2^2}{\epsilon(h(\mb{x}))} \right. \right\}.$%\nonumber
    }
\end{align}
The main theoretical result in \cite{alan2021safe} relates properties of the function $\epsilon$ and controllers synthesized via an ISSf-CBF to the input-to-state safety of the set $\C$:

\begin{theorem}[\cite{alan2021safe}]
\label{theo:ISSf}
Let ${\C\subset\R^n}$ be the 0-superlevel set of a continuously differentiable function ${h:\R^n\to\R}$. Let $h$ be an ISSf-CBF for \eqref{eq:eom_dist} on $\C$ with corresponding functions ${\alpha\in\K_{\infty}^{\rm e}}$ and ${\epsilon:\R\to\R_{>0}}$ such that $\epsilon$ and ${\alpha^{-1}\in \K_{\infty}^{\rm e}}$ are continuously differentiable and $\epsilon$ satisfies:
\begin{equation}
    \label{eq:eps_geq0}
     \der{\epsilon}{r}(h(\mb{x})) \geq 0,
\end{equation}
for all ${\mb{x}\in\R^n}$. Then the set $K_{\rm ISSf}(\mb{x})$ is non-empty for each ${\mb{x}\in\R^n}$, and if a continuous controller ${\mb{k}:\R^n\to\R^m}$ satisfies ${\mb{k}(\mb{x})\in K_{\rm ISSf}}(\mb{x)}$ for all ${\mb{x}\in\R^n}$, then for any ${\delta\geq 0}$, the system \eqref{eq:cloop_dist} is safe with respect to the set $\C_\delta$ defined as in \eqref{eq:Cdelta}-\eqref{eq:IntCdelta} with ${\gamma}$ defined as:
\begin{equation}    \label{eq:issf_gamma}
    \gamma(h(\mb{x}), {\delta}) \triangleq -\alpha^{-1}\left(-\frac{\epsilon(h(\mb{x}))  {\delta}^2}{4}\right),
\end{equation}
for all $\mb{d}$ satisfying ${\Vert\mb{d}\Vert_\infty \leq \delta}$. This implies $\C$ is an ISSf set.
\end{theorem}

\begin{remark}
The original definition of ISSf presented in \cite{kolathaya2018input} differs from Definition 3 in the function $\gamma$. We allow $\gamma$ to be a function of $h$ in addition to $\delta$. This leads to a generalization of the ISSf-CBF definition in \cite{kolathaya2018input}, which reduces to the definition given in \cite{kolathaya2018input} if ${\epsilon(r) = c >0}$ for all $r\in\R$. The definitions presented here provide a factor of flexibility in controller design as detailed in \cite{alan2021safe}.
\end{remark}

\begin{figure}[t]
    \vspace{-2mm}
    \centering
    \includegraphics[width=0.3\textwidth]{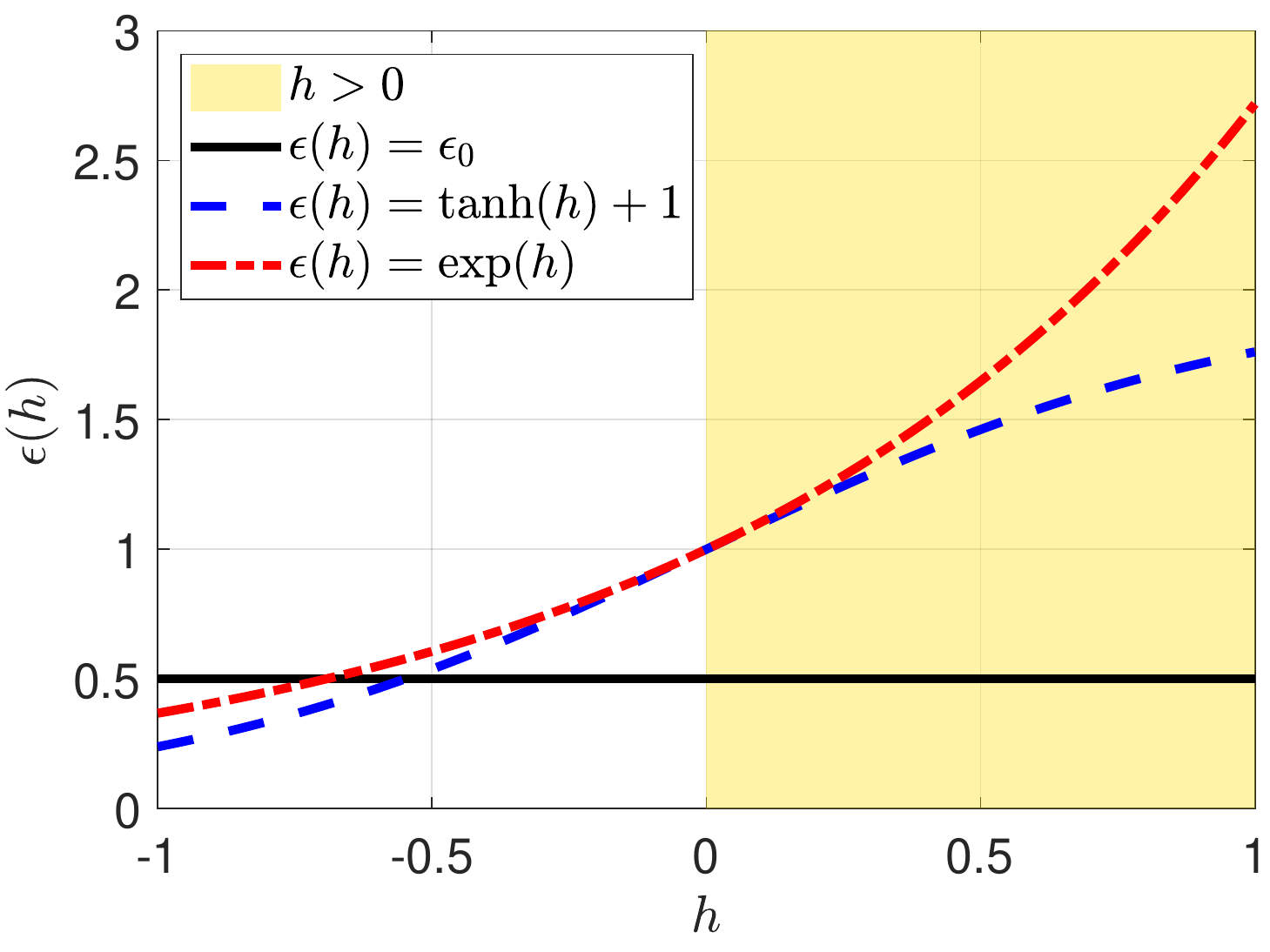}
    \caption{Examples of the function $\epsilon$ that satisfy the condition in \eqref{eq:eps_geq0}.}
    \label{fig:epsilon_alt}
    \vspace{-3.5 mm}
\end{figure}

The boundary of the set $\C_{\delta}$ that is rendered forward invariant is defined as a level-set of the ISSf-CBF $h$ as in \eqref{eq:BoundCdelta}. Given a ${\delta\geq 0}$, the value of $h$ on this level set, denoted as ${h^*\leq 0}$, can be found by solving the equation:
\begin{equation}
\label{eq:solveforCdelta}
    h^*  \underbrace{-\alpha^{-1}\left(-\frac{\epsilon(h^*)  {\delta}^2}{4} \right)}_{\gamma(h^*,\delta)} = 0.
\end{equation}
By definition, ${\gamma(h^*,\delta)}$ must be strictly positive for ${\delta > 0}$, implying that ${h^* < 0}$ in the presence of disturbances. The safety-critical controllers designed in the next section will guarantee ${h(\mb{x}(t))\geq h^*}$.
Moreover, as $\delta$ increases, $h^*$ must get more negative, implying that the boundary of $\C_\delta$ falls farther from the boundary of $\C$. Control over this degradation in safety can be achieved by modifying the function $\epsilon$ to yield different values of $h^*$ as specified in \eqref{eq:solveforCdelta}. Various functions that satisfy the necessary conditions for $\epsilon$ can be seen in Fig.~\ref{fig:epsilon_alt}.

\begin{figure*}[t]
    \centering
    \begin{subfloat}
    {\includegraphics[width=0.32\textwidth, valign = t]{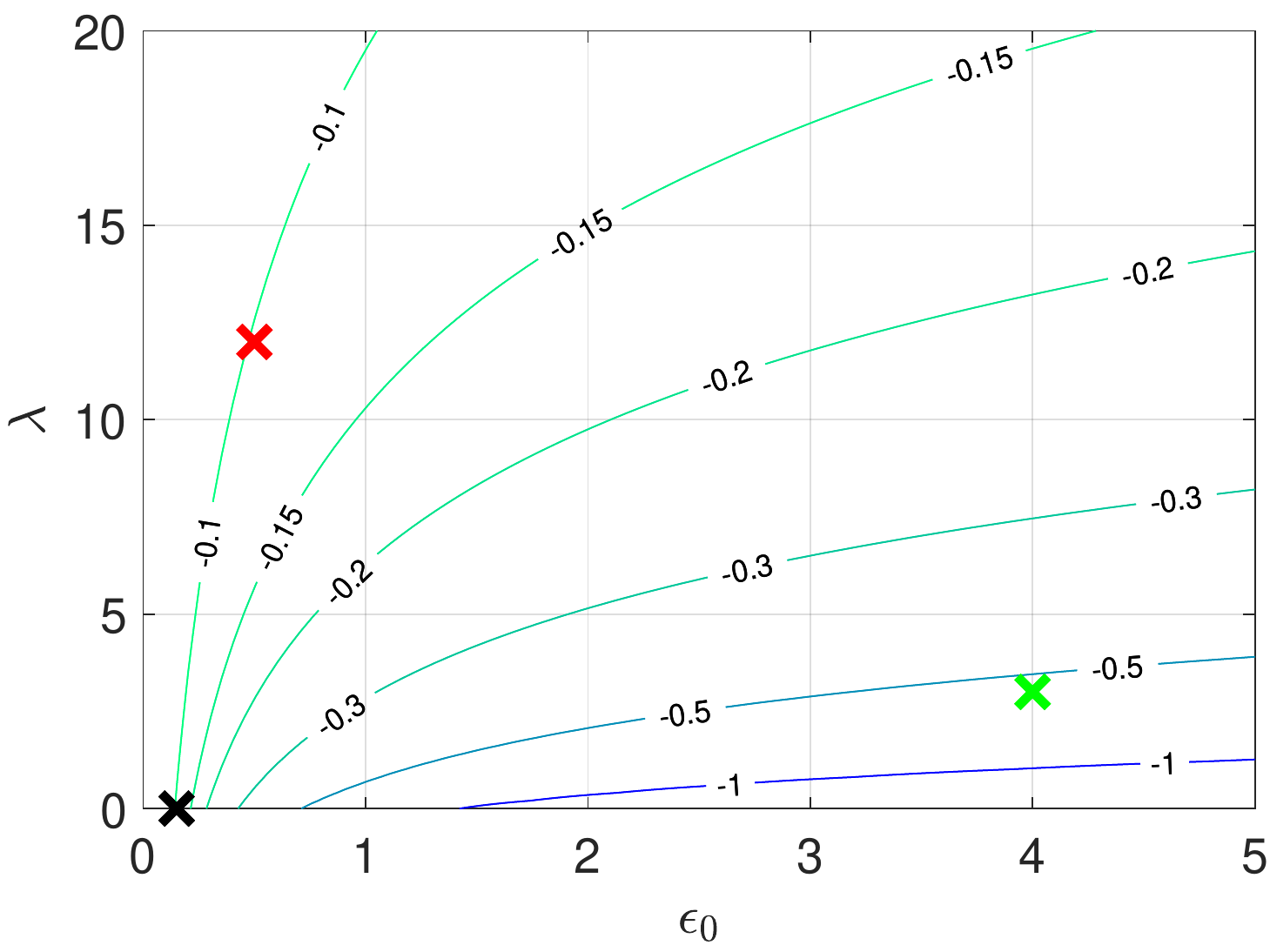}}
    \end{subfloat}
    \hfill
    \begin{subfloat}
     {\includegraphics[width =0.32\textwidth, valign =t]{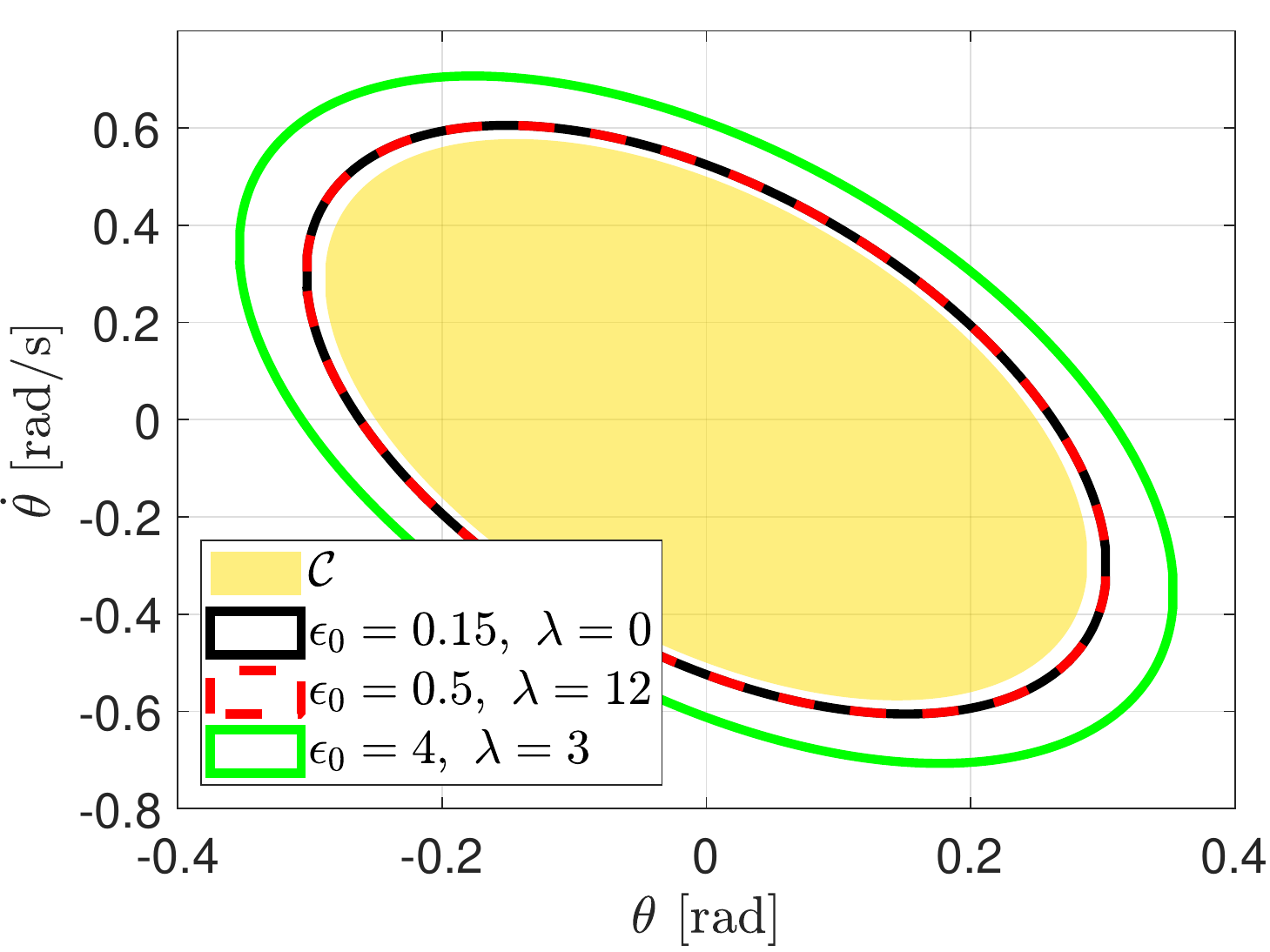}}
    \end{subfloat}
    \hfill
    \begin{subfloat}
     {\includegraphics[width =0.32\textwidth, valign =t]{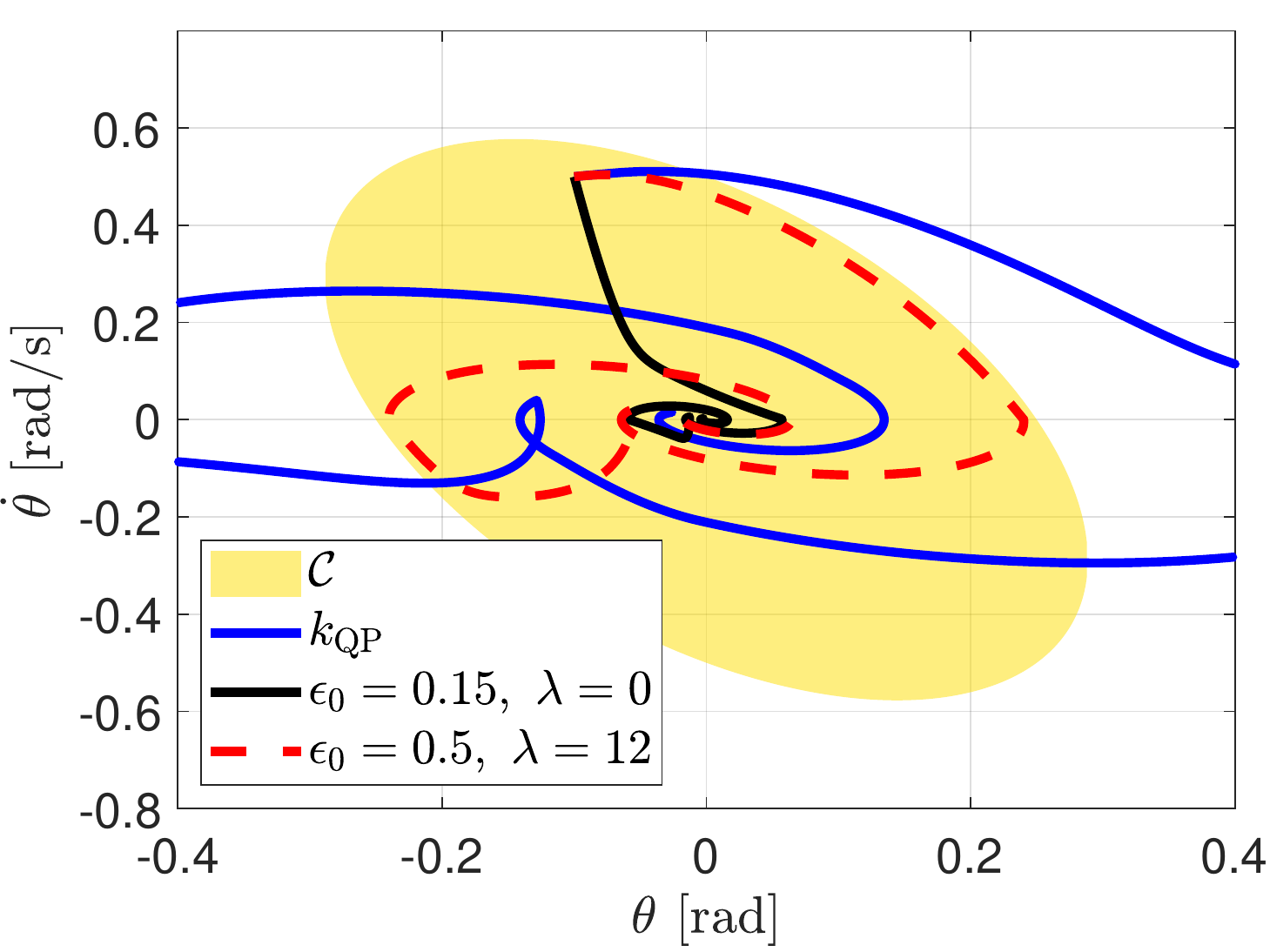}}
    \end{subfloat}
    \caption{(Left) Curves corresponding to the value of $h^*$ solving \eqref{eq:ivp_hstar} across the ${(\epsilon_0,\lambda)}$  parameter space for the inverted pendulum example. (Center) The boundary of the set $\C_{\delta}$ rendered forward invariant for different choices of the parameters $\epsilon_0$ and $\lambda$ for the inverted pendulum example. Note that the $\C_{\delta}$ contains the set $\C$ for each parameter set. (Right) Simulation results for the inverted pendulum system with disturbances. The gold ellipse is the safe set $\C$ defined in \eqref{eq:ivp_C}. The blue line is the trajectory of the system evolving under $k_{\rm QP}$ defined in \eqref{eq:switchstrucmax}-\eqref{eq:switchstrucmin}, which is not robust to disturbances and leaves the safe set. The black and dashed red lines are the trajectory of the system evolving under $\overline{k}_{\rm QP}$ defined in \eqref{eq:switchTISSfmax}-\eqref{eq:switchTISSfmin} with different values for $\epsilon_0$ and $\lambda$. While both parameter sets yield the same forward invariant set $\C_{\delta}$, the red parameter set is less conservative and allows the system to approach the boundary.}
    \label{fig:eps0_lambda_hstar}
    \vspace{-2 mm}
\end{figure*}

\begin{example}   
\label{ex:ISSfSet}
Given our choice of $\alpha$ for the inverted pendulum system, we have that:
\begin{equation}
\label{eq:ivp_gamma}
    \gamma(h(\theta,\dot{\theta}),{\delta}) = \frac{\epsilon(h(\theta,\dot{\theta}))  {\delta}^2}{4 \alpha_{\rm c}}.
\end{equation}
As our disturbance signal is bounded by $M$, determining the set kept forward invariant is done by considering ${\delta=M}$. Thus, we use the parameter value ${\delta = 0.75}$~[N$\cdot$m]. We choose the  exponential function:
\begin{equation}
    \label{eq:ivp_eps}
    \epsilon(r)=\epsilon_0 {\rm e}^{\lambda r},
\end{equation}
with parameters ${\epsilon_0>0}$ and ${\lambda\geq0}$. With this choice we have that \eqref{eq:solveforCdelta} reduces to:
\begin{equation}
    \label{eq:ivp_hstar}
    h^* + \frac{\epsilon_0 {\rm e}^{\lambda h^*} \delta^2}{4 \alpha_{\rm c}} = 0.
\end{equation}
Once $\epsilon_0$ and $\lambda$ are specified, \eqref{eq:ivp_hstar} can be solved for $h^*$ to find the value of $h$ that corresponds to the boundary $\partial\C_\delta$. Fig.~\ref{fig:eps0_lambda_hstar} (left) shows the value of $h^*$ for the different choices of $\epsilon_0$ and $\lambda$ specified in Table~\ref{tab:eps0_lambda}. The boundary $\partial\C_\delta$ corresponding to each set of parameters is shown in Fig.~\ref{fig:eps0_lambda_hstar} (right) using the same color code. The black and red parameter sets return the same value of $h^*$, and thus the produce the same boundary $\partial\C_{\delta}$. In contrast, the green parameter set yields a larger set $\C_\delta$ as indicated by the smaller value of $h^*$ in Table~\ref{tab:eps0_lambda}.
\begin{table}[h]
\vspace{-2 mm}
\centering
\begin{tabular}{|c||c|c|c|}
\hline
\textbf{Color}        & Black & Red & Green \\ \hline
$\epsilon_0\, [\frac{1}{{\rm N}^{2} {\rm m}^{2} {\rm s}}]$ & 0.15  & 0.5 & 4     \\ \hline
$\lambda$    & 0     & 12  & 3     \\ \hline
$h^*$        & $-$0.1  & $-$0.1   & $-$0.55     \\ \hline 
\end{tabular}
\caption{Parameter sets for \eqref{eq:ivp_eps} in the inverted pendulum example.}
\label{tab:eps0_lambda}
\vspace{-3 mm}
\end{table}
\end{example}

\vspace{-5 mm}
\subsection{Robust Safety-Critical Controller}

As we saw with CBFs, it is possible to use an ISSf-CBF to synthesize controllers that render a set $\C_\delta$ forward invariant, thus rendering the set $\C$ ISSf. Suppose that we have an ISSf-CBF $h$ for \eqref{eq:eom_dist} on $\C$ with corresponding functions ${\alpha\in\mathcal{K}_\infty^{\rm e}}$ and ${\epsilon:\R\to\R_{>0}}$ that meet the requirements of Theorem \ref{theo:ISSf}.
%It is noted that ${\alpha^{-1}\in \K_{\infty}^{\rm e}}$ is continuously differentiable and $\epsilon$ satisfies \eqref{eq:eps_geq0}. 
Furthermore, suppose we have a continuous nominal controller ${\mb{k}_{\rm n}:\R^n\to\R^m}$ that is not necessarily safe. Motivated by the optimization-based safety-critical formulation given in \eqref{eq:SC-QP}, we define a controller ${\overline{\mb{k}}_{\rm QP}:\R^n\to\R^m}$ as:
\begin{align}
\label{eq:ISSf-QP}
\overline{\mb{k}}_{\rm QP}(\mb{x}) =  \,\,\underset{\mb{u} \in \R^m}{\argmin}  &  \quad \frac{1}{2} \| \mb{u}-\mb{k}_{\rm n}(\mb{x}) \|_2^2  \\
\mathrm{s.t.} \quad & \quad \dot{h}(\mb{x},\mb{u})\geq-\alpha(h(\mb{x})) + \frac{\Vert L_{\mb{g}}h(\mb{x}) \Vert_2^2}{\epsilon(h(\mb{x}))}. \nonumber
\end{align}
The following theorem provides a closed-form solution to the optimization problem defining this controller and specify the continuity and safety properties of the resulting controller:

\begin{theorem}   
\label{theo:Theo4}
Let $\C$ be the 0-superlevel set of a function ${h:\R^n\to\R}$, and let ${\mb{k}_{\rm n}:\R^n\to\R^m}$ be a continuous controller. If $h$ is an ISSf-CBF for \eqref{eq:eom_dist} on the set $\C$ with corresponding functions ${\alpha\in\K_{\infty}^{\rm e}}$ with continuously differentiable inverse ${\alpha^{-1}\in\K_{\infty}^{\rm e}}$, and continuously differentiable ${\epsilon:\R\to\R_{>0}}$ satisfying \eqref{eq:eps_geq0}, then the optimization problem in \eqref{eq:ISSf-QP} is feasible for any ${\mb{x}\in\R^n}$ and has a closed-form solution given by:
\begin{equation}
\label{eq:ISSf-CF}
\overline{\mb{k}}_{\rm QP}(\mb{x}) = \mb{k}_{\rm n}(\mb{x}) + \max\left\{0, \overline{\eta}(\mb{x}) \right\}L_\mb{g}h(\mb{x})^\top, 
\end{equation}
where the function ${\overline{\eta}:\R^n\to\R}$ is defined as:
\begin{equation}
\label{eq:overlinelambdastar}
    \overline{\eta}(\mb{x}) = \begin{cases}  -\frac{ \dot{h}\left(\mb{x},\mb{k}_{\rm n}(\mb{x})\right)+\alpha(h(\mb{x}))}{\Vert L_{\mb{g}}h(\mb{x}) \Vert_2^2} + \frac{1}{\epsilon(h(\mb{x}))} \quad & \textrm{if}~ L_\mb{g}h(\mb{x}) \neq \mb{0}, \\  0 & \textrm{if}~ L_\mb{g}h(\mb{x}) = \mb{0}.
    \end{cases}
\end{equation}
Furthermore, $\overline{\mb{k}}_{\rm QP}$ is continuous and ${\overline{\mb{k}}_{\rm QP}(\mb{x})\in K_{\textrm{\rm ISSf}}(\mb{x})}$ for all ${\mb{x}\in\R^n}$.
\end{theorem}
\noindent The proof of this theorem is performed similarly to the proof of Theorem \ref{thm:equiv} with simple modifications for the introduction of $\epsilon$, and thus, it is omitted.

\begin{remark}
For a single input ($m=1)$, if ${{L_\mb{g}h(\mb{x}) > 0}}$ for a particular $\mb{x}\in\R^n$, the controller \eqref{eq:ISSf-CF} can be expressed as:
\begin{equation}
\label{eq:switchTISSfmax}
\overline{k}_{\rm QP}(\mb{x}) = \max\left\{k_{\rm n}(\mb{x}), -\frac{L_\mb{f}h(\mb{x})+\alpha(h(\mb{x}))}{L_\mb{g}h(\mb{x})} + \frac{L_\mb{g}h(\mb{x})}{\epsilon(h(\mb{x}))} \right\}.
\end{equation}
Similarly, if ${L_\mb{g}h(\mb{x}) < 0}$ for a particular $\mb{x}\in\R^n$, the controller \eqref{eq:ISSf-CF} reduces to:
\begin{equation}
\label{eq:switchTISSfmin}
\overline{k}_{\rm QP}(\mb{x}) = \min\left\{k_{\rm n}(\mb{x}), -\frac{L_\mb{f}h(\mb{x})+\alpha(h(\mb{x}))}{L_\mb{g}h(\mb{x})} + \frac{L_\mb{g}h(\mb{x})}{\epsilon(h(\mb{x}))}\right\}.
\end{equation}
These controllers can be switched between depending on the sign of $L_\mb{g}h(\mb{x})$, with ${\overline{k}_{\rm QP}(\mb{x})=k_{\rm n}(\mb{x})}$ when ${L_\mb{g}h(\mb{x})=0}$.
\end{remark}

\begin{example}
We deploy the safety-critical controller $k_{\rm QP}$ defined in \eqref{eq:switchstrucmax}-\eqref{eq:switchstrucmin} with the nominal controller $k_{\rm n}$ as in \eqref{eq:ivp_kn} to the inverted pendulum system without considering the disturbance $d$ defined in \eqref{eq:ivp_dist}. We see in the right panel of Fig.~\ref{fig:eps0_lambda_hstar} that this controllers fails to keep the system in the safe set $\C$, and  deviates from it significantly.
We next deploy the safety-critical controller $\overline{k}_{\rm QP}$ defined in \eqref{eq:switchTISSfmax}-\eqref{eq:switchTISSfmin} with the nominal controller $k_{\rm n}$ as in \eqref{eq:ivp_kn}. The exponential function given in \eqref{eq:ivp_eps} is utilized with the black and red parameter pairs as specified in Table~\ref{tab:eps0_lambda}. We see in the right panel of Fig.~\ref{fig:eps0_lambda_hstar} that for both parameter sets, the controller keep the trajectories within $\C$, and thus, within $\C_{\delta}$, that is, it guarantees ${h(\mb{x}(t))\geq h^*}$. Despite having the same values of $h^*$, we see the red parameter sets allows the system to more closely approach the boundary of the safe set, while the black parameter set forces the system to the equilibrium more directly. A detailed discussion about the effect of the parameter $\lambda$ on conservativeness of the controller is provided in \cite{alan2021safe}.  %The added term in second switch condition in \eqref{eq:switchTISSfmax}-\eqref{eq:switchTISSfmin} reduces as $\epsilon$ gets larger inside $\C$.
%The green parameter set, however, ends up with a larger set $\C_\delta$ that is guaranteed to be forward invariant.
\end{example}

% \begin{table}[h]
% \centering {\small
% \begin{tabular}{|l|l|l|}
% \hline
%       $m = 2$ [kg] & $a = 0.25$ [rad] & $K_{\rm p} = 0.6$ [1/s$^2$]  \\ \hline
%       $l = 1$ [m] & $b = 0.5$ [rad/s] & $K_{\rm d} = 0.6$ [1/s] \\ \hline
%       $g = 10$ [m/s$^2$] & $\alpha_{\rm c} = 0.2$ [1/s] & $\delta = M = 0.75$ [N$\cdot$m] \\ \hline
% \end{tabular}
% }
% \caption{Inverted Pendulum Example Simulation Parameters}
% \label{tab:ivtpndparam}
% \vspace{-2mm}
% \end{table}

% \begin{figure}[h]
%      \includegraphics[width =0.45\textwidth, valign =t]{Cleaned_Figures/ivtpndsim_fig2.pdf}
%     \caption{Simulation results for the inverted pendulum system with disturbances. The gold ellipse is the safe set $\C$ defined in \eqref{eq:ivp_C}. The blue line is the trajectory of the system evolving under $k_{\rm QP}$ defined in \eqref{eq:switchstrucmax}-\eqref{eq:switchstrucmin}, which is not robust to disturbances and leaves the safe set. The black and dashed red lines are the trajectory of the system evolving under $\overline{k}_{\rm QP}$ defined in \eqref{eq:switchTISSfmax}-\eqref{eq:switchTISSfmin} with different values for $\epsilon_0$ and $\lambda$. While both parameter sets yield the same forward invariant set $\C_{\delta}$, the red parameter set is less conservative and allows the system to approach the boundary.}
%     \label{fig:ivtpnd_simresults}
% \end{figure}

%%%%%%%%%%%%%%%%%%%%%%%%%%%%%%%%%%%%%%%%%%%%%%%%%%%%%%%%%%%%%%%%%%%%%%%%%%%%%%%%%%%%%%%
\section{Safety-Critical Controller Design for a Connected Automated Truck} \label{sec:truck}
%%%%%%%%%%%%%%%%%%%%%%%%%%%%%%%%%%%%%%%%%%%%%%%%%%%%%%%%%%%%%%%%%%%%%%%%%%%%%%%%%%%%%%%
In this section, we go through the process of designing a safety-critical longitudinal controller for a connected automated truck. We first introduce the physical system and define a safe set via a Control Barrier Function. We then present a nominal performance-based controller, and synthesize a safety-critical controller that modifies this nominal controller in a minimally invasive way while ensuring safety.

%%%%%%%%%%%%%%%%%%%%%%%%%%%%%%%%%%%%%%%%%%%%%%%%%%%%%%%%%%%%%%%%%%%%%%%%%%%%%%%%%%%%%%%
\subsection{Modeling Longitudinal Dynamics}  \label{sec:truckmodel}
%%%%%%%%%%%%%%%%%%%%%%%%%%%%%%%%%%%%%%%%%%%%%%%%%%%%%%%%%%%%%%%%%%%%%%%%%%%%%%%%%%%%%%%

\begin{figure}[b]
    \centering
    \vspace{-4mm}
    \includegraphics[trim=0 0 0 0,clip, width=0.9\linewidth]{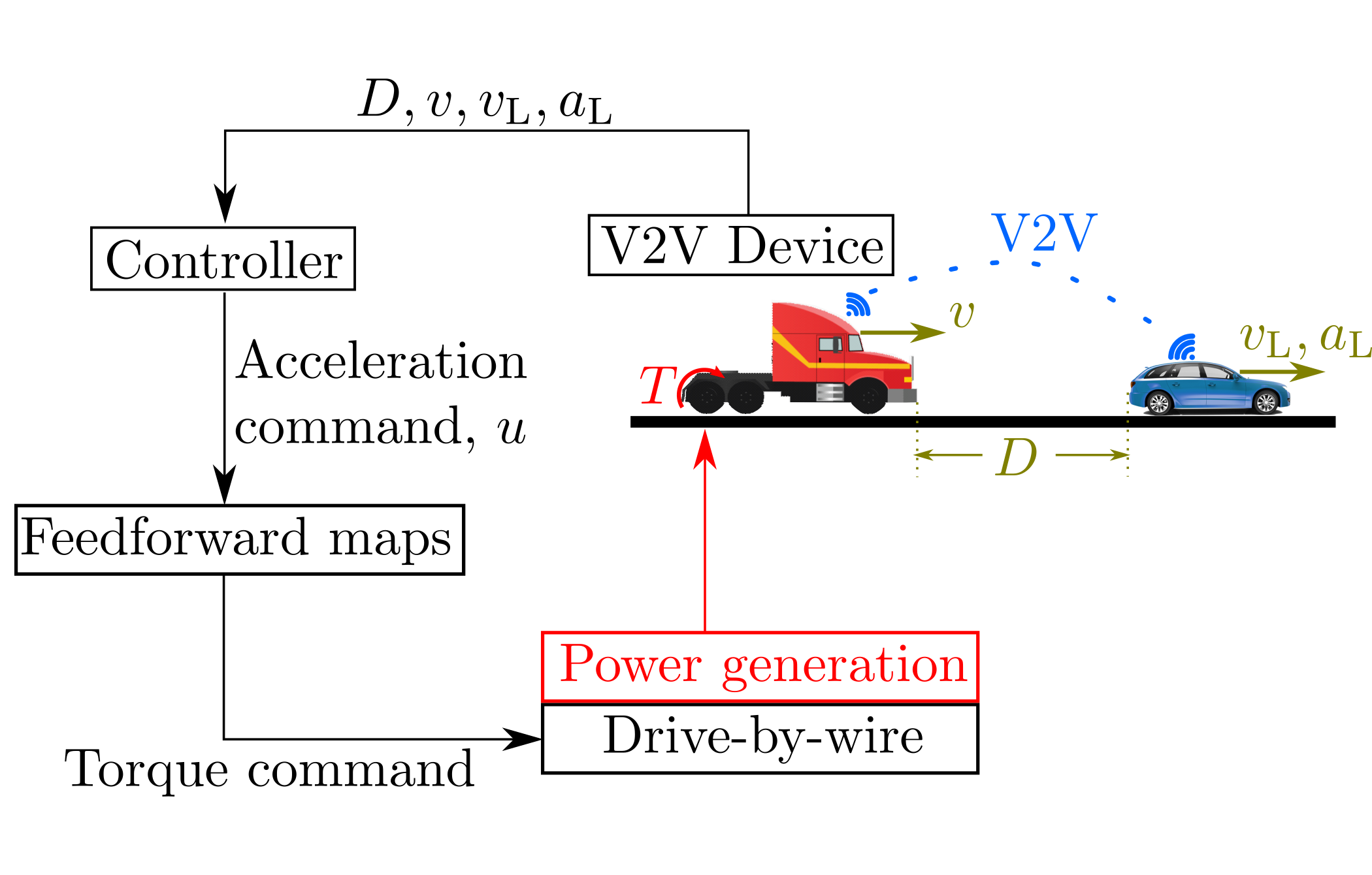}
    \caption{A connected automated truck following a connected vehicle.}
    \label{fig:CarFollow}
    \vspace{-0mm}
\end{figure}

In this work we consider a rear-axle-driven truck without a trailer. Assuming the truck's tires roll without slipping and the truck travels on a flat road with no headwind, the longitudinal dynamics of the truck are described by the following model:
\begin{equation}
\label{eq:truck_mechmodel}
   \dot{v} = \frac{T}{m_{\rm eff} R} - \frac{k v^2 + mg \gamma}{m_{\rm eff}}.
\end{equation}
Here the state is given by the truck's longitudinal speed ${v\in\R}$, the input is the torque applied on the rear axle ${T\in\R}$, and the parameters in the model are the mass of the truck $m$, the mass moment of inertia of the rotating elements $I$, the tire radius $R$, the effective mass ${m_{\rm eff}=m+\frac{I}{R^2}}$, the air drag constant $k$, gravitational acceleration $g$, and rolling resistance coefficient $\gamma$.
Note that the second term in \eqref{eq:truck_mechmodel} is dissipative in nature, and slows down the vehicle when it has a positive velocity. This term may be directly accounted for in the control design through via feedback linearization techniques \cite{sastry1999nonlinear}, or may be ignored as its omission simply introduces a factor of conservativeness to the controller in terms of safety. The torque input commanded of the system is computed from a desired longitudinal acceleration command ${u\in\R}$ via feed-forward maps. This torque input command is provided by a drive-by-wire system to the low-level power generation systems that produce the actual torque $T$; see Fig.~\ref{fig:CarFollow}. With these feed-forward maps in mind, we may simplify the model of the longitudinal dynamics of the truck to:
\begin{equation}
\label{eq:truck_simpmodel}
    \dot{v} = u.
\end{equation}

Now let us consider the scenario when the truck follows a connected vehicle as depicted in Fig.~\ref{fig:CarFollow}. Using the truck model in \eqref{eq:truck_simpmodel}, the dynamics of this connected system are given by:
\begin{equation}   
\label{eq:truck_model2}
    \begin{split}
        \dot{D} &=v_{\rm L}-v, \\
        \dot{v} &=u, \\
        \dot{v}_{\rm L} &= a_{\rm L},
    \end{split}
\end{equation}
where ${v_{\rm L}, a_{\rm L}\in\R}$ are the speed and acceleration of the leading vehicle, respectively, and ${D\in\R}$ denotes the bumper-to-bumper headway distance between the truck and the lead vehicle, yielding the state ${\mb{x}=\left[ D,v,v_{\rm L} \right]^\top\in\R^3}$. The truck and lead vehicle are outfitted with vehicle-to-vehicle (V2V) communication systems, permitting the truck to receive motion information from the lead vehicle such as its GPS position which yields the distance $D$, its speed $v_{\rm L}$, and its acceleration $a_{\rm L}$. We assume that the leader's behavior satisfies:
\begin{equation} 
    \label{eq:truck_limits}
    a_{\rm L} \in [-\underline{a}_{\rm L}, \overline{a}_{\rm L}], \qquad v_{\rm L} \in [0,\overline{v}_{\rm L}],
\end{equation}
where the parameters ${\underline{a}_{\rm L}, \overline{a}_{\rm L}, \overline{v}_{\rm L}>0}$ reflect a city-driving scenario; see Table~\ref{tab:swtichparam}.

\begin{table}[b]
\centering {\small
\begin{tabular}{|l|l|l|}
\hline
    $\overline{a}_{\rm L} = 5$ [m/s$^2$]  &  $c_0 = 2$ [m]  & $\kappa=0.8$ [1/s]  \\  \hline
    $\underline{a}_{\rm L} = 10$ [m/s$^2$] &  $c_1 = 1.1$ [s] & $\alpha_{\rm c}= 0.1$ [1/s] \\ \hline
    $\overline{v}_{\rm L} = 20$ [m/s]       &  $c_2 = 0.6$ [s]  & $D_{\rm st} = 5$ [m]  \\ \hline
    $\delta = 4.5$ [m/s$^2$]                   &  $c_3 = 0.03$ [s$^2$/m] & $D_{\rm go}= 30$ [m]  \\ \hline
   $\epsilon_0 = 0.5$ [s$^3$/m] & $c_4 = -0.03$ [s$^2$/m] & $A = 0.4$ [1/s] \\ \hline
   $\lambda = 0.4$ [1/m] & $c_5 = -0.03$ [s$^2$/m] & $B = 0.5$ [1/s]  \\ \hline
\end{tabular}
}
\caption{Parameter values used in controller design.}
\label{tab:swtichparam}
\vspace{-3 mm}
\end{table}

%%%%%%%%%%%%%%%%%%%%%%%%%%%%%%%%%%%%%%%%%%%%%%%%%%%%%%%%%%%%%%%%%%%%%%%%%%%%%%%%%%%%%%%

\subsection{Safety and Control Barrier Function}   \label{sec:truck_safety}
%%%%%%%%%%%%%%%%%%%%%%%%%%%%%%%%%%%%%%%%%%%%%%%%%%%%%%%%%%%%%%%%%%%%%%%%%%%%%%%%%%%%%%%

% \begin{figure*}[t]
%     \centering
% 	{\includegraphics[width=0.4\textwidth, valign = t]{Cleaned_Figures/rho_surf.pdf}}
% 	\qquad
% 	{\includegraphics[width=0.4\textwidth, valign = t]{Cleaned_Figures/truck_cbf.pdf}}
%     \caption{(Left) The value of the function $\rho$ defined in \eqref{eq:truck_hhat}, which defines the minimum safe following distance as a function of the leader's velocity $v_{\rm L}$ and truck velocity $v$. (Right) The value of ${L_{\mb{f}}h(\mb{x})+\alpha(h(\mb{x}))}$ as defined in \eqref{eq:truckLfh0} when ${L_{\mb{g}}h(\mb{x}) = 0}$ as given in \eqref{eq:truckLgh0}. }
% 	\label{fig:CarFollowBarrier}
% \end{figure*}

\begin{figure}[t]
\centering
	{\includegraphics[width=0.35\textwidth, valign = t]{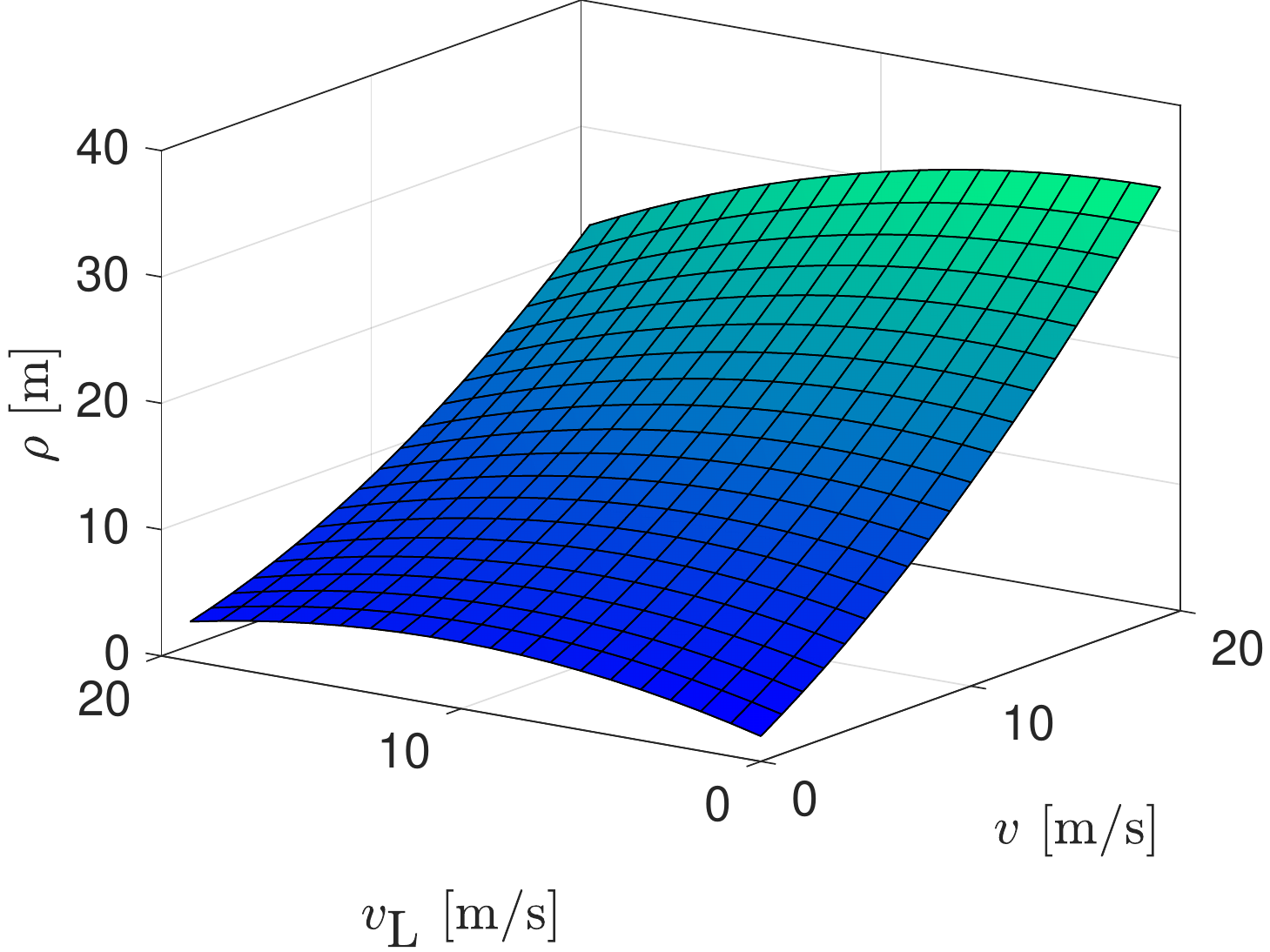}}
    \caption{The value of the function $\rho$ defined in \eqref{eq:truck_hhat}, which defines the minimum safe following distance as a function of the leader's velocity $v_{\rm L}$ and truck velocity $v$.}
	\label{fig:CarFollowBarrier}
	\vspace{-4 mm}
\end{figure}

The safety task for the truck is to maintain a safe distance behind the leader. This task motivates a Control Barrier Function of the form:
\begin{equation}  
\label{eq:truck_h}
    h(D,v,v_{\rm L}) = D - \rho(v,v_{\rm L}),
\end{equation}
where the headway function ${\rho:\R^2 \to \R}$ describes the minimum safe distance between the vehicles given their current velocities, $v$ and $v_{\rm L}$. Motivated by \cite{ames2017control} and \cite{he2018safety}, we define the headway function as:
\begin{equation} 
    \label{eq:truck_hhat}
    \rho(v,v_{\rm L}) = c_0 + c_1 v + c_2 v_{\rm L} + c_3 v^2 + c_4 v v_{\rm L} + c_5 v_{\rm L}^2,
\end{equation}
with parameters ${c_i\in\R}$ for ${i=0,\hdots,5}$; see Table~\ref{tab:swtichparam}.
The value of the function $\rho$ is visualized in the left panel of Fig.~\ref{fig:CarFollowBarrier}. The corresponding safe set defined by $h$ is given by:
\begin{equation}  
\label{eq:truck_safeset}
\C = \left\{\left. \begin{bmatrix} D \\ v \\ v_{\rm L} \end{bmatrix} \in \R^3 ~\right|~ D \geq \rho(v,v_{\rm L}) \right\}.
\end{equation}

To verify that the function $h$ is a CBF for \eqref{eq:truck_model2}, observe that:
\begin{equation}\label{eq:truckLgh0}
    L_{\mb{g}}h(D_0,v_0,v_{\textrm{L}, 0}) = 0 \implies  c_1+2c_3v_0+c_4v_{\textrm{L},0} = 0,
\end{equation}
which describes a line in ${(v,v_{\rm L})}$ space where the condition \eqref{eq:cbf_alt} must be met for all ${D\in\R}$.
We consider ${\alpha(r)=\alpha_{\rm c} r}$ with ${\alpha_{\rm c}>0}$, yielding:
\begin{equation}\label{eq:truckLfh0}
\begin{split}
    L_{\mb{f}}h(D,v_0,v_{\textrm{L},0}) &+  \alpha(h(D,v_0,v_{\textrm{L},0}))  
    \\ 
    =  &~v_{\textrm{L},0}-v_0 - a_{\rm L}(c_2+c_4v_0+2c_5v_{\textrm{L},0})
    \\ &~+ \alpha_{\rm c}(D - \rho(v_0,v_{\textrm{L},0})).
\end{split}
\end{equation}
Since checking the condition \eqref{eq:cbf_alt} analytically may be cumbersome using \eqref{eq:truckLfh0}, we graphically evaluate it over a range of $D$ and $v_{\textrm{L},0}$ (and $v_0$ defined implicitly through \eqref{eq:truckLgh0}) while taking the worst case value of $a_{\rm L}$ making \eqref{eq:truckLfh0} as negative as possible; see the right panel in Fig.~\ref{fig:CarFollowBarrier}. This shows that for $\alpha_{\rm c} = 0.1$~[1/s], the value of \eqref{eq:truckLfh0} is strictly positive, ensuring the condition \eqref{eq:cbf_alt} is satisfied.

\begin{figure}[t]
\centering
	{\includegraphics[width=0.35\textwidth, valign = t]{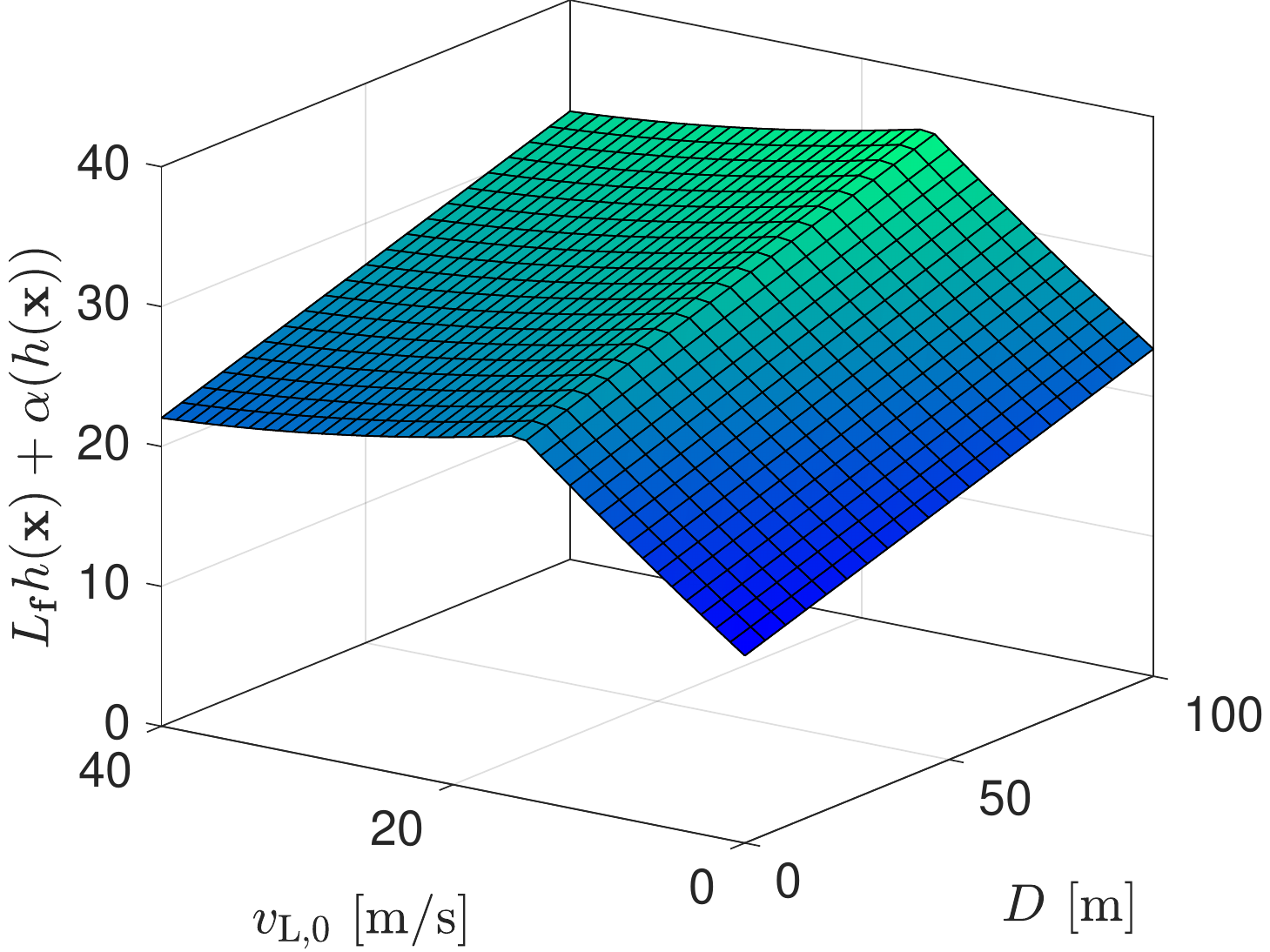}}
	\caption{The value of ${L_{\mb{f}}h(\mb{x})+\alpha(h(\mb{x}))}$ as defined in \eqref{eq:truckLfh0} when ${L_{\mb{g}}h(\mb{x}) = 0}$ for various distances. As the function is strictly positive over the domain of interest, $h$ is a CBF on $\C$ for \eqref{eq:truck_model2}.}  
	\label{fig:CarFollowBarrierCheck}
	\vspace{-4 mm}
\end{figure}

%%%%%%%%%%%%%%%%%%%%%%%%%%%%%%%%%%%%%%%%%%%%%%%%%%%%%%%%%%%%%%%%%%%%%%%%%%%%%%%%%%%%%%%
\subsection{Controller Design}\label{sec:truck_controller}
%%%%%%%%%%%%%%%%%%%%%%%%%%%%%%%%%%%%%%%%%%%%%%%%%%%%%%%%%%%%%%%%%%%%%%%%%%%%%%%%%%%%%%%

\begin{figure}[b]
\vspace{-3 mm}
    \centering
    \includegraphics[width = 0.32\textwidth]{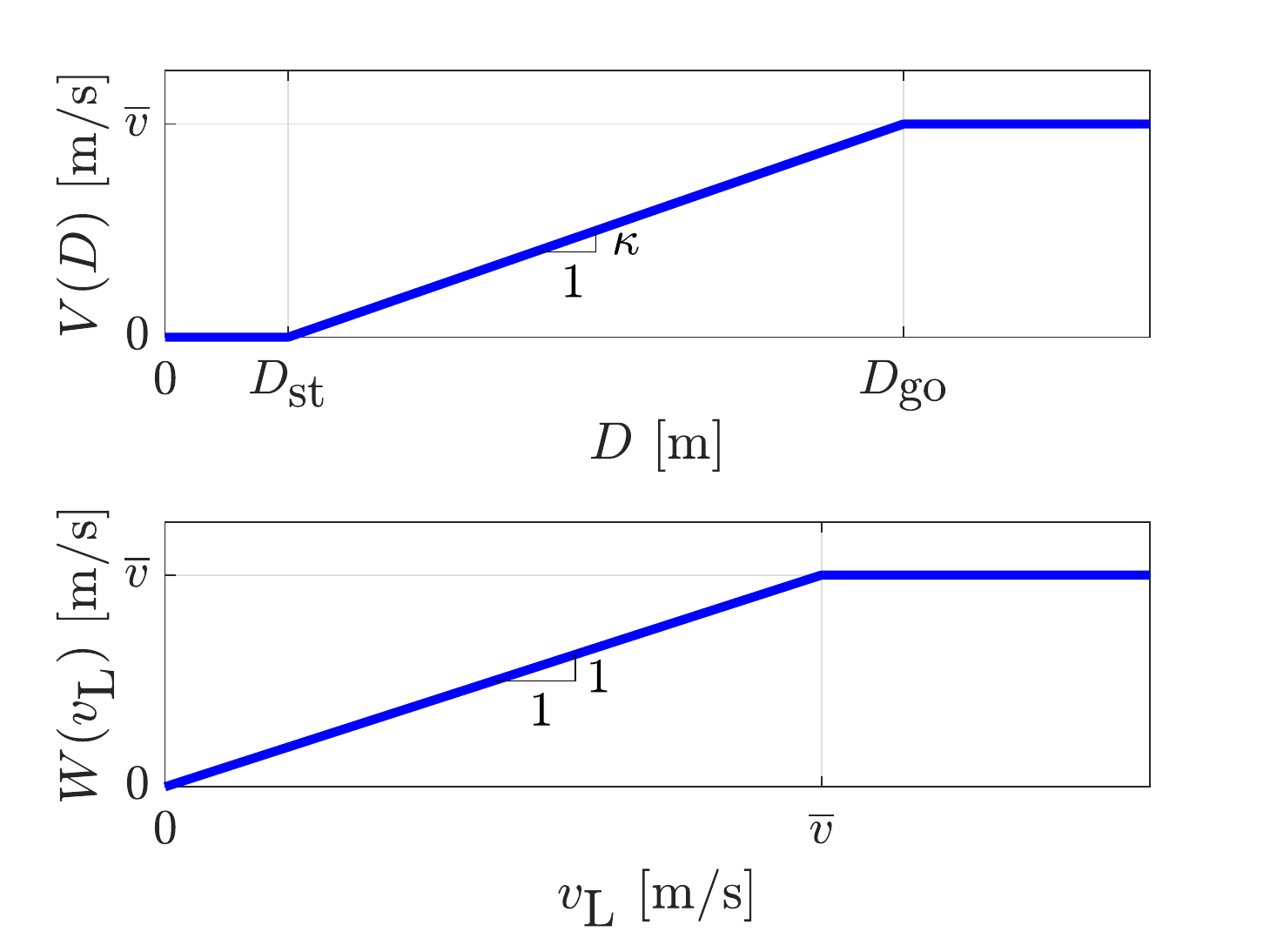}
    \caption{(Top) Distance based range policy $V$ defined in \eqref{eq:truck_V}. (Bottom) Speed policy $W$ defined in \eqref{eq:truck_W}.}
    \label{fig:VWfuncs}
\vspace{-3 mm}
\end{figure}

Beyond the task of safety, we wish for our controller to maximize other performance criteria such as ride comfort, fuel economy, and string stability \cite{orosz2016connected, ge2018experimental}. To accomplish this goal, we first design a nominal controller that prioritizes performance. In particular, we will design a connected cruise controller (CCC) for the truck that utilizes information about the lead vehicle available through V2V connectivity. We propose the following controller structure:
\begin{equation}\label{eq:truck_nominal}
k_{\rm n}(D,v,v_{\rm L}) = A(V(D)-v) + B(W(v_{\rm L})-v),
\end{equation} 
with parameters ${A,B>0}$, functions ${V:\R_{\geq 0}\to\R_{\geq 0}}$, and ${W:\R_{\geq 0}\to\R_{\geq 0}}$. 
The first term in \eqref{eq:truck_nominal} specifies the distance based speed error with the range policy:
\begin{equation} 	\label{eq:truck_V}
V(D) = \begin{cases}
0 	& {\rm if} \quad D<D_{\rm st}, \\
\kappa(D-D_{\rm st}) 	& {\rm if} \quad D_{\rm st}
\leq D \leq D_{\rm go}, \\
\overline{v}_{\rm L} 	& {\rm if} \quad D>D_{\rm go},
\end{cases}
\end{equation}
depicted in the top panel of Fig.~\ref{fig:VWfuncs}, producing a desired speed based on the distance $D$. Here ${D_{\rm st}> 0}$ is the desired stopping distance, ${1/\kappa >0}$ is the desired time headway, and ${D_{\rm go} = \overline{v}_{\rm L}/\kappa + D_{\rm st}}$. 
The second term in \eqref{eq:truck_nominal} specifies the error related to the relative speed with the speed policy:
\begin{equation}	\label{eq:truck_W}
W(v_{\rm L}) = \begin{cases} 
v_{\rm L} 	& {\rm if} \quad v_{\rm L} \leq \overline{v}_{\rm L}, \\
\overline{v}_{\rm L} 	& {\rm if} \quad v_{\rm L} > \overline{v}_{\rm L},
\end{cases}
\end{equation} 
depicted in the bottom panel of Fig.~\ref{fig:VWfuncs}, which bounds the speed error if the lead vehicle violates $v_{\rm L} \leq \overline{v}_{\rm L}$.

\begin{figure*}[t]
	\centering
	\begin{subfloat}
	{\includegraphics[width=0.3\textwidth, valign = t]{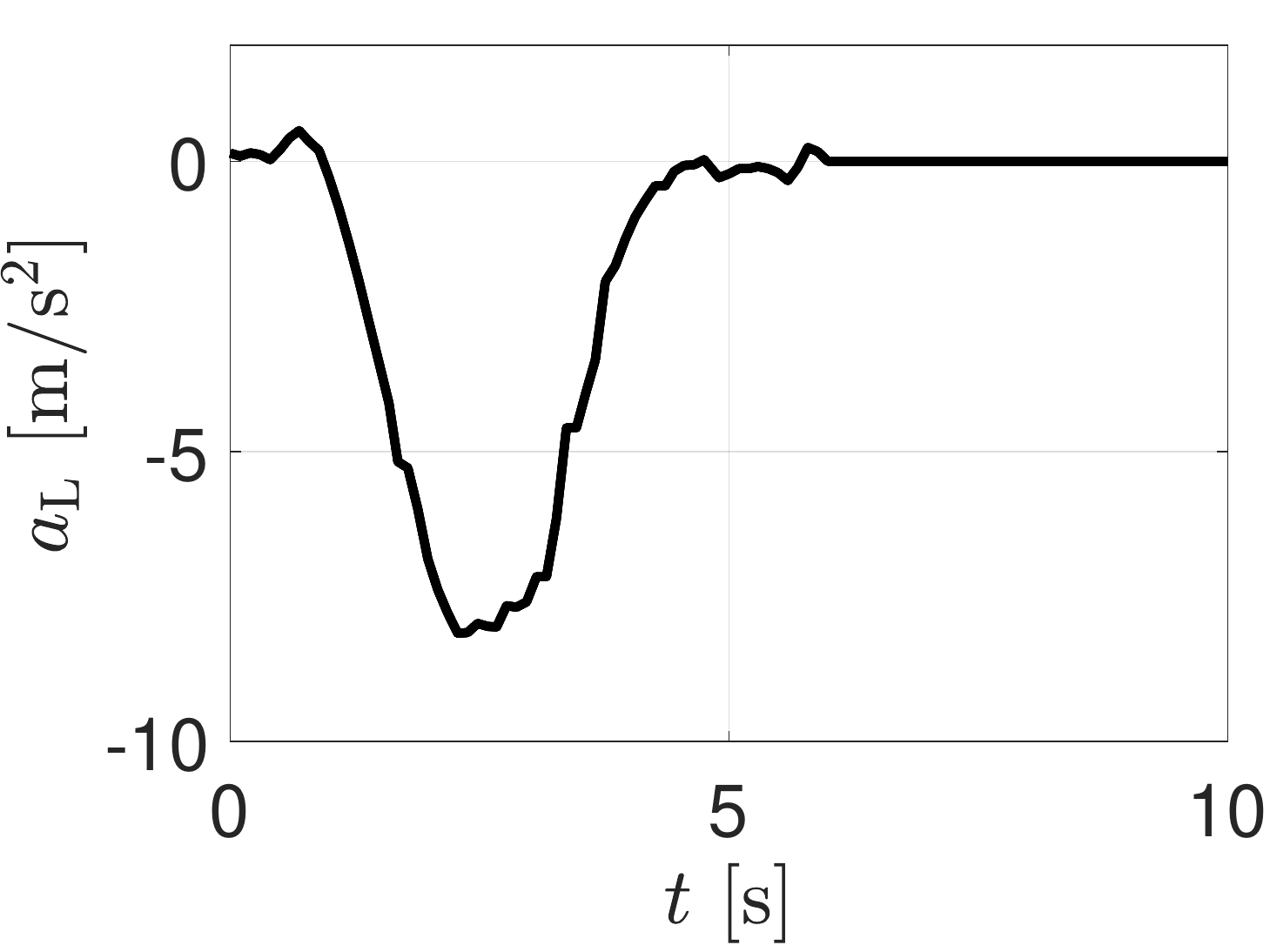}}
	\end{subfloat}
	\hfill
	\begin{subfloat}
	{\includegraphics[width=0.3\textwidth, valign = t]{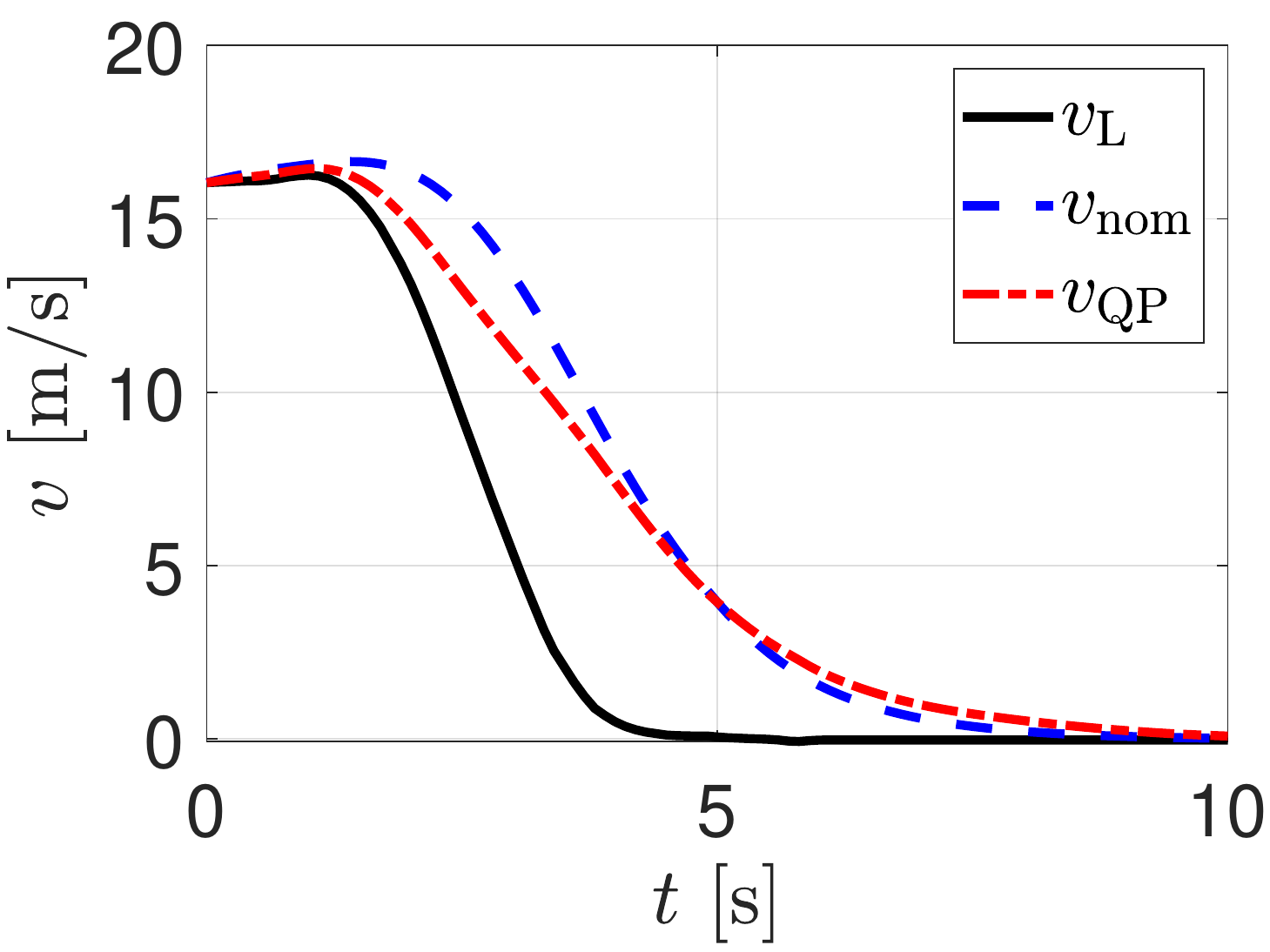}}
	\end{subfloat}
	\hfill
	\begin{subfloat}
	{\includegraphics[width=0.3\textwidth, valign = t]{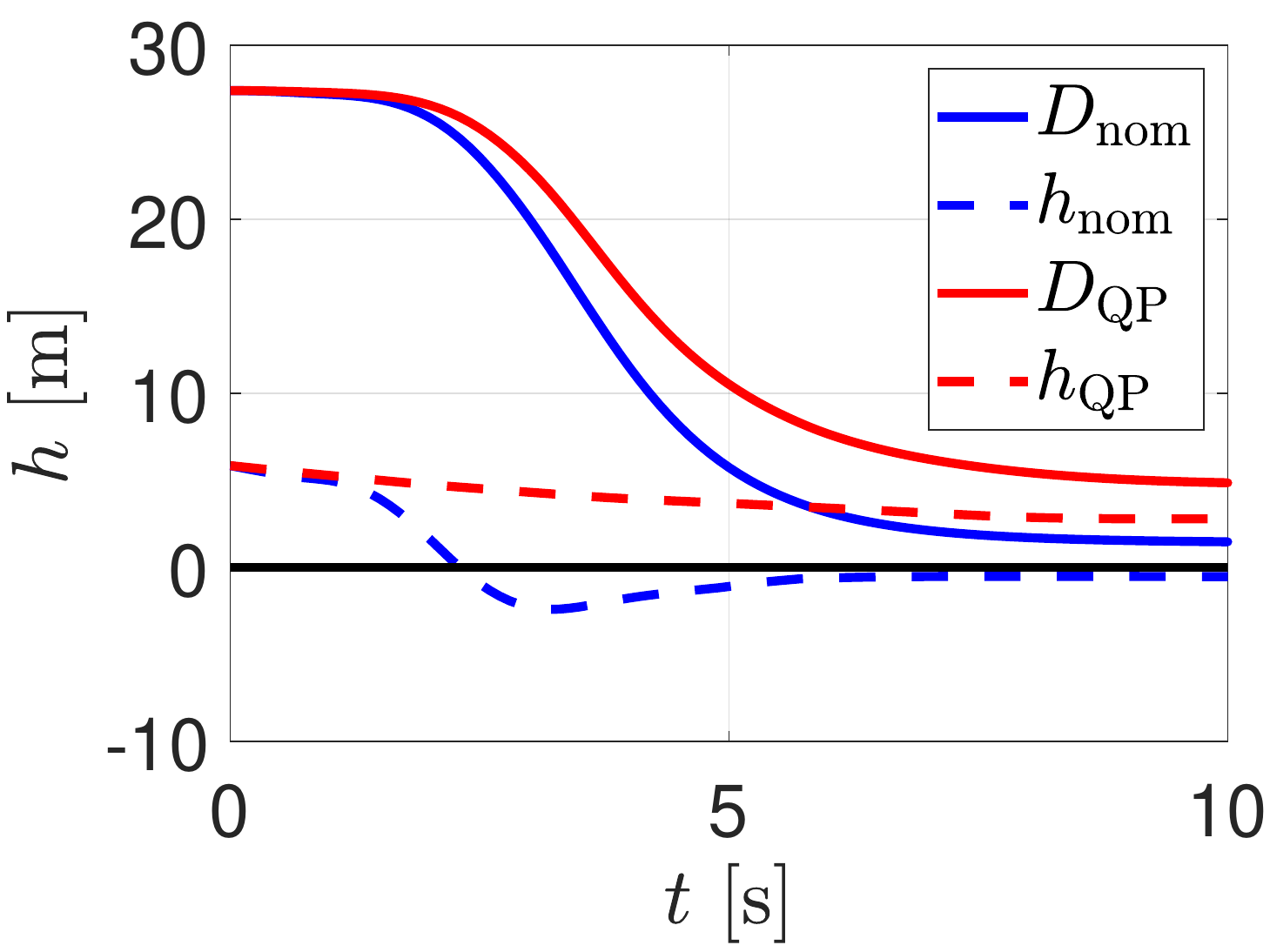}}
	\end{subfloat}
	\caption{(Left) Example profile for acceleration $a_{\rm L}$ of lead vehicle used in numerical simulation. (Center) Velocity  $v_{\rm L}$ of lead vehicle (black) and velocity of the truck using the nominal controller \eqref{eq:truck_nominal} (blue) and safety-critical controller \eqref{eq:truck_kQPmin} (red). (Right) Following distance $D$ and value of CBF $h$ using the nominal controller and safety-critical controller.}
	\label{fig:truck_sim}
	\vspace{-2 mm}
\end{figure*}

Having designed the CBF $h$ and the performance based nominal controller $k_{\rm n}$, we can unify them via the safety-critical controller formulation for single input systems given in \eqref{eq:switchstrucmax}-\eqref{eq:switchstrucmin}. Here we have:
\begin{equation}\label{eq:LgLf}
\begin{split}
    L_\mb{f}h(D,v,v_{\rm L}) &= v_{\rm L}-v - a_{\rm L}\left(c_2+c_4v+2c_5v_{\rm L} \right), \\
    L_\mb{g}h(D,v,v_{\rm L}) &= -c_1-2c_3v-c_4v_{\rm L},
\end{split}
\end{equation}
where  ${L_\mb{g}h(D,v,v_{\rm L})<0}$ for  ${v\geq0}$ and ${v_{\rm L}\in[0,\overline{v}_{\rm L}]}$. Then one may utilize the switch structure  \eqref{eq:switchstrucmin}:
\begin{equation}\label{eq:truck_kQPmin}
k_{\rm QP}(D,v,v_{\rm L}) = \min\left\{k_{\rm n}(D,v,v_{\rm L}), k_{\rm s}(D,v,v_{\rm L}) \right\},
\end{equation}
where the second term is defined as:
\begin{equation}\label{eq:truck_ks}
k_{\rm s}(D,v,v_{\rm L}) = -\frac{ L_\mb{f}h(D,v,v_{\rm L}) + \alpha_{\rm c} h(D,v,v_{\rm L}) }{ L_\mb{g}h(D,v,v_{\rm L})}.
\end{equation}
This controller utilizes the nominal controller $k_{\rm n}$ to optimize the performance when it is safe. Otherwise, the provably safe controller $k_{\rm s}$ becomes smaller than $k_{\rm n}$ and intervenes to ensure safety.
Note that ${L_\mb{g}h(D,v,v_{\rm L})>0}$ for sufficiently large ${v_{\rm L}>\overline{v}_{\rm L}}$ (as $c_4$ is negative) as well as sufficiently negative ${v<0}$, yielding the switch structure \eqref{eq:switchstrucmax}, but this is outside the domain of interest in this application.

We simulate both the nominal controller and safety-critical controller via numerical integration of the model \eqref{eq:truck_model2} from the initial condition ${\mb{x}(0) = [27.4, 16, 16]^\top\in\C}$. We use parameter values as specified in Table~\ref{tab:swtichparam}. The acceleration $a_{\rm L}$ of the lead vehicle is given by a time profile reflecting a hard braking event, as seen in Fig.~\ref{fig:truck_sim} (left). The velocity of the truck converges to zero and a crash does not occur for both controllers, but only the safety-critical controller ensures the truck maintains a safe distance (indicated by $h_{\rm QP}(\mb{x}(t)) \geq 0$) as seen in Fig.~\ref{fig:truck_sim} (center, right). We see that the nominal controller brakes less aggressively than the safety-critical controller, and thus does not react quickly enough to avoid violating the safe following distance requirement.

%%%%%%%%%%%%%%%%%%%%%%%%%%%%%%%%%%%%%%%%%%%%%%%%%%%%%%%%%%%%%%%%%%%%%%%%%%%%%%%%%%%%%%%
\section{Experimental Results \& Robust Design}\label{sec:exp}
%%%%%%%%%%%%%%%%%%%%%%%%%%%%%%%%%%%%%%%%%%%%%%%%%%%%%%%%%%%%%%%%%%%%%%%%%%%%%%%%%%%%%%%

% \begin{figure}[b]
% 	\centering
%     \includegraphics[trim=0 0 0 0,clip,width=\linewidth]{Cleaned_Figures/Hardware_v2.png}
% 	\caption{Connected Automated Truck used in the experiment.}  
% 	\label{fig:truck_hardware}
% \end{figure}

\begin{figure}
	\centering
    \includegraphics[trim=0 0 0 0,clip,width=.95\linewidth]{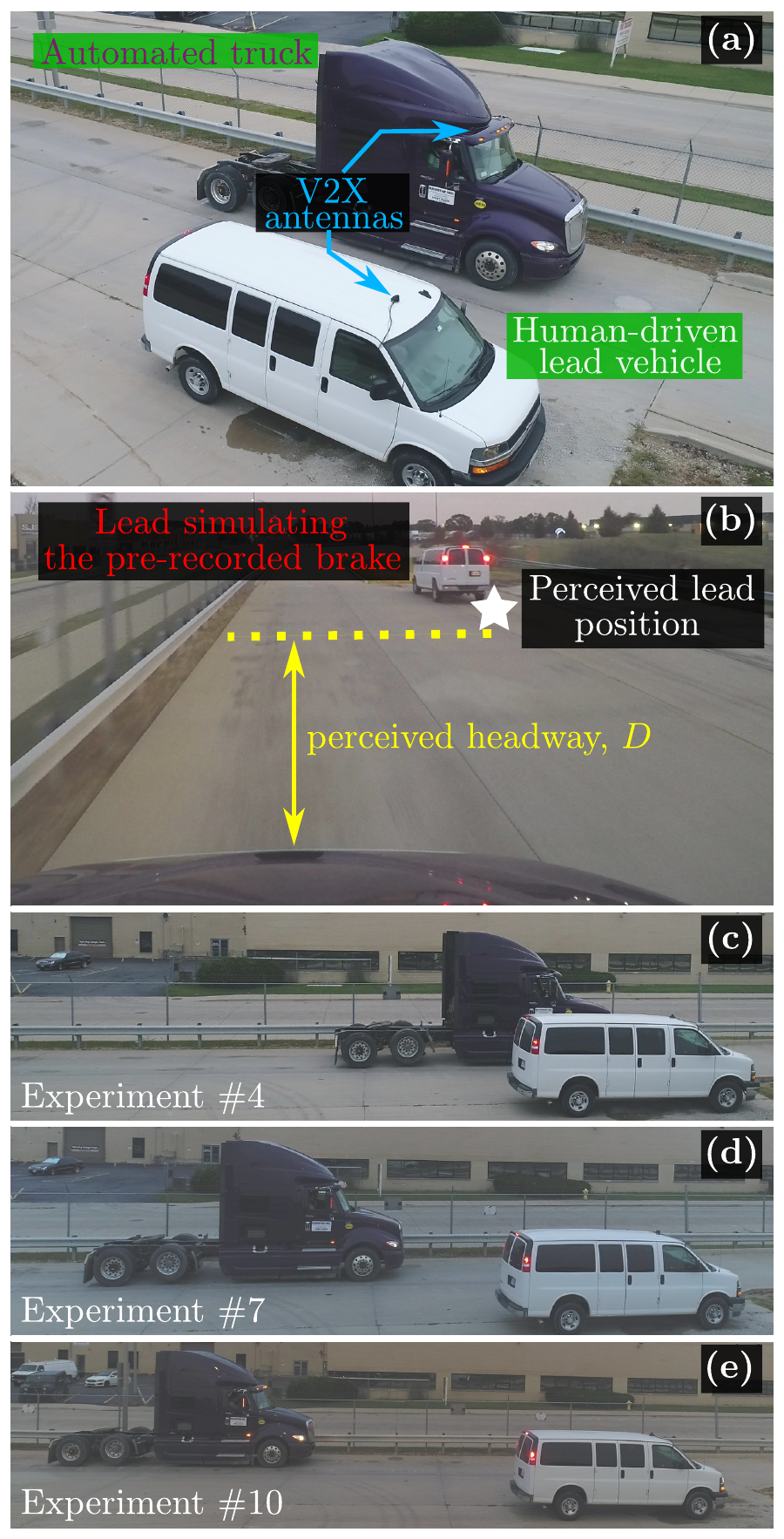}
	\caption{(a) Vehicles used in experiments. (b) Image from the dashboard of the truck during an experimental run. (c,d,e) Final configurations of separate experiments. See \cite{SupplementVideo} for a video.}
	\label{fig:ExperimentsConfiguration}
\end{figure}

In this section we provide a description of the automated truck experimental configuration and present results using the nominal and safety-critical controllers. Furthermore, we deploy the method of robust control design using ISSf developed in Section \ref{sec:issf}, and demonstrate its advantages experimentally.

\begin{figure*}[t]
	\centering
	\begin{subfloat}
	{\includegraphics[width=0.32\textwidth, valign = t]{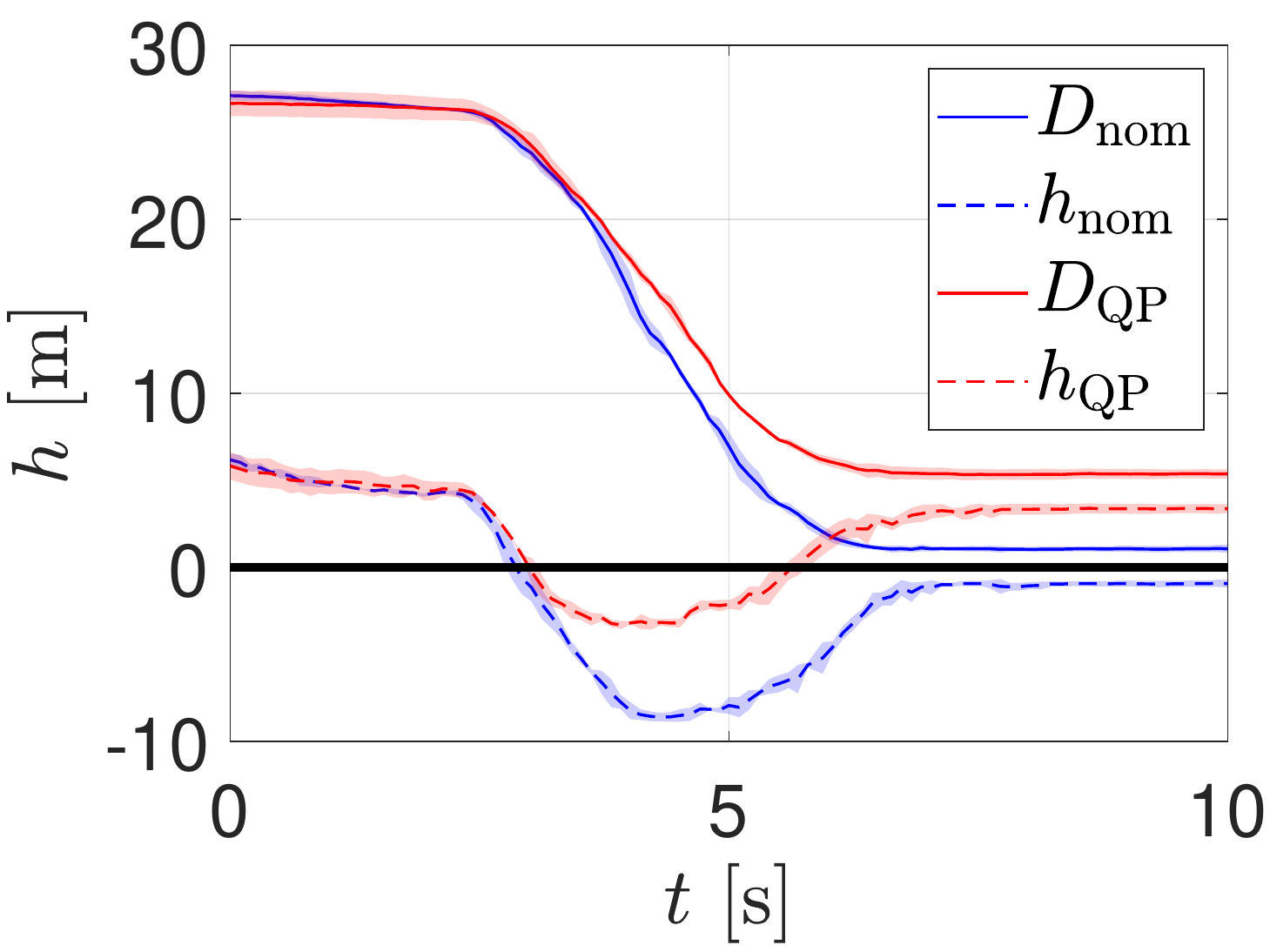}}
	\end{subfloat}
	\hfill
	\begin{subfloat}
	{\includegraphics[width=0.32\textwidth, valign = t]{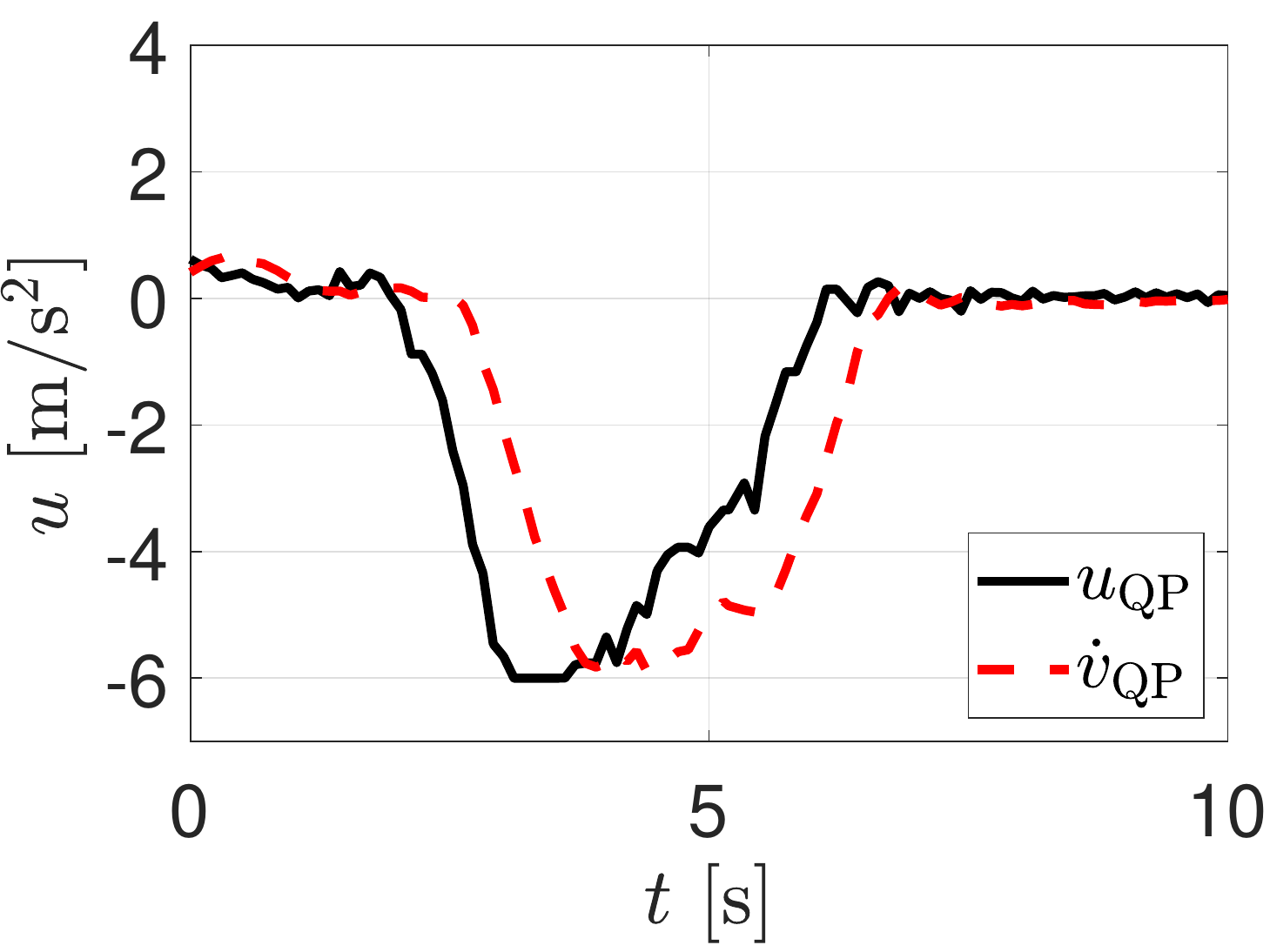}}
	\end{subfloat}
	\hfill
	\begin{subfloat}
	{\includegraphics[width=0.32\textwidth, valign = t]{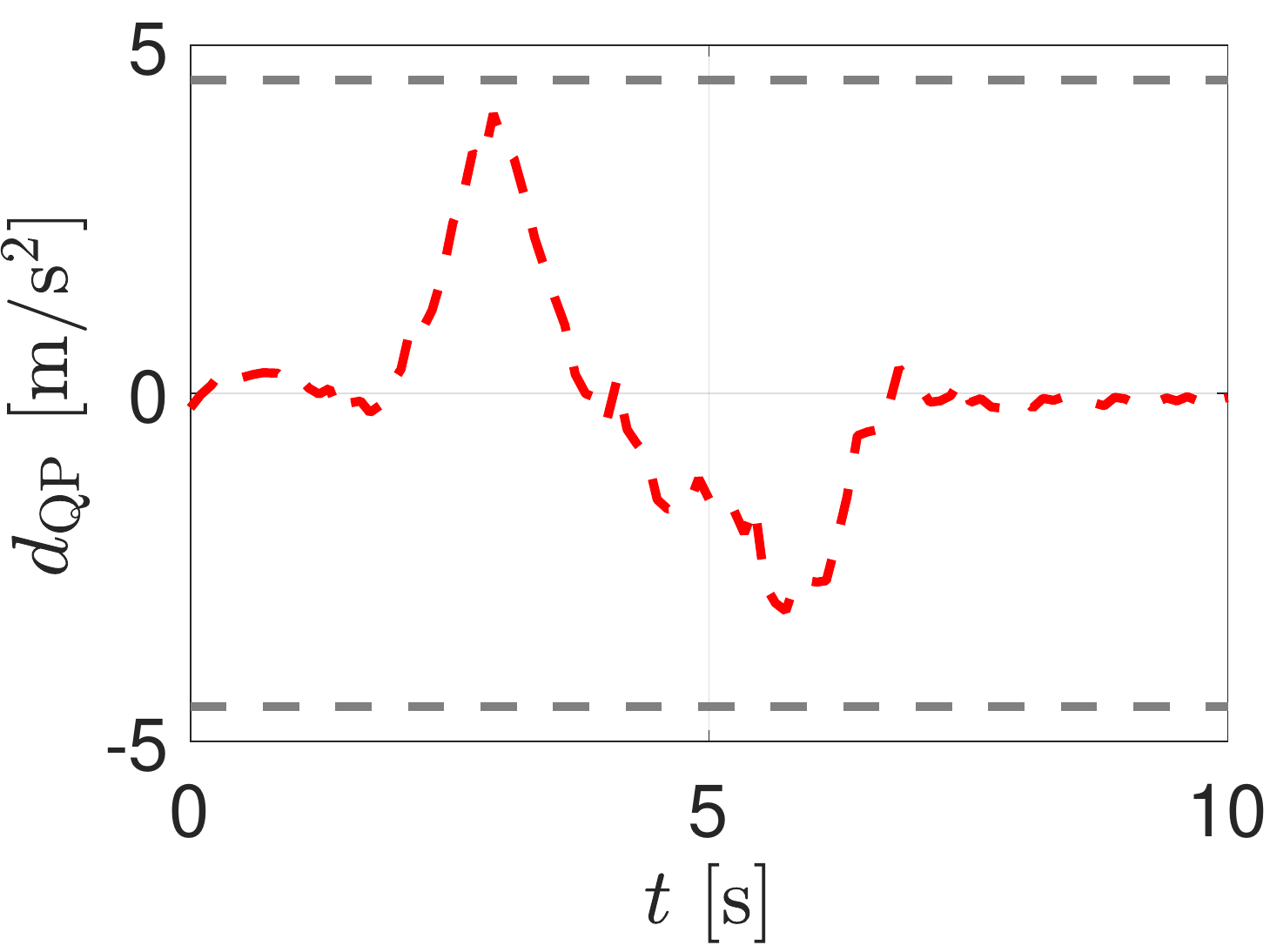}}
	\end{subfloat}
	\caption{(Left) Mean value (lines) and standard deviations (fills) of the distance $D$ and the CBF $h$ when using the nominal controller defined in \eqref{eq:truck_nominal} (blue) and the safety-critical controller defined in \eqref{eq:truck_kQPmin} (red) in the truck experiment. The repeated experiments with these controllers are highly consistent. (Center) Discrepancy between acceleration commanded by safety-critical controller (black) and actual acceleration of the automated truck (red). (Right) Disturbance signal in input seen by the truck used to define the model \eqref{eq:truck_simpmodeldist}. 
	}  
	\label{fig:truck_baseline_exps}
	\vspace{-3 mm}
\end{figure*}

%%%%%%%%%%%%%%%%%%%%%%%%%%%%%%%%%%%%%%%%%%%%%%%%%%%%%%%%%%%%%%%%%%%%%%%%%%%%%%%%%%%%%%%
\subsection{Experimental Setup and Procedure}     \label{sec:setupandprocedure}
%%%%%%%%%%%%%%%%%%%%%%%%%%%%%%%%%%%%%%%%%%%%%%%%%%%%%%%%%%%%%%%%%%%%%%%%%%%%%%%%%%%%%%%

The automated truck used in our experiments is an International ProStar+ Class-8 truck developed by the Navistar~\cite{navistar2021prostar}; see Fig.~\ref{fig:ExperimentsConfiguration}(a). Both the automated truck and the lead vehicle are equipped with a V2X Onboard Unit (OBU) developed by Commsignia~\cite{commsignia2021v2x}. These units are equipped with an accelerometer, gyroscope, magnetometer, and GPS unit.
Furthermore, these OBUs support peer-to-peer communication such that the automated truck may receive position, velocity, and acceleration data from the lead vehicle through V2X antennas shown in Fig.~\ref{fig:ExperimentsConfiguration}(a). The automated truck is additionally equipped with a Mobile Real-Time Targeting Machine developed by Speedgoat \cite{speedgoat2021mrtu}, which interfaces with the V2X OBU and the truck's Engine Controller Unit (ECU) through a Controller Area Network (CAN) bus. The Speedgoat runs the controller for the system given a measurement stream of values for $D, v, v_{\rm L}$, and $a_{\rm L}$ coming from the V2X OBUs. It computes a desired acceleration input and converts it to a corresponding torque value through a feed-forward map. A drive-by-wire system on the truck controls the engine and the brake torques accordingly. The steering of the truck is done manually by a human driver in the experiments.

In an effort to evaluate the repeatability of our experiments, it is necessary to eliminate variation in the lead vehicle's behavior, which is being driven by a human. To achieve this, we use a pre-recorded time profile of position, velocity, and acceleration of the lead vehicle while it performs a hard braking event. This profile for $a_{\rm L}$ and $v_{\rm L}$ is seen in the left and center panels of Fig.~\ref{fig:truck_sim}, and was used to produce our simulation results. 
We stream this data to the truck controller as the \textit{perceived lead vehicle} in our experiments.
Experiments also include a physical lead vehicle simulating the pre-recorded motion for visualization purposes; see Fig.~\ref{fig:ExperimentsConfiguration}(b).
Importantly, the evaluation of safety is derived from evaluating the CBF using the recorded time profiles rather than simply detecting collisions such as Fig.~\ref{fig:ExperimentsConfiguration}(c). A video of the experiments are available online \cite{SupplementVideo}.

%%%%%%%%%%%%%%%%%%%%%%%%%%%%%%%%%%%%%%%%%%%%%%%%%%%%%%%%%%%%%%%%%%%%%%%%%%%%%%%%%%%%%%%
\vspace{-2 mm}
\subsection{Input Disturbances}
%%%%%%%%%%%%%%%%%%%%%%%%%%%%%%%%%%%%%%%%%%%%%%%%%%%%%%%%%%%%%%%%%%%%%%%%%%%%%%%%%%%%%%%

We deploy both the nominal and safety-critical controller on the automated truck with results as seen in the left panel of Fig.~\ref{fig:truck_baseline_exps}. We see that not only does the nominal controller consistently fails to meet the safety requirements imposed by the CBF $h$, but the safety-critical controller also consistently fails to meet the safety requirements. The top row in Fig.~\ref{fig:intro_truck} illustrates an experimental run with the nominal controller. 

To understand why the safety-critical controller fails, we examine the discrepancy between the commanded acceleration and actual acceleration of the automated truck, as seen in the center panel of Fig.~\ref{fig:truck_baseline_exps}. One may observe a delay between the commanded acceleration and the achieved acceleration. This delay in acceleration is due to the fact that the power generation of the truck is a complex nonlinear dynamical system that has been imperfectly abstracted away by the feed-forward maps that allow the simplified model in \eqref{eq:truck_simpmodel}. Rather than attempting to work with this complex nonlinear dynamic system and improving the feed-forward maps, we describe the discrepancy in commanded and actual acceleration as a disturbance in the simplified model:
\begin{equation}
\label{eq:truck_simpmodeldist}
    \dot{v} = u + d(t),
\end{equation}
where ${d:\R_{\geq 0}\to\R}$ reflects the difference between commanded acceleration and actual acceleration.

As the disturbance $d$ is caused by the complicated interactions of the drive-by-wire system and power generation dynamics, it may be difficult to use model-based techniques to construct a meaningful bound $\delta$ for the worst-case disturbance. Instead, we estimate the worst-case disturbance empirically by comparing the actual acceleration $\dot{v}(t)$ to the commanded acceleration $u(t)$. In the right panel of Fig.~\ref{fig:truck_baseline_exps}, we see that the largest difference in the commanded and actual acceleration is around $4$~[m/s$^2$]. Thus, we study the degradation of safety of the system taking a slightly larger value $\delta = 4.5$~[m/s$^2$].

% There is a restriction imposed on $u$ due to the physical limitation of the engine and brakes:
% \begin{equation} 
%     \label{eq:truck_limit_u}
%     u \in [-\underline{a}, \overline{a}].
% \end{equation}
% The unknown and time varying disturbance in the experimental setup is illustrated in Fig.~\ref{fig:d_exp}, where a time series of acceleration command $u$ and corresponding measured acceleration $\dot{v}$ to a braking event are depicted.

% \begin{figure}[h]
%     \vspace{-0 mm}
%     \centering
%     \begin{subfloat}
%       {\includegraphics[trim=5 0 30 8,clip, width=.95\linewidth]{Figures/disturbance_v3.eps}}
%     \end{subfloat} \\
%     %\hfill
%     \begin{subfloat}
%       {\includegraphics[trim=5 0 30 8,clip, width=.95\linewidth]{Figures/disturbance_v4.eps}}
%     \end{subfloat}
%     \caption{Input disturbance $d$ in the experimental system.}  
%     \vspace{-0 mm}
%     \label{fig:d_exp}
% \end{figure}

%%%%%%%%%%%%%%%%%%%%%%%%%%%%%%%%%%%%%%%%%%%%%%%%%%%%%%%%%%%%%%%%%%%%%%%%%%%%%%%%%%%%%%%
\vspace{-2 mm}
\subsection{Robust Design}
%%%%%%%%%%%%%%%%%%%%%%%%%%%%%%%%%%%%%%%%%%%%%%%%%%%%%%%%%%%%%%%%%%%%%%%%%%%%%%%%%%%%%%%

\begin{figure*}[t]
\centering
	{\includegraphics[width=0.4\textwidth, valign = t]{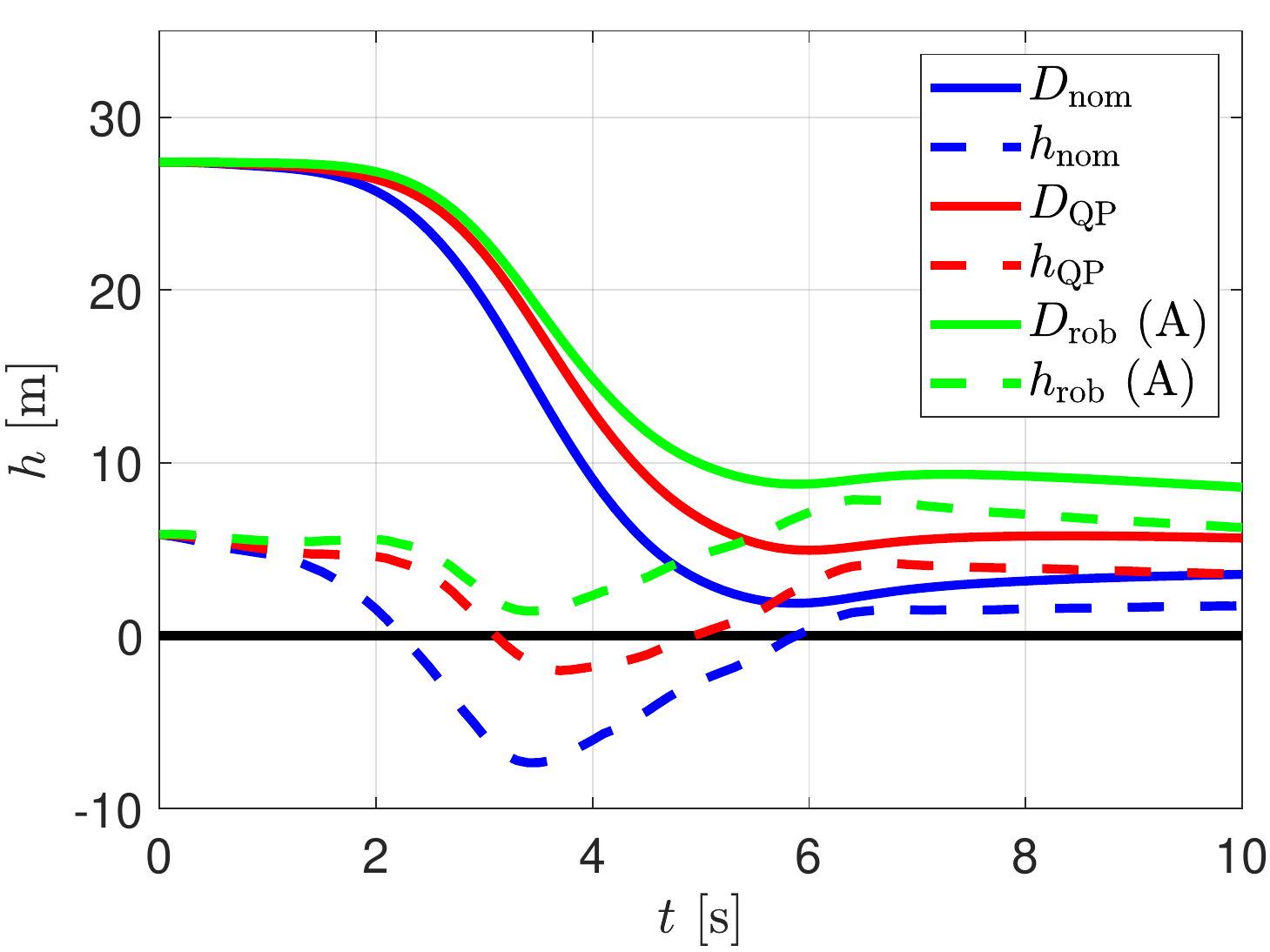}}
	\qquad
	{\includegraphics[width=0.4\textwidth, valign = t]{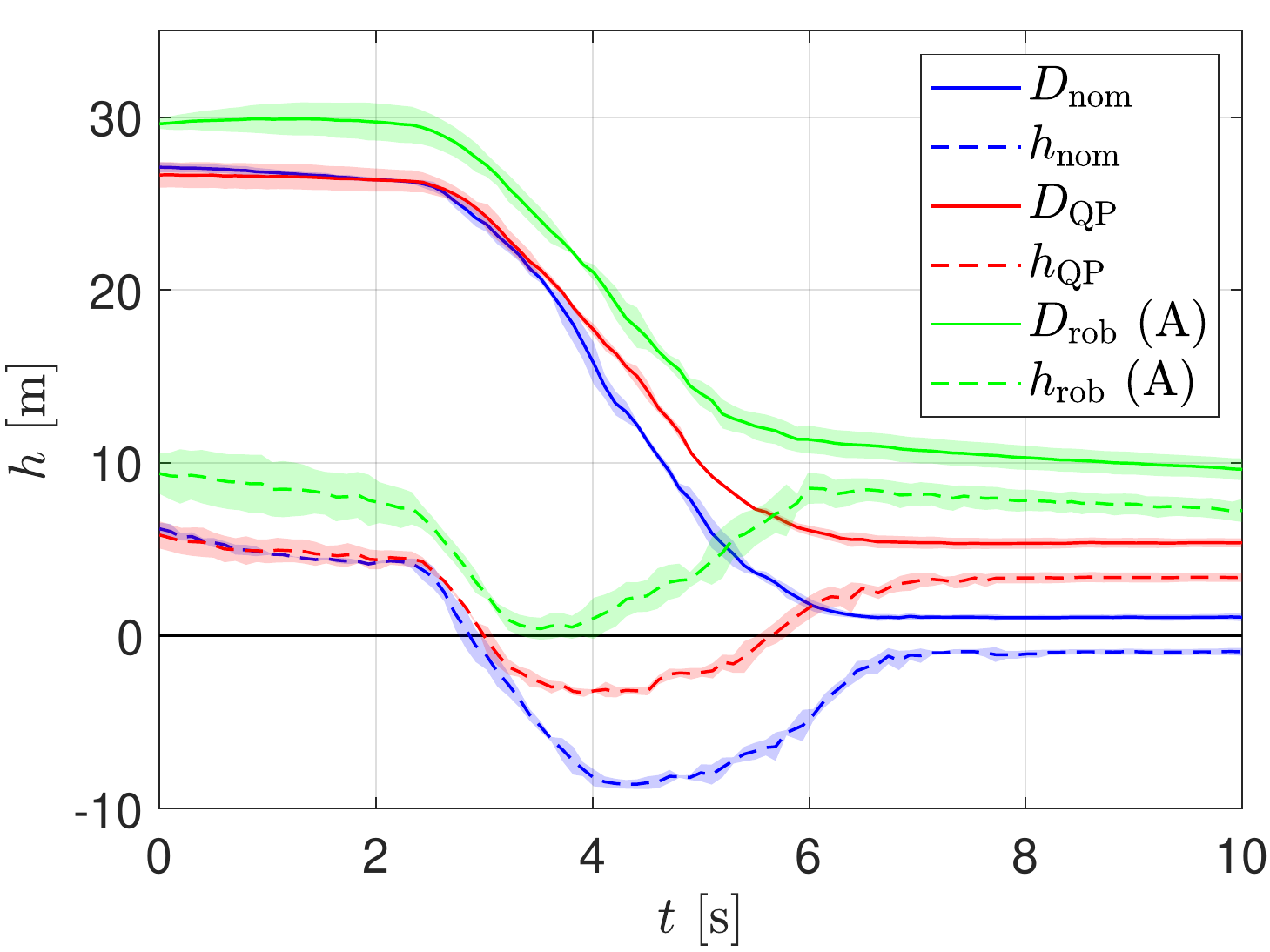}}
	\caption{(Left) Following distance and value of ISSf-CBF using the nominal controller (blue), safety-critical controller (red), and robust safety-critical controller (green) in the disturbed simulation. (Right) Mean value (lines) and standard deviations (fills) of the distance $D$ and the ISSf-CBF $h$  using the nominal controller (blue), safety-critical controller (red), and robust safety-critical controller (green) in  experiment.}  
	\label{fig:truck_dist_sim}
	\vspace{-3 mm}
\end{figure*}

To overcome this disturbance and improve the safe behavior of the truck, we deploy the tools of ISSf-CBFs described in Section \ref{sec:issf}. As $h$ satisfies the CBF condition \eqref{eq:cbf_alt}, it also satisfies the ISSf-CBF condition \eqref{eq:issf-cbf}, where we take:
\begin{equation}\label{eq:expeps}
    \epsilon(r) = \epsilon_0\mathrm{e}^{\lambda r},
\end{equation}
with ${\epsilon_0> 0}$ and ${\lambda\geq 0}$. The parameter $\lambda$ introduces a measure of flexibility by allowing one to require a greater degree of robustness when the truck is close to the leading vehicle, and less robustness when the distance is greater. Given \eqref{eq:expeps}, the forward invariant set is given by:
\begin{align}\label{eq:truck_Cdelta}
\C_\delta = \left\{\left. \begin{bmatrix} D \\ v \\ v_{\rm L} \end{bmatrix} \in \R^3 ~\right|~ h(D,v,v_{\rm L}) \geq - \frac{ \epsilon_0 {\rm e}^{ {\lambda} h(D,v,v_{\rm L}) } \delta^2}{4\alpha_{\rm c}} \right\}.
\end{align}
As discussed in the inverted pendulum example, the set $C_{\delta}$ being forward invariant implies that ${h(\mb{x}(t))\geq h^*}$, where $h^*$ is the value of the ISSf-CBF $h$ on the boundary of $C_{\delta}$, which can be calculated by solving \eqref{eq:solveforCdelta}.
The value of $h^*$ for different choices of $\epsilon_0$ and $\lambda$ can be seen in Table~\ref{tab:eps0_lambda_truck}. We then construct an optimization-based controller giving the switch structure  \eqref{eq:switchTISSfmin}, since ${L_\mb{g}h(D,v,v_{\rm L})<0}$ for ${v\geq0}$ and ${v_{\rm L}\in[0,\overline{v}_{\rm L}]}$ (cf.~\eqref{eq:LgLf}). This results in:
\begin{equation}\label{eq:truck_krobmax}
k_{\rm rob}(D,v,v_{\rm L}) = \min\left\{k_{\rm n}(D,v,v_{\rm L}), \overline{k}_{\rm s}(D,v,v_{\rm L}) \right\},
\end{equation}
where:
\begin{equation}
    \overline{k}_{\rm s}(D,v,v_{\rm L}) = k_{\rm s}(D,v,v_{\rm L})+\frac{ L_\mb{g}h(D,v,v_{\rm L}) }{ \epsilon_0 {\rm e}^{ \lambda h(D,v,v_{\rm L})} },
\end{equation}
and ${k_{\rm n}}$ and ${k_{\rm s}}$ are given by \eqref{eq:truck_nominal} and \eqref{eq:truck_ks}, respectively.

We simulate the nominal controller, safety-critical controller, and robust safety-critical controller via numerical integration of the model \eqref{eq:truck_model2} from the initial condition ${\mb{x}(0) = [27.4, 16, 16]^\top\in\C}$ while disturbing the input using the signal in shown in the right panel of Fig.~\ref{fig:truck_baseline_exps}. We use parameter values as specified in Table~\ref{tab:swtichparam}. We see in the left panel of Fig. \ref{fig:truck_dist_sim} that introducing the disturbance signal into our simulation allows us to recreate the failures of the nominal controller and safety-critical controller that we saw experimentally in Fig. \ref{fig:truck_baseline_exps}. Furthermore, we see that the robust safety-critical controller maintains the safety of the system even in the presence of the disturbance.

%%%%%%%%%%%%%%%%%%%%%%%%%%%%%%%%%%%%%%%%%%%%%%%%%%%%%%%%%%%%%%%%%%%%%%%%%%%%%%%%%%%%%%%
\subsection{Robust Experimental Results}
\label{sec:expresults}
%%%%%%%%%%%%%%%%%%%%%%%%%%%%%%%%%%%%%%%%%%%%%%%%%%%%%%%%%%%%%%%%%%%%%%%%%%%%%%%%%%%%%%%

\begin{table}[b]
\vspace{-3 mm}
\centering
\begin{tabular}{|c|c|c|c|c|c|}
\hline
 & $\epsilon_0$ & $\lambda$ & $h^*$ & $h_{\rm min}$& $\tilde{D}_{\rm ss}$ \\ 
 Label & [s$^3$/m] & [1/m] & [m] & [m] & [m] \\ \hline
 (B) & 0.8          & 0     &  $-$40.50   & 22.09             & 25.43            \\ \hline
 & 3          & 0      &  $-$151.88  & 2.99              & 6.19            \\ \hline
 (D) & 4          & 0      &  $-$202.50  & 1.02              & 4.69            \\ \hline
 & 5          & 0      &  $-$253.13  & $-$0.45             & 3.54            \\ \hline
 (A) & 0.5          & 0.4     &  $-$4.38    & 0.35              & 4.70            \\ \hline
 & 0.5          & 0.5      &  $-$3.80    & $-$1.27             & 2.22            \\ \hline
 (C) & 0.8          & 0.25      &  $-$7.01    & 0.78              & 4.34            \\ \hline
 & 0.8          & 0.35      &  $-$5.64    & $-$1.03             & 2.22            \\ \hline
 & 1.0          & 0.25      &  $-$7.59    & $-$0.86             & 3.07            \\ \hline
\end{tabular}
\caption{Sets of parameter values used for the exponential function \eqref{eq:expeps} in the automated truck experiments with theoretical safety guarantee $h^*$, minimum experimental value of the ISSf-CBF $h_{\rm min}$, and shift in the steady-state tracking distance $\tilde{D}_{\rm ss}$ by \eqref{eq:Dss_tilde}.}
\label{tab:eps0_lambda_truck}
\end{table}

Here we show the results when the robust safety-critical controller is deployed on the connected automated truck. Sets of three experimental runs were conducted using each parameter pair $\epsilon_0$ and $\lambda$ shown in Table~\ref{tab:eps0_lambda_truck}. The experimental results using the parameter set ${\epsilon_0 = 0.5}$~[s$^3$/m] and ${\lambda = 0.4}$~[1/m] (labeled as parameter pair (A)) can be seen in the right panel of Fig.~\ref{fig:truck_dist_sim} and are visualized at the bottom row of Fig.~\ref{fig:intro_truck}. With these parameters the system is rendered safe, as the value of $h$ does not drop below $0$. Although the robust safety-critical controller displays a larger standard deviation across the three experimental runs compared to the nominal and safety-critical controllers, it consistently satisfies the original safety requirement.

% \begin{figure*}[t]
%     \vspace{-0mm}
%     \centering
%     \includegraphics[ width=0.4\textwidth]{Cleaned_Figures/truck_eps0_lambda.pdf}
%     \qquad
%     \includegraphics[ width=0.4\textwidth]{Cleaned_Figures/hstar_Dss_trade.pdf}
%     \caption{(Left) Parameter values for $\epsilon_0$ and $\lambda$ used in the truck experiments, with contours showing theoretical values of $h^*$. Green markers denote parameter sets which achieve the original safety goal (${h\geq 0}$), while red markers denote parameter sets for which the original safety goal is violated. (Right) Theoretical values of $h^*$ and the shift in steady-state tracking distance, denoted by $\tilde{D}_{\rm ss}$, for the parameter sets used in the truck experiments. The blue markers denote parameter sets with ${\lambda> 0}$, while the black markers denote parameter sets with ${\lambda = 0}$.}
%     \label{fig:truck_eps0_lambda}
%     \vspace{-3 mm}
% \end{figure*}

\begin{figure*}[t]
	\centering
	\begin{subfloat}
	{\includegraphics[width=0.32\textwidth, valign = t]{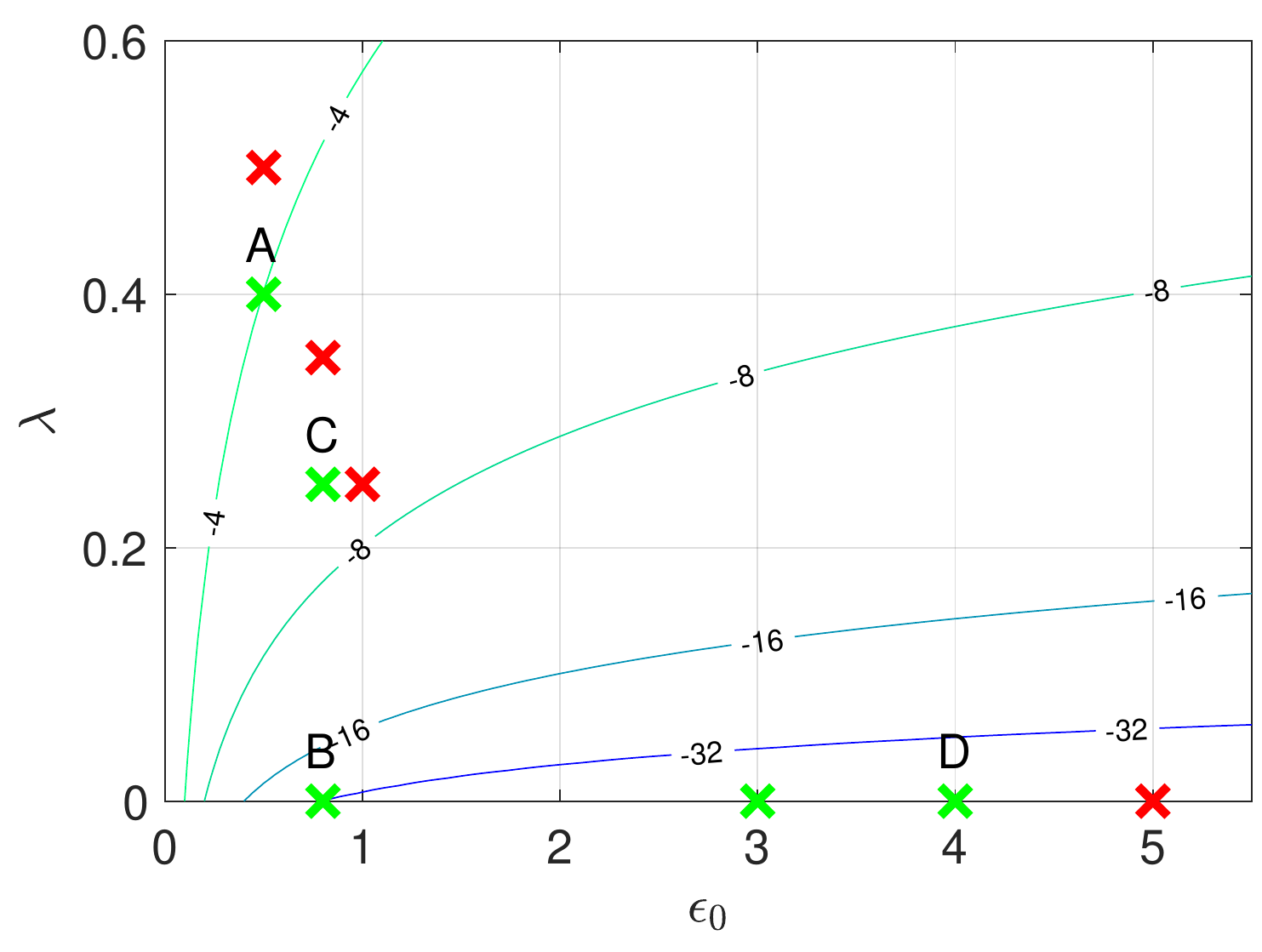}}
	\end{subfloat}
	\hfill
	\begin{subfloat}
	{\includegraphics[width=0.32\textwidth, valign = t]{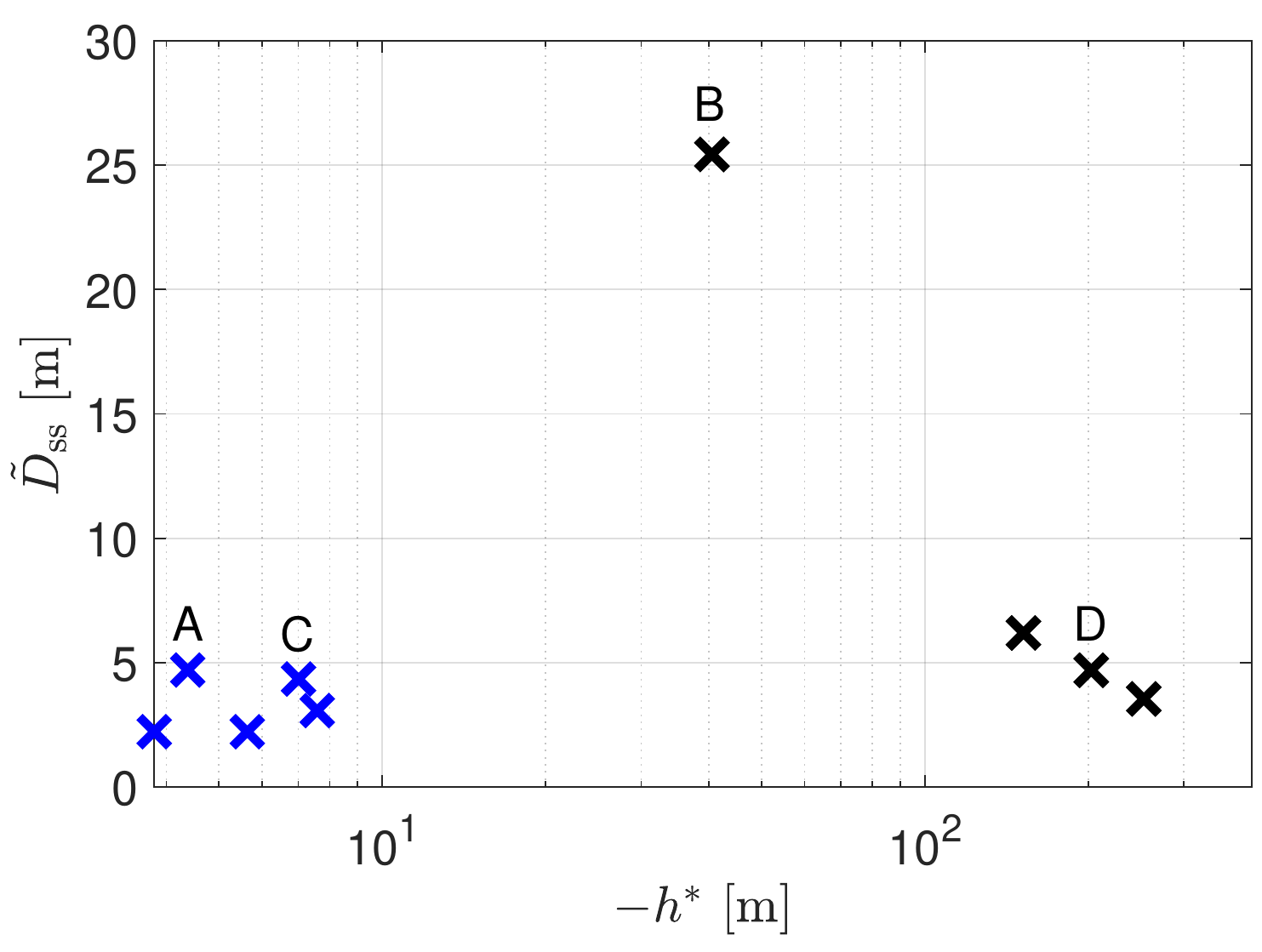}}
	\end{subfloat}
	\hfill
	\begin{subfloat}
	{\includegraphics[width=0.32\textwidth, valign = t]{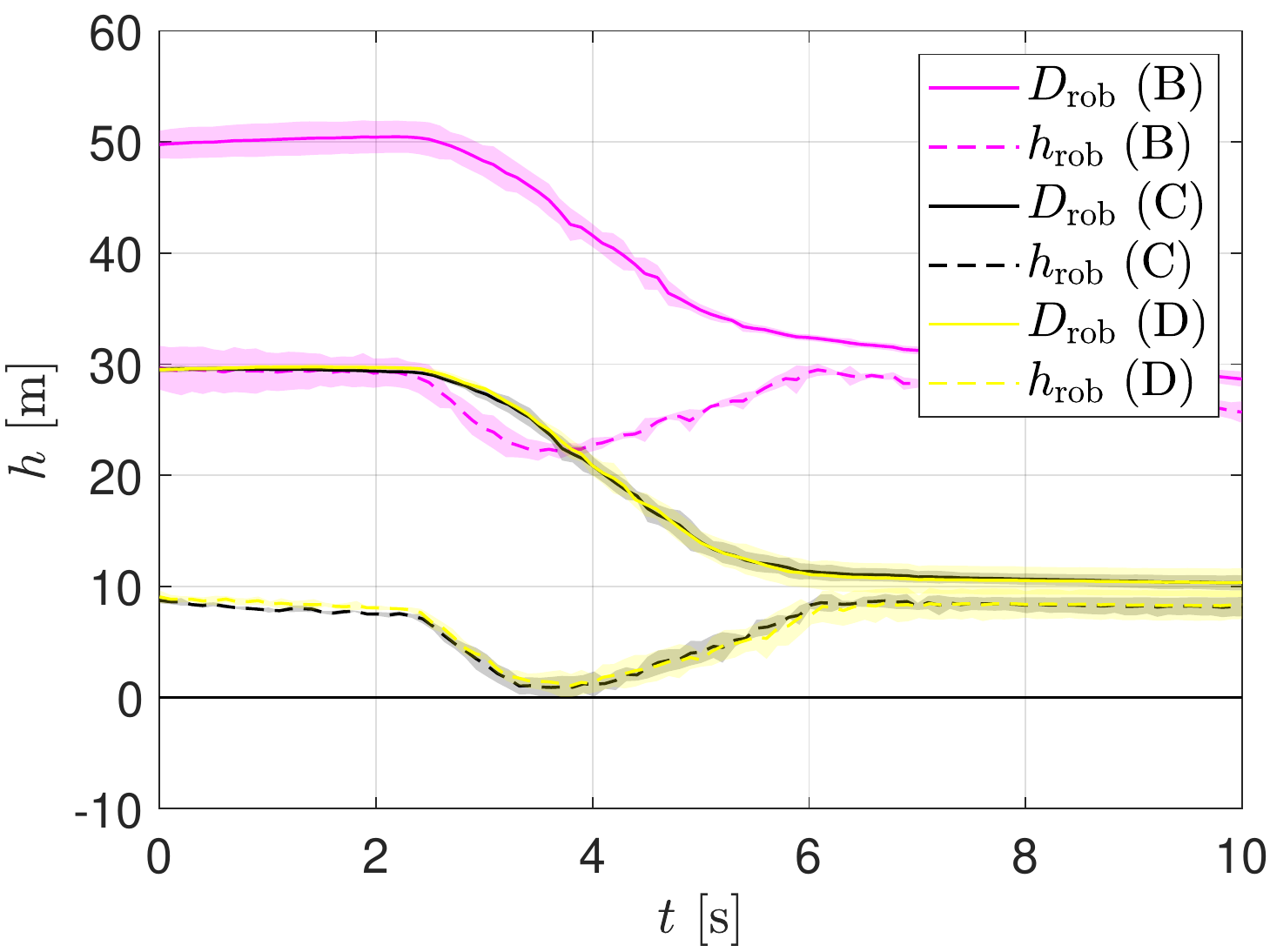}}
	\end{subfloat}
	\caption{(Left) Parameter values for $\epsilon_0$ and $\lambda$ used in the truck experiments, with contours showing theoretical values of $h^*$. Green markers denote parameter sets which achieve the original safety goal (${h\geq 0}$), while red markers denote parameter sets for which the original safety goal is violated. (Center) Theoretical values of $h^*$ and the shift in steady-state tracking distance, denoted by $\tilde{D}_{\rm ss}$, for the parameter sets used in the truck experiments. The blue markers denote parameter sets with ${\lambda> 0}$, while the black markers denote parameter sets with ${\lambda = 0}$.
	(Right) Experimental results for parameter pairs (B), (C) and (D) in Table~\ref{tab:eps0_lambda_truck}. Case (B) is highly conservative as indicated by the large steady-state tracking distance error. Cases (C) and (D) display nearly identical behavior, though case (C) possesses a much stronger theoretical guarantee. 
	}  
	\label{fig:truck_eps0_lambda}
	\vspace{-3 mm}
\end{figure*}

When evaluating how the system behavior depends on the values of the parameters $\epsilon_0$ and $\lambda$, we first consider whether the original safety requirement is met, i.e., whether or not the value of $h$ remains positive. While the robust-safety critical controller does not provide a theoretical guarantee that $h$ will remain non-negative (it only guarantees that ${h(\mb{x}(t)) \geq h^*}$), for certain values of $\epsilon_0$ and $\lambda$ the original safety requirement are still met, as seen in the inverted pendulum example as well as the connected automated truck experiments. The minimum value $h_{\rm min}$ of the barrier function, observed during the experimental runs, is shown in Table~\ref{tab:eps0_lambda_truck}. This is also visualized in the left panel of Fig.~\ref{fig:truck_eps0_lambda}, where green markers indicate sets of parameter values for which the safety requirement is met, and red markers indicate those for which it is not met. We see that safety can be achieved using the original ISSf-CBF formulation in \cite{kolathaya2018input} (where ${\lambda = 0}$) for sufficiently small values of $\epsilon_0$, but may also be achieved using small values of $\lambda$. 

We remark that when changing the controller from $k_{\rm QP}$ (cf.~\eqref{eq:truck_kQPmin}) to $k_{\rm rob}$ (cf.~\eqref{eq:truck_krobmax}) the equilibrium of the system is shifted
as can be noticed once comparing the runs on the right panel of Fig.~\ref{fig:truck_dist_sim}. We characterize this by the shift in the steady-state tracking distance error defined as
\begin{equation}
    \label{eq:Dss_tilde}
    \tilde{D}_{\rm ss} \triangleq D^{\rm exp}_{\rm ss} - D^*.
\end{equation}
Here $D^{\rm exp}_{\rm ss}$ is the steady-state distance captured in experiments when the leader is moving with the steady-state speed ${v^*\in(0,\overline{v}_{\rm L})}$ before braking. The term ${D^*=V^{-1}(v^*)}$ captures the desired steady-state distance given by the inverse of the range policy \eqref{eq:truck_V}. In the experiments we have ${v^*=16}$~[m/s], yielding ${D^*=25}$~[m]. The values of $\tilde{D}_{\rm ss}$ corresponding to different parameter pairs are given
in Table~\ref{tab:eps0_lambda_truck}.
In the right panel of Fig.~\ref{fig:truck_eps0_lambda} we visualize the theoretical values of $h^*$ and the experimental values of $\tilde{D}_{\rm ss}$ for different parameter sets. The black markers indicate parameter sets with ${\lambda = 0}$, while the blue markers show parameter sets with ${\lambda > 0}$. With ${\lambda = 0}$, the theoretical guarantees are nearly meaningless (observe the large negative values of $h^*$), and improving them requires dramatically increasing $\tilde{D}_{\rm ss}$. In contrast, the parameter sets with ${\lambda> 0}$ allow us to obtain significantly (an order of magnitude) stronger theoretical guarantees without greatly increasing $\tilde{D}_{\rm ss}$, thereby also achieve good performance.
In the right panel of Fig.~\ref{fig:truck_baseline_exps} we give experimental results of three other parameter pairs labeled as (B), (C) and (D) in Table~\ref{tab:eps0_lambda_truck}.
The poor performance of case (B) is indicated by the large value of $\tilde{D}_{\rm ss}$. The results for cases (C) and (D) nearly overlap,  but the introduction of $\lambda$ allows strong theoretical guarantee for case (C) which is missing for case (D).

\section{Conclusion} \label{sec:conclusion}
%%%%%%%%%%%%%%%%%%%%%%%%%%%%%%%%%%%%%%%%%%%%%%%%%%%%%%%%%%%%%%%%%%%%%%%%%%%%%%%%%%%%%%%

In conclusion, this work has developed a theoretically rigorous approach for safety-critical control synthesis through Control Barrier Functions (CBFs). The notion of Input-to-State Safety (ISSf) is utilized to capture the impact of disturbances in the input to the system. A simple parametric modification to CBFs enabled the formulation of ISSf-CBFs as a practical tool for achieving both performant behavior and meaningful theoretical safety guarantees. We provided a tutorial on these tools in the context of an inverted pendulum system, and carried out a practical design problem of a safety-critical controller for a connected automated truck. Moreover, we demonstrated the tangible benefits of the design using ISSf-CBFs by deploying this controller experimentally on an automated truck.

%%%%%%%%%%%%%%%%%%%%%%%%%%%%%%%%%%%%%%%%%%%%%%%%%%%%%%%%%%%%%%%%%%%%%%%%%%%%%%%%%%%%%%%
\vspace{-3 mm}
\appendix

%%%%%%%%%%%%%%%%%%%%%%%%%%%%%%%%%%%%%%%%%%%%%%%%%%%%%%%%%%%%%%%%%%%%%%%%%%%%%%%%%%%%%%%
\subsection{Proof of Theorem \protect \ref{thm:equiv}}     \label{app:ProofTheo2}
%%%%%%%%%%%%%%%%%%%%%%%%%%%%%%%%%%%%%%%%%%%%%%%%%%%%%%%%%%%%%%%%%%%%%%%%%%%%%%%%%%%%%%%
\begin{proof}
We first prove that the optimization problem in \eqref{eq:SC-QP} has a closed-form solution given by \eqref{eq:SC-CF} and \eqref{eq:lambdastar}, thereby proving it is feasible for any ${\mb{x}\in\R^n}$ and satisfies ${\mb{k}_{\rm QP}(\mb{x})\in K_{\rm CBF}(\mb{x})}$ for all ${\mb{x}\in\R^n}$. Then we prove that $\mb{k}_{\rm QP}$ is a continuous function.

Let us first consider an ${\mb{x}\in\R^n}$ such that ${L_{\mb{g}}h(\mb{x}) = \mb{0}}$. By assumption, the function $h$ is a CBF for \eqref{eq:eom} on the set $\C$ with corresponding function ${\alpha\in\mathcal{K}^{\rm e}_\infty}$. Thus, we know from the condition in \eqref{eq:cbf_alt} that
\begin{equation}
    L_{\mb{f}}h(\mb{x}) + \alpha(h(\mb{x})) > 0, 
\end{equation}
such that the inequality constraint in \eqref{eq:SC-QP} is satisfied for any choice of $\mb{u}$. The definition of a norm requires that for any ${\mb{y}\in\R^m}$, we have  ${\Vert\mb{y}\Vert_2\geq 0}$ and ${\Vert\mb{y}\Vert_2 = 0}$ implies ${\mb{y} = \mb{0}}$. Thus, we may conclude that the minimizing choice of $\mb{u}$ is given by ${\mb{u}=\mb{k}_{\rm n}(\mb{x})}$, such that ${\mb{k}_{\rm QP}(\mb{x}) = \mb{k}_{\rm n}(\mb{x})}$ as required by the closed-form solution in \eqref{eq:SC-CF} and \eqref{eq:lambdastar}.

Next let us consider an ${\mb{x}\in\R^n}$ such that ${L_{\mb{g}}h(\mb{x})\neq\mb{0}}$. Observe that the cost function and constraint function defining \eqref{eq:SC-QP} are both convex and continuously differentiable with respect to the decision variable $\mb{u}$. Thus the optimization problem is convex, and the \textit{Karush-Kuhn Tucker} (KKT) conditions provide a necessary and sufficient\footnote{An additional \textit{constraint qualification} is necessary for the KKT conditions to be necessary and sufficient conditions for optimality. One such qualification is \textit{Slater's Condition} \cite[\S 5.2.3]{boyd2004convex}, which is easily verified to hold in our setting.} condition for optimality \cite[\S 5.5.3]{boyd2004convex}. More precisely, the KKT conditions state that for an optimal solution ${\mb{u}^*\in\R^m}$ to \eqref{eq:SC-CF}, we must have a ${\mu^*\in\R}$ such that:
\begin{align}
    L_{\mb{f}}h(\mb{x})+L_{\mb{g}}h(\mb{x})\mb{u}^*+\alpha(h(\mb{x})) &\geq 0, \label{eqn:primefeas} \\
    \mu^* &\geq 0, \label{eqn:dualfeas} \\
    \mu^*( L_{\mb{f}}h(\mb{x})+L_{\mb{g}}h(\mb{x})\mb{u}^*+\alpha(h(\mb{x}))) &= 0, \label{eqn:compslack} \\
    \mb{u}^*-\mb{k}_{\rm n}(\mb{x})-\mu^*L_{\mb{g}}h(\mb{x})^\top &= 0. \label{eqn:station}
\end{align}
The first and second conditions are referred to as \textit{primal} and \textit{dual feasibility}, respectively. The third condition is referred to as \textit{complementary slackness}, and the fourth condition is referred to as \textit{stationarity}.

Rearranging the stationarity condition \eqref{eqn:station} yields
\begin{equation}
\label{eqn:station2}
     \mb{u}^*=\mb{k}_{\rm n}(\mb{x})+\mu^*L_{\mb{g}}h(\mb{x})^\top.
\end{equation}
To solve for the value of $\mu^*$ (and consequently $\mb{u}^*$), we use the primal feasibility condition \eqref{eqn:primefeas} and the complementary slackness condition \eqref{eqn:compslack}. In particular, suppose that
\begin{equation}
\label{eqn:case1}
     L_{\mb{f}}h(\mb{x})+L_{\mb{g}}h(\mb{x})\mb{u}^*+\alpha(h(\mb{x})) > 0.
\end{equation}
The complementary slackness condition \eqref{eqn:compslack} then implies ${\mu^* = 0}$, and thus, we have from \eqref{eqn:station2} that ${\mb{u}^*=\mb{k}_{\rm n}(\mb{x})}$. Combining this with \eqref{eqn:case1}, we obtain
\begin{equation}
\label{eqn:case1outcome}
     L_{\mb{f}}h(\mb{x})+L_{\mb{g}}h(\mb{x})\mb{k}_{\rm n}(\mb{x})+\alpha(h(\mb{x})) > 0.
\end{equation}
Next let us suppose that
\begin{equation}
\label{eqn:case2}
     L_{\mb{f}}h(\mb{x})+L_{\mb{g}}h(\mb{x})\mb{u}^*+\alpha(h(\mb{x})) = 0.
\end{equation}
Using the expression for $\mb{u}^*$ in \eqref{eqn:station2} yields
\begin{equation}
     L_{\mb{f}}h(\mb{x})+L_{\mb{g}}h(\mb{x})\mb{k}_{\rm n}(\mb{x})+\mu^*\Vert L_{\mb{g}}h(\mb{x})\Vert_2^2+\alpha(h(\mb{x})) = 0,
\end{equation}
which may be solved for $\mu^*$, yielding
\begin{equation}
\label{eqn:mustar}
    \mu^* = -  \frac{L_{\mb{f}}h(\mb{x})+L_{\mb{g}}h(\mb{x})\mb{k}_{\rm n}(\mb{x})+\alpha(h(\mb{x}))}{\Vert L_{\mb{g}}h(\mb{x})\Vert_2^2}.
\end{equation}
Given this expression, the dual feasibility condition \eqref{eqn:dualfeas}
requires that, if the equality in \eqref{eqn:case2} holds, we must have
\begin{equation}
\label{eqn:case2outcome}
     L_{\mb{f}}h(\mb{x})+L_{\mb{g}}h(\mb{x})\mb{k}_{\rm n}(\mb{x})+\alpha(h(\mb{x})) \leq 0.
\end{equation}
Substituting \eqref{eqn:mustar} into \eqref{eqn:station2} yields
\begin{equation}
\label{eqn:case2control}
     \mb{u}^*=\mb{k}_{\rm n}(\mb{x})-  \frac{L_{\mb{f}}h(\mb{x})+L_{\mb{g}}h(\mb{x})\mb{k}_{\rm n}(\mb{x})+\alpha(h(\mb{x}))}{\Vert L_{\mb{g}}h(\mb{x})\Vert_2^2}L_{\mb{g}}h(\mb{x})^\top.
\end{equation}
Noting that \eqref{eqn:case1} and \eqref{eqn:case1outcome} are equivalent, and \eqref{eqn:case2} and \eqref{eqn:case2outcome} are equivalent, we may combine these results with the preceding results obtained for ${L_{\mb{g}}h(\mb{x}) = \mb{0}}$ and conclude that
\begin{equation}
    \mb{k}_{\rm QP}(\mb{x}) = \mb{k}_{\rm n}(\mb{x})+\max\{0,\eta(\mb{x})\}L_{\mb{g}}h(\mb{x})^\top.
\end{equation}

% In the case that \eqref{eqn:case1outcome} holds, then we have that:
% \begin{equation}
%     \dot{h}(\mb{x},\mb{k}_{\rm QP}(\mb{x})) =  L_{\mb{f}}h(\mb{x})+L_{\mb{g}}h(\mb{x})\mb{k}_{\rm n}(\mb{x})>-\alpha(h(\mb{x})) ,
% \end{equation}
% such that $\mb{k}_{\rm QP}(\mb{x})\in K_{\rm CBF}(\mb{x})$. In the case that \eqref{eqn:case2outcome} holds, we have that:
% \begin{equation}
%     \dot{h}(\mb{x},\mb{k}_{\rm QP}(\mb{x})) = -\alpha(h(\mb{x})) ,
% \end{equation}
% such that $\mb{k}_{\rm QP}(\mb{x})\in K_{\rm CBF}(\mb{x})$. Furthermore, we see that the constraint holds with equality, such that the control action deviates from $\mb{k}_{\rm n}$ only until the prescribed safety requirement is met (and not further).

To show the function $\mb{k}_{\rm QP}$ is continuous, let us first define a function ${\psi:\R^n\to\R}$ as
\begin{equation}
    \psi(\mb{x}) = L_{\mb{f}}h(\mb{x}) + L_{\mb{g}}h(\mb{x})\mb{k}_{\rm n}(\mb{x}) + \alpha(h(\mb{x})).
\end{equation}
As $h$ is continuously differentiable, and $\mb{f}, \mb{g}$, and $\alpha$ are continuous, we may conclude that the function $\psi$ is continuous. Consider an arbitrary state ${\mb{x}\in\R^n}$ such that ${\psi(\mb{x})>0}$, noting that we may have ${L_{\mb{g}}h(\mb{x}) = \mb{0}}$. By \eqref{eqn:case1outcome}, we have ${\mb{k}_{\rm QP}(\mb{x})=\mb{k}_{\rm n}(\mb{x})}$. By continuity of $\psi$, we may conclude that there exists ${\delta>0}$ such that ${\psi(\mb{y})>0}$ for all ${\mb{y}\in B_{\delta}(\mb{x})}$ (the open ball of radius $\delta$ centered at $\mb{x}$). By \eqref{eqn:case1outcome} we then have that ${\mb{k}_{\rm QP}(\mb{y}) = \mb{k}_{\rm n}(\mb{y})}$ for all ${\mb{y}\in B_{\delta}(\mb{x})}$. As $\mb{k}_{\rm n}$ is continuous, we may conclude that $\mb{k}_{\rm QP}$ is continuous at $\mb{x}$. 

Next consider an arbitrary state ${\mb{x}\in\R^n}$ such that ${\psi(\mb{x}) < 0}$, noting that we must have ${L_{\mb{g}}h(\mb{x}) \neq \mb{0}}$ at this state. By \eqref{eqn:case2outcome} we have
\begin{equation}
\label{eqn:kqpxactive}
    \mb{k}_{\rm QP}(\mb{x}) = \mb{k}_{\rm n}(\mb{x})-  \frac{\psi(\mb{x})}{\Vert L_{\mb{g}}h(\mb{x})\Vert_2^2}L_{\mb{g}}h(\mb{x})^\top.
\end{equation}
By the continuity of $L_{\mb{g}}h$ and $\psi$, we may conclude that there exists a ${\delta>0}$ such that ${L_{\mb{g}}h(\mb{y})\neq \mb{0}}$ and ${\psi(\mb{y})<0}$ for all ${\mb{y}\in B_{\delta}(\mb{x})}$. We then have
\begin{equation}
\label{eqn:kqpyactive}
    \mb{k}_{\rm QP}(\mb{y}) = \mb{k}_{\rm n}(\mb{y})-  \frac{\psi(\mb{y})}{\Vert L_{\mb{g}}h(\mb{y})\Vert_2^2}L_{\mb{g}}h(\mb{y})^\top.
\end{equation}
for all ${\mb{y}\in B_{\delta}(\mb{x})}$. As $L_{\mb{g}}h$, $\mb{k}_{\rm n}$, and $\psi$ are continuous and ${L_{\mb{g}}h(\mb{x}) \neq \mb{0}}$, we may conclude that $\mb{k}_{\rm QP}$ is continuous at $\mb{x}$.

Lastly, let us consider a state ${\mb{x}\in\R^n}$ such that ${\psi(\mb{x}) = 0}$, noting that we must have ${L_{\mb{g}}h(\mb{x}) \neq \mb{0}}$ at this state. By \eqref{eqn:case2outcome} and the fact ${\psi(\mb{x}) = 0}$, we have that ${\mb{k}_{\rm QP}(\mb{x}) = \mb{k}_{\rm n}(\mb{x})}$. Let ${\epsilon>0}$ be arbitrary. By the continuity of $L_{\mb{g}}h$, we may conclude that there exists ${\delta_1>0}$ such that ${L_{\mb{g}}h(\mb{y}) \neq \mb{0}}$ for all ${\mb{y}\in B_{\delta_1}(\mb{x})}$. Let ${\mb{y}\in B_{\delta_1}(\mb{x})}$ be such that ${\psi(\mb{y}) > 0}$. As before, we then have ${\mb{k}_{\rm QP}(\mb{y}) = \mb{k}_{\rm n}(\mb{y})}$. By the continuity of $\mb{k}_{\rm n}$, there exists a ${\delta_2>0}$ with ${\delta_2 < \delta_1}$ such that if ${\mb{y}\in B_{\delta_2}(\mb{x})}$ and ${\psi(\mb{y})>0}$, then
\begin{equation}
    \Vert\mb{k}_{\rm QP}(\mb{y})-\mb{k}_{\rm QP}(\mb{x}) \Vert_2 = \Vert\mb{k}_{\rm n}(\mb{y})-\mb{k}_{\rm n}(\mb{x}) \Vert_2 < \epsilon.
\end{equation}
Next let ${\mb{y}\in B_{\delta_1}}$ be such that ${\psi(\mb{y})\leq 0}$, such that $\mb{k}_{\rm QP}(\mb{y})$ is given in \eqref{eqn:kqpyactive}. Although ${\mb{k}_{\rm QP}(\mb{x}) = \mb{k}_{\rm n}(\mb{x})}$, we may use the fact that ${\psi(\mb{x}) = 0}$ to write $\mb{k}_{\rm QP}(\mb{x})$ as in \eqref{eqn:kqpxactive}. By the continuity of $\mb{k}_n$, $\psi$, and $L_{\mb{g}}h$, there exists a ${\delta_3>0}$ with ${\delta_3 < \delta_1}$ such that if ${\mb{y}\in B_{\delta_3}(\mb{x})}$ and ${\psi(\mb{y})\leq 0}$, then
\begin{equation}
    \Vert\mb{k}_{\rm n}(\mb{y})-\mb{k}_{\rm n}(\mb{x}) \Vert_2 < \frac{\epsilon}{2},
\end{equation}
and
\begin{equation}
    \left\Vert \frac{\psi(\mb{y})}{\Vert L_{\mb{g}}h(\mb{y})\Vert_2^2}L_{\mb{g}}h(\mb{y})^\top - \frac{\psi(\mb{x})}{\Vert L_{\mb{g}}h(\mb{x})\Vert_2^2}L_{\mb{g}}h(\mb{x})^\top \right\Vert_2 < \frac{\epsilon}{2}.
\end{equation}
Therefore we have that if ${\mb{y}\in B_{\delta_3}(\mb{x})}$ and ${\psi(\mb{y})\leq 0}$, then:
\begin{equation}
    \Vert\mb{k}_{\rm QP}(\mb{y})-\mb{k}_{\rm QP}(\mb{x}) \Vert_2 < \frac{\epsilon}{2}+\frac{\epsilon}{2} = \epsilon.
\end{equation}
Taking ${\delta = \min\{\delta_2,\delta_3\}}$, we have that ${\mb{y}\in B_{\delta}(\mb{x})}$ implies:
\begin{equation}
    \Vert\mb{k}_{\rm QP}(\mb{y})-\mb{k}_{\rm QP}(\mb{x}) \Vert_2 < \epsilon,
\end{equation}
proving $\mb{k}_{\rm QP}$ is continuous at $\mb{x}$. As we considered the three cases that ${\psi(\mb{x}) > 0}$, ${\psi(\mb{x})<}0$, and ${\psi(\mb{x}) = 0}$, we have shown the function $\mb{k}_{\rm QP}$ is continuous.

% first consider an $\mb{x}\in\R^n$ such that $L_{\mb{g}}h(\mb{x}) = \mb{0}$. By the CBF condition \eqref{eq:cbf_alt}, we have that:
% \begin{equation}
%     L_{\mb{f}}h(\mb{x}) + \alpha(h(\mb{x})) > 0.
% \end{equation}
% As $h$ is continuously differentiable and $\mb{f}$ and $\alpha$ are continuous, we have that $L_{\mb{f}}h$ and $\alpha\circ h$ are continuous. Thus there exists an open neighborhood $U\subseteq\R^n$ of $\mb{x}$ such that:
% \begin{equation}
%       L_{\mb{f}}h(\mb{y}) + \alpha(h(\mb{y})) > 0. 
% \end{equation}
% for all $\mb{y}\in U$. 

\end{proof}


\vspace{-5 mm}
\begin{IEEEbiography}[{\includegraphics[width=1in,height=1.25in,clip,keepaspectratio]{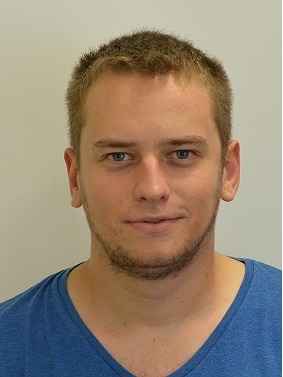}}]{Anil~Alan} received the BSc degree in mechanical engineering from Middle East Technical University, Turkey, in 2012 and the MSc degree in Bilkent University, Turkey, in 2017. He is currently pursuing the PhD degree in mechanical engineering with the University of Michigan, Ann Arbor, MI, USA. His current research interests include control of connected autonomous vehicles, safety-critical control, nonlinear control, vehicle dynamics. 
\end{IEEEbiography}

\vspace{-5 mm}
\begin{IEEEbiography}[{\includegraphics[width=1in,height=1.25in,clip,keepaspectratio]{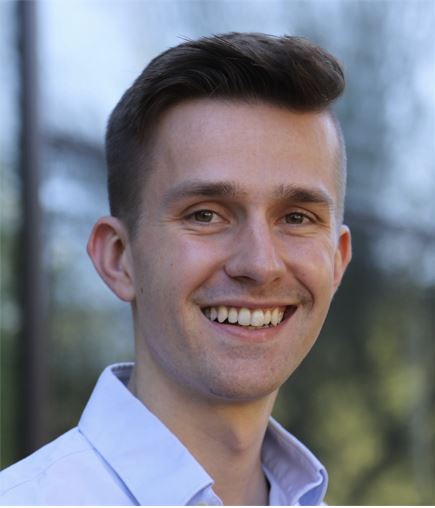}}]{Andrew~J.~Taylor} received the B.S. and M.S. degrees in aerospace engineering from the University of Michigan, Ann Arbor, in 2016 and 2017, respectively. He is currently pursuing a Ph.D. degree at California Institute of Technology in Control and Dynamical Systems. His research interests include safety-critical control for robotic systems and data-driven control techniques for nonlinear systems.
\end{IEEEbiography}

\vspace{-5 mm}
\begin{IEEEbiography}[{\includegraphics[width=1in,height=1.25in,clip,keepaspectratio]{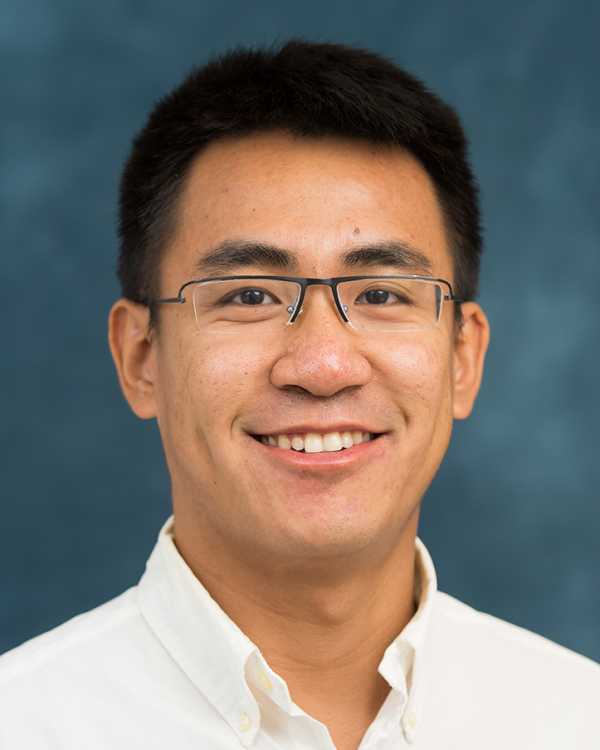}}]{Chaozhe~R.~He} received the BSc degree in applied mathematics from the Beijing University of Aeronautics and Astronautics in 2012, the MSc and PhD in Mechanical Engineering from the University of Michigan, Ann Arbor, USA, in 2015 and 2018 respectively. Dr.~He is with Plus.ai Inc.\ and is working on planning and control algorithm development. His research interests include dynamics and control of connected automated vehicles, optimal and nonlinear control theory, and data-driven control.
\end{IEEEbiography}

\vspace{-5 mm}
\begin{IEEEbiography}
[{\includegraphics[width=1in,height=1.25in,clip,keepaspectratio]{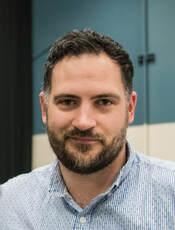}}]
{Aaron D. Ames} is the Bren Professor of Mechanical and Civil Engineering and Control and Dynamical Systems at Caltech. Prior to joining Caltech in 2017, he was an Associate Professor at Georgia Tech in the Woodruff School of Mechanical Engineering and the School of Electrical {\&} Computer Engineering. He received a B.S. in Mechanical Engineering and a B.A. in Mathematics from the University of St. Thomas in 2001, and he received a M.A. in Mathematics and a Ph.D. in Electrical Engineering and Computer Sciences from UC Berkeley in 2006. He served as a Postdoctoral Scholar in Control and Dynamical Systems at Caltech from 2006 to 2008, and began his faculty career at Texas A{\&}M University in 2008. At UC Berkeley, he was the recipient of the 2005 Leon O. Chua Award for achievement in nonlinear science and the 2006 Bernard Friedman Memorial Prize in Applied Mathematics, and he received the NSF CAREER award in 2010, the 2015 Donald P. Eckman Award, and the 2019 IEEE CSS Antonio Ruberti Young Researcher Prize.  His research interests span the areas of robotics, nonlinear, safety-critical control and hybrid systems, with a special focus on applications to dynamic robots -— both formally and through experimental validation.
\end{IEEEbiography}

\vspace{-5 mm}
\begin{IEEEbiography}
[{\includegraphics[width=1in,height=1.25in,clip,keepaspectratio]{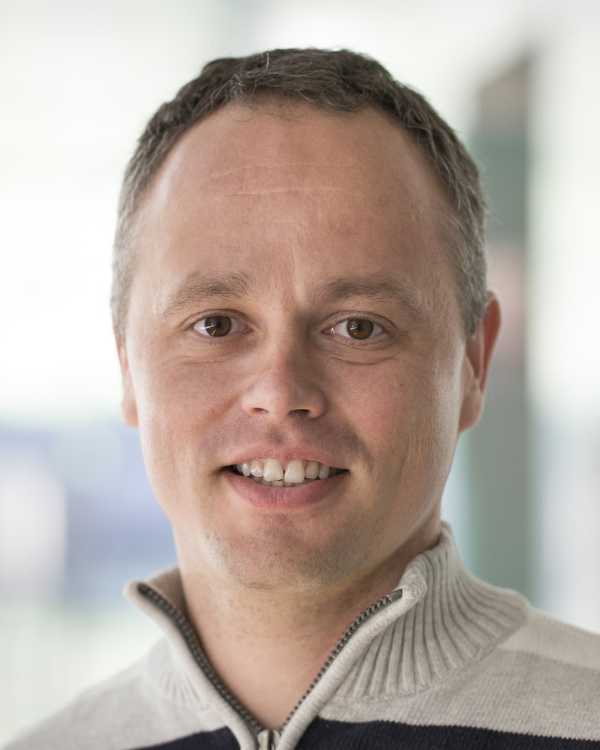}}]
{G{\'{a}}bor Orosz} received the M.Sc. degree in Engineering Physics from the Budapest University of Technology, Hungary, in 2002 and the Ph.D. degree in Engineering Mathematics from University of Bristol, UK, in 2006. He held postdoctoral positions at the University of Exeter, UK, and at the University of California, Santa Barbara. In 2010, he joined the University of Michigan, Ann Arbor where he is currently an Associate Professor in Mechanical Engineering and in Civil and Environmental Engineering. His research interests include nonlinear dynamics and control, time delay systems, and reinforcement learning with applications to connected and automated vehicles, traffic flow, and biological networks.
\end{IEEEbiography}

\end{document}